\documentclass[12pt,fleqn]{article}
\usepackage{lscape}
\usepackage{amsfonts,ulem,url}
\usepackage{xr-hyper}
\usepackage{amssymb}
\usepackage{amsmath}
\usepackage{amstext,verbatim,graphicx,ifthen,multirow,amsthm,algorithm2e, mathtools}
\usepackage{bm}
\usepackage[round]{natbib}
\usepackage[bf]{caption}
\usepackage{subcaption}
\usepackage{setspace}
\usepackage{enumitem}
\usepackage{placeins}
\usepackage{xcolor}

\usepackage{rotating}
\usepackage{adjustbox}

\usepackage{appendix}

\newcommand{\trace}[1]{\textbf{tr}\left(#1\right)}
\newcommand{\wrob}{\omega^{rob}}
\newcommand{\wrobn}{\tilde\omega^{rob}}
\newcommand{\wlim}{\omega^{\star}}
\newcommand{\wlimn}{\tilde\omega^{\star}}
\newcommand{\wdet}{\omega^{det}}
\newcommand{\wdetn}{\tilde\omega^{det}}

\renewcommand{\thepage}{}

\usepackage{float}

\newtheorem{assumption}{Assumption}[section]
\newtheorem{theorem}{Theorem}
\newtheorem{lemma}{Lemma}
\newtheorem{prop}{Proposition}
\newtheorem{corollary}{Corollary}

\theoremstyle{definition}

\numberwithin{equation}{section}

\usepackage{apptools}
\AtAppendix{\counterwithin{lemma}{section}}
\AtAppendix{\counterwithin{theorem}{section}}
\AtAppendix{\counterwithin{corollary}{section}}

\DeclareMathOperator*{\argmin}{arg\,min}

\newcommand{\appendixpagenumbering}{
  \break
  \pagenumbering{arabic}
  \renewcommand{\thepage}{\thesection-\arabic{page}}
}

\graphicspath{{./figures/}}

\usepackage[pdftex,colorlinks=true,unicode,bookmarksnumbered=false, hyperfootnotes=true]{hyperref}
\hypersetup{citecolor = blue}

\begin{document}

\title{\textbf{On Policy Evaluation With Aggregate Time-Series Instruments}\thanks{{\small This paper benefited greatly from our discussions with Manuel Arellano, St\'{e}phane Bonhomme, David Hirshberg, Guido Imbens, Dmitry Mukhin, Emi Nakamura, and Jon Steinsson. We also want to thank Isaiah Andrews, Kirill Borusyak, Peter Hull, Xavier Jaravel, Alexey Makarin, Anna Mikusheva, Eduardo Morales, Davide Viviano, Jesse Shapiro, and Liyang Sun, as well as seminar participants at Carlos III, CEMFI, CUHK, Harvard, MIT, Northwestern University, UCL, and conference participants at ASSA meeting, NBER Summer Institute 2022 Monetary Economics, and World Congress of Econometric Society for helpful comments and suggestions. Asya Evdokimova and Gleb Kurovskiy provided excellent research assistance. Dmitry Arkhangelsky gratefully acknowledges financial support from the Mar\'{i}a de Maeztu Units of Excellence Programme MDM-2016-0684. Vasily Korovkin acknowledges financial support from the Czech Science Foundation grant (19-25383S) and the European Union's Horizon 2020 research and innovation program under the Marie Sk\l{}odowska-Curie grant agreement No. 870245.}} }
\author{Dmitry Arkhangelsky\thanks{{\small  CEMFI, CEPR, darkhangel@cemfi.es.}} \and Vasily Korovkin\thanks{{\small Universitat Pompeu Fabra, Barcelona School of Economics, CEPR, vasily.korovkin@upf.edu.}} }
\date{
\today }
\maketitle\thispagestyle{empty}

\begin{abstract}
\singlespacing
We develop an estimator for applications where the variable of interest is endogenous and researchers have access to aggregate instruments. Our method addresses the critical identification challenge -- unobserved confounding, which renders conventional estimators invalid. Our proposal relies on a new data-driven aggregation scheme that eliminates the unobserved confounders. We illustrate the advantages of our algorithm using data from \cite{nakamura2014fiscal} study of local fiscal multipliers. We introduce a finite population model with aggregate uncertainty to analyze our estimator. We establish conditions for consistency and asymptotic normality and show how to use our estimator to conduct valid inference.
\end{abstract}

\begin{footnotesize}

\end{footnotesize}

\baselineskip=20pt
\setcounter{page}{1}
\renewcommand{\thepage}{\arabic{page}}
\renewcommand{\theequation}{\arabic{section}.\arabic{equation}}

\section{Introduction} \label{sec:intro}

Aggregate events are often used to identify policy-relevant quantities. A classic example is \cite{card1990impact}, where a single aggregate policy change -- the Mariel Boatlift of 1980 -- allows the author to quantify the effect of immigration on the Miami labor market by comparing it to unaffected locations before and after the shock. Since then, this difference in differences (DiD) strategy has become the standard tool for empirical analysis in similar settings and spurred a vast methodological literature; see \citet{arkhangelsky2023causal}  for a recent survey.  However, many problems in empirical economics do not conform with the standard DiD setting because aggregate events occur frequently and affect all units simultaneously. Developing methodology for such applications is the primary goal of this paper. 

We focus on environments where the aggregate variation is used as an instrument for unit-specific policy variables. In applications we have in mind the variables of interest are often equilibrium objects, and external instruments are necessary to solve the endogeneity problem. We develop a new estimator for such applications, establish its statistical properties, and show its superiority to the benchmark that dominates current empirical practice -- the two-stage least-squares (TSLS) regression with unit and time fixed effects.


Formally, suppose we observe outcome $Y_{it}$ and treatment $W_{it}$ for $n$ units over $T$ periods. We also have access to aggregate instrument $Z_t$, and our goal is to quantify the contemporaneous effect of  $W_{it}$ on $Y_{it}$. For example, in \cite{nakamura2014fiscal}, $W_{it}$ is the military procurement spending in state $i$ and year $t$, $Y_{it}$ is a measure of state-level growth, $Z_t$ is the national military spending in the US, and the contemporaneous effect describes a local fiscal multiplier. It is natural to view both $W_{it}$ and $Y_{it}$ as jointly determined in equilibrium, forcing the authors to address the endogeneity problem to identify the multiplier. Having access to an aggregate instrument suggests an easy solution: as long as $Z_t$ satisfies conventional assumptions of \cite{imbens1994identification}, we can establish a causal link between $Y_{it}$ and $W_{it}$ by constructing an instrumental variables (IV) estimator separately for each unit $i$ and reporting a summary of these estimators, e.g., the average. 

In practice, however, this naive approach is unlikely to produce a credible answer. Researchers suspect that $Z_t$ correlates with other unobserved aggregate variables that affect the outcomes, and thus, the assumptions of \cite{imbens1994identification} are not satisfied. For example, in \cite{nakamura2014fiscal}, the authors are concerned that aggregate military spending is correlated with fiscal and monetary policies, directly affecting local outcomes. Similar problems are evident in simpler environments: a comparison of Miami's labor market before and after the Mariel Boatlift is likely to be corrupted by aggregate confounders. 

To address these concerns, researchers are adopting a different strategy. In particular, the following equation is estimated by the TSLS:
\begin{equation}\label{eq:TSLS_intro}
Y_{it} = \alpha_{i} + \mu_t + \tau W_{it} + \epsilon_{it},
\end{equation}
 using $D_iZ_t$ as an instrument.\footnote{Applications abound from different domains. See, e.g., studies of the effect of Airbnb on the housing market \cite{barron2021effect}, plantation owners' power on wages and incarceration rates \cite{dippel2020outside}, commodities prices on violent conflict \cite{dube2013commodity}, dams on poverty and development \cite{duflo2007dams}, housing wealth effects \cite{guren2018housing}, local fiscal multipliers \cite{nakamura2014fiscal}, foreign aid on conflict \cite{nunn2014us} and a few others.} 
 Here, $D_i$ is a measure of ``exposure'' of unit $i$ to the aggregate variable, and $\tau$ is the parameter of interest. The critical part of equation \eqref{eq:TSLS_intro} are time-fixed effects that are meant to capture the unobserved aggregate confounders. To understand the mechanics of this estimator, suppose we have only two units $i  =1,2$, and we know that the first unit is affected by $Z_t$, $D_1 =1$, while the second one is unaffected at all, $D_2 =0$. Then, by looking at differences across units, we eliminate the aggregate confounder as long as it affects both units in the same way. This procedure is a natural extension of the conventional DiD strategy, and it is precisely what the TSLS estimation of \eqref{eq:TSLS_intro} amounts to for the case with two units. Arguably, not much else can be done in this setting. 

The situation changes once we have access to multiple units. In this case, the TSLS estimator first averages units with high and low values of $D_i$ and then takes the difference between the resulting averages. This particular aggregation scheme is valid and can even be statistically efficient as long as we believe that the unobserved confounder affects all units in the same way. While unavoidable in the case of two units, this assumption becomes increasingly dissatisfying once we have multiple units. Researchers recognize this threat and commonly view it as the main danger to the validity of the TSLS identification strategy (e.g., \citealp{chodorowstock,guren2018housing}). The estimator we propose directly addresses this threat. 

To construct our estimator, we reserve a part of the sample to learn weights $\wrob_i$, which we then use to aggregate the rest of the data and to build a single IV estimator. To produce $\wrob_i$,  we first project out the effect of the aggregate instrument and use residuals to construct a combination of units with higher values of $D_i$, which resembles a combination of units with lower values of $D_i$. The identification assumption that justifies this approach is the existence of a particular aggregation scheme (unit-specific weights), which eliminates the unobserved confounding when applied to the data. This condition is valid whenever $D_i$ cannot be fully explained by systematic unit-level variation in outcomes, i.e., a no-multicollinearity restriction.  We show that this weak restriction holds in various models, all of which generalize the one that justifies the TSLS estimation. A similar identification assumption is behind the Synthetic DiD (SDiD) estimator introduced in \cite{arkhangelsky2019synthetic} (see also the recent analysis in \citealp{imbens2023identification}). The development of the new method that extends the SC ideas to settings with aggregate instruments and thus vastly expands the applicability of the SC approach constitutes the first major contribution of this paper.

The second major contribution lies in the formal framework we use to analyze our method and the statistical properties we establish within this framework. In particular, we combine insights from two strands of the literature: the one on quasi-experimental designs and the one on the properties of the SC-type estimators. 

We consider a finite population model where the sole source of uncertainty comes from aggregate variables: the observed instrument and unobserved confounders. This approach takes its roots in the experimental literature \citep[e.g.][]{neyman1923applications,rubin1977assignment}, but it has also been successfully applied to observational data with the recent methodological literature emphasizing the importance of this framework both for estimation \citep[e.g.][]{borusyak2022quasi,borusyak2023nonrandom} and inference \citep[e.g.][]{abadie2020sampling,abadie2023should,adao2019shift}. The key new ingredient we bring to this setup is the presence of unobserved confounders, which we explicitly model.  We expect this idea to be useful more broadly because confounding is a natural threat in all observational studies. The finite population framework is especially appealing in our setting for two reasons. First, it allows us to focus explicitly on the key identifying variation, which comes from the aggregate variables. Second, in many relevant empirical applications, it is natural to treat the observed units as the whole population rather than a sample. For example, this is the case when units represent geographic entities, e.g., states or counties. 

We use this framework to establish the statistical properties of the new estimator. We analyze it in a high-dimensional regime where $n$ is similar or larger than $T$. In many applications of interest, $n$ and $T$ are comparable, making this regime natural. We prove that our algorithm delivers consistent and $\sqrt{T}$-convergent estimators, even in the presence of aggregate confounding.  We also show how to use our method to conduct valid inference as long as there is enough variation in the baseline outcomes. To derive these results, we build on recent insights in SC literature \citep[e.g.][]{ arkhangelsky2019synthetic,ferman2021synthetic}, and connect empirical weights $\omega_i^{rob}$ to a specific deterministic analog. However, our analysis is necessarily different for two reasons. First, the policy variable in our environment is endogenous, and the construction of weights relies on aggregate instruments, which are not present in the conventional SC literature. Second, the only source of uncertainty in our model comes from the aggregate variation. To the best of our knowledge, this paper is the first to derive any statistical properties of SC-type estimators in such models. More recently, \cite{imbens2023identification} follow a related path when analyzing a class of SC estimators. 

The TSLS estimator is widely adopted for a reason -- the transparency and flexibility of the method compensate for the underlying strong assumptions. Moreover, practitioners might expect \eqref{eq:TSLS_intro} to provide a good approximation in many applications, rendering confounding concerns less relevant. To investigate this, 
we demonstrate the benefits of our approach using the data from \cite{nakamura2014fiscal}. First, we reevaluate their study using our method and find fiscal multipliers larger in magnitude than the original ones. One of the goals of this exercise is to demonstrate to the applied-minded reader that our method is highly transparent and, by using it, one can gain insights about the TSLS estimator itself. We then construct a simulation that mimics the properties of the original dataset of \cite{nakamura2014fiscal}. We use this simulation to show that our estimator remains competitive in simple designs, can outperform the TSLS even when the latter is consistent, and is a clear winner in more realistic situations with unobserved aggregate confounders. An important practical message that we infer from these results is that there are reasons to prefer our method even in applications where the TSLS is a valid option. 

Our approach addresses the major shortcoming of the conventional TSLS estimation of (\ref{eq:TSLS_intro}): its invalidity in the presence of unobserved aggregate confounders correlated with the instrument. In practice, there are other reasons why equation (\ref{eq:TSLS_intro}) can be problematic, for example, nonlinearity or dynamic treatment effects. In these cases, the TSLS might be the wrong tool to start with, and by extension, the same holds for our method. As a result, researchers should use our estimator in applications where the TSLS can be a priori reasonable, but they are worried about potential aggregate confounders. These concerns are not unique to our setting and equally apply to methods that aim to relax the DiD assumptions; see \cite{arkhangelsky2023causal} for a discussion. 

Our setup is related to the literature on invalid instruments (e.g., \citealp{andrews1999consistent,kolesar2015identification,lewbel2012using,windmeijer2019use}). In contrast to this literature, we do not need to assume that a sufficient number of instruments or their known combination, such as average, is valid. Instead, we focus on situations with unobserved aggregate confounders, which allows us to impose a factor structure on the omitted-variable bias. This structure connects our setup to the literature on interactive fixed effects (e.g., \citealp{bai2009panel,moon2015linear}), which focuses on exogenous regressors and thus is not directly relevant to applications we have in mind. An important exception is \cite{moon2017dynamic}, which considers endogenous regressors but does not allow for low-rank instrumental variables we are interested in. More substantially, we depart from this literature by changing both the estimator and the approach to the analysis. First, we do not estimate the underlying factors but look for the appropriate aggregation scheme. Recent theoretical results in the SC literature illustrate the benefits of this approach even in environments where direct factor-based estimators are available \citep[e.g., see the discussion in][]{imbens2023identification}. Second, we establish the properties of our estimator in the finite population framework with aggregate uncertainty.

Our model is also related to the recent econometric literature on shift-share designs (\citealp{adao2019shift,borusyak2022quasi,goldsmith2020bartik,jaeger2018shift}). First, our finite population analysis is motivated by the approach in \citet{adao2019shift} and the related earlier work by \cite{abadie2023should}. Second, similar to this literature, we consider situations where an instrument has a particular product structure. However, our goal is quite different: we propose and analyze a new estimator, while this literature has focused on the properties of the standard IV estimator under alternative assumptions. Crucially, we relax the exogeneity assumption made in the shift-share literature and allow for unobserved aggregate variables that affect different units differently. 

The paper proceeds as follows: in Section \ref{sec:sec_2}, we discuss the mechanics of TSLS regression (\ref{eq:TSLS_intro}) in more detail, present our algorithm, apply it to \cite{nakamura2014fiscal}, and discuss informally when we expect it to be valid. In Section \ref{sec:formal_an}, we introduce the causal model along with statistical restrictions, and we demonstrate the formal properties of our algorithm. Section \ref{sec:discussion} discusses possible extensions of our algorithm, heterogeneous treatment effects, and connections to the literature on shift-share designs. Section \ref{sec:sim} demonstrates the properties of our estimator in simulations, and Section \ref{sec:concl} concludes.

We use $\mathbb{E}[\cdot]$ and $\mathbb{V}[\cdot]$ to denote expectation and variance operators, respectively. We use $\|\cdot\|_2$ to denote the $l_2$-norm of a vector, and $\| \cdot\|_{op}$ to denote the operator norm of a matrix. We use $\trace{A}$ to denote the trace of a square matrix $A$. For two sequences $a_k$ and $b_k$, we write $a_k \lesssim b_k$ if $\frac{a_k}{b_k}$ is bounded and $a_k \sim b_k$ if $a_k \lesssim b_k$ and $b_k \lesssim a_k$. We use $O_p(1)$ and $o_p(1)$ for sequences of random variables that are bounded in probability and converge to zero in probability, respectively. 

\section{Empirical Example}\label{sec:sec_2}

This section introduces our estimator in the context of an empirical example. We describe the data from \cite{nakamura2014fiscal} and replicate their baseline results. Next, we propose a new estimator and apply it to this dataset. Our estimates are larger in magnitude, though still within the range reported by \cite{nakamura2014fiscal} in various specifications.  We tie our method to a particular econometric model in Section \ref{sec:formal_an}, where we establish its theoretical properties. We also demonstrate the performance of our method in simulations in Section \ref{sec:sim}. 


\subsection{Original Analysis}

In \cite{nakamura2014fiscal}, the authors investigate the relationship between government spending and state GDP growth. They use state data on total military procurement for 1966 through 2006 and combine it with U.S. Bureau of Economic Analysis state GDP and state employment datasets. The authors complement these data with the oil prices data from the St. Louis Federal Reserve’s FRED database, the state-level inflation series constructed by \cite{del1998aggregate}, and their own inflation calculations after 1995.

By estimating the growth-spending relationship, \cite{nakamura2014fiscal} want to capture the open-economy fiscal multiplier. The authors compare different U.S. states and study their reaction to fluctuations in aggregate military spending. They argue that this strategy allows them to control for other aggregate variables (such as monetary policy). It also allows them to account for the potential endogeneity of local procurement spending.

To illustrate their approach, we introduce some notation. For a generic observation -- a state $i$, and a generic period $t$, -- denote per capita output growth in state $i$ from year $t-2$ to $t$ by $Y_{it}$. Similarly, denote two-year growth in per capita military procurement spending in state $i$ and year $t$, normalized by output in year $t-2$ by $W_{it}$. Finally, let $Z_t$ be the change in total national procurement from year $t-2$ to $t$. The final dataset has $n =51$ states and $T =39$ periods.

The main object of interest -- the fiscal multiplier -- is estimated using the TSLS regression (\ref{eq:TSLS_intro}) with $D_iZ_t$ as the instrument. The authors construct $D_i$ by estimating $n$ state-level first-stage OLS coefficients
\begin{equation}\label{eq:un_spec_fs}
W_{it} = \alpha_i^{(w)}  + \pi_i Z_t + u^{(w)}_{it},
\end{equation}
and setting $D_i := \hat \pi_{i}$. As expected, for 49 states, $D_i$ is positive, with Mississippi and North Dakota being the exceptions. In the analysis below, we drop these states and the state of Alaska, where the output growth is exceptionally responsive to the changes in national procurement. This choice leaves us with $n=48$ states.\footnote{Results for the whole sample are similar in magnitude but are estimated less precisely. We report the full sample results in Appendix \ref{ap:rob_check}.} 

\begin{figure}[t] 
    \begin{center}
    \caption{Reduced-Form and First-Stage Coefficients for \cite{nakamura2014fiscal} Data} \label{fig:cs_graph} 
     \begin{subfigure}[b]{0.49\textwidth}
         \caption*{\textbf{Panel A:} Nakamura and Steinsson weights}
         \centering
         \includegraphics[width=\textwidth]{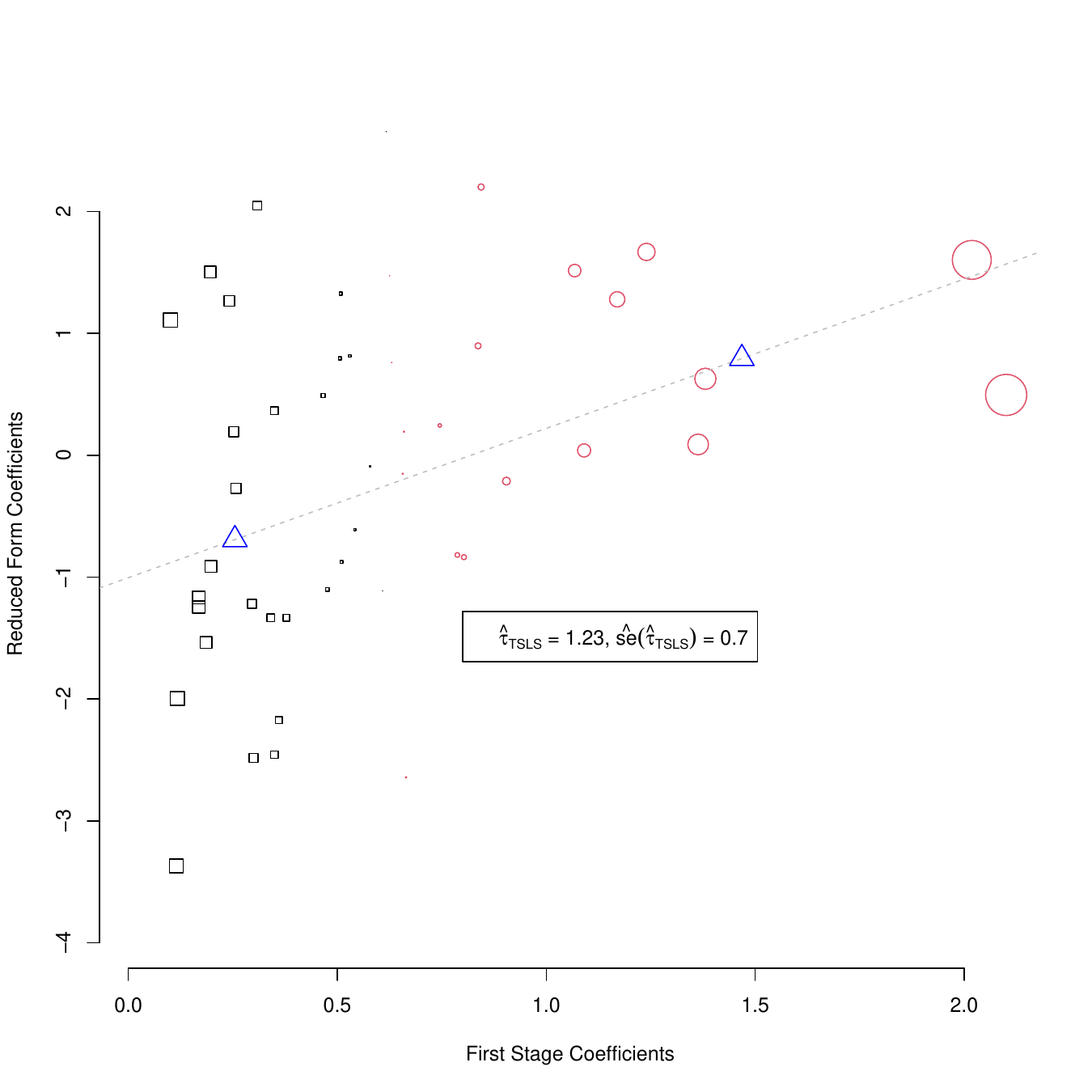}
     \end{subfigure}
     \begin{subfigure}[b]{0.49\textwidth}
        \caption*{\textbf{Panel B:} Robust weights
        }
         \centering
         \includegraphics[width=\textwidth]{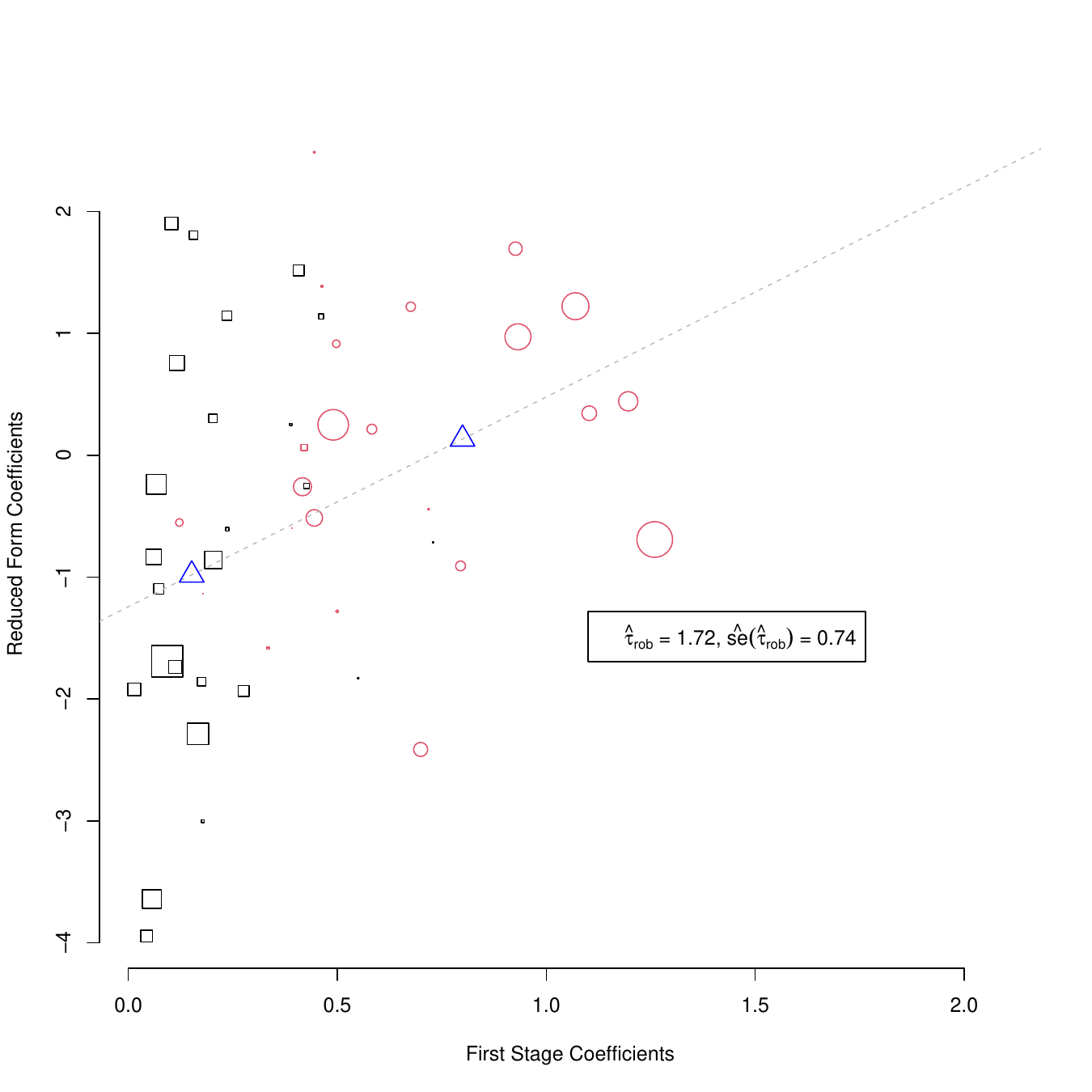}
     \end{subfigure}
     \end{center}
    \renewcommand{\baselinestretch}{0.7}
    \footnotesize{\textit{Notes}: Shape sizes reflect the absolute value of weights; negative weights are printed in black squares, and positive are in red circles. Blue triangles are centers of mass for negative and positive weights. 
Panel A presents the results using the whole period of 1968 to 2006 for $n=48$ states. Panel B shows the results from our estimation algorithm. Under our data-splitting procedure, Panel B reports the results for 1978 to 2006, as we use the first third of the data for weight estimation.}
\end{figure}

The TSLS estimator for $\tau$ is equal to 
\begin{equation*}
 \hat \tau_{TSLS} = \frac{\sum_{i\le n} \sum_{t\le T}Y_{it}(Z_t - \frac{1}{T}\sum_{l\le T}Z_l)(D_i - \frac{1}{n}\sum_{j\le n}D_j) }{\sum_{i\le n} \sum_{t\le T}W_{it}(Z_t - \frac{1}{T}\sum_{l\le T}Z_l)(D_i  - \frac{1}{n}\sum_{j\le n}D_j) },    
\end{equation*}
and can be interpreted in two different ways. First, it is a combination of the state-level coefficients,
\begin{equation}\label{eq:cs_int}
    \hat \tau_{TSLS} = \frac{\sum_{i\le n}\hat \delta_{i}(D_i -  \frac{1}{n}\sum_{j\le n}D_j) }{\sum_{i\le n}\hat \pi_{i}(D_i - \frac{1}{n}\sum_{j\le n}D_j) },
\end{equation}
where $\hat \delta_{i}$ is the state-specific reduced-form coefficient, analogous to $\hat \pi_{i}$. Panel A of Figure \ref{fig:cs_graph} plots $\{(\hat \pi_i,\hat\delta_i)\}_{i\le n}$, with the size of each point being proportional to $\left|D_i -  \frac{1}{n}\sum_{j\le n}D_j\right|$, and the colors reflecting the sign. Representation (\ref{eq:cs_int}) shows that $\hat \tau_{TSLS}$ is equal to the slope of the line that connects the centers of mass of points with negative and positive weights (blue triangles). We see that the coefficients vary a lot, but the association is positive, which results in $\hat \tau_{TSLS} = 1.23$. 

\begin{figure}[t!]
        \begin{center}
        \caption{Aggregate Time-Series for \cite{nakamura2014fiscal} Data} \label{fig:ts_graph}
     \begin{subfigure}{0.85\textwidth}
         \centering
         \caption*{\textbf{Panel A:} Aggregation Over $n = 48$ States with Original Weights}
         \includegraphics[width=\textwidth]{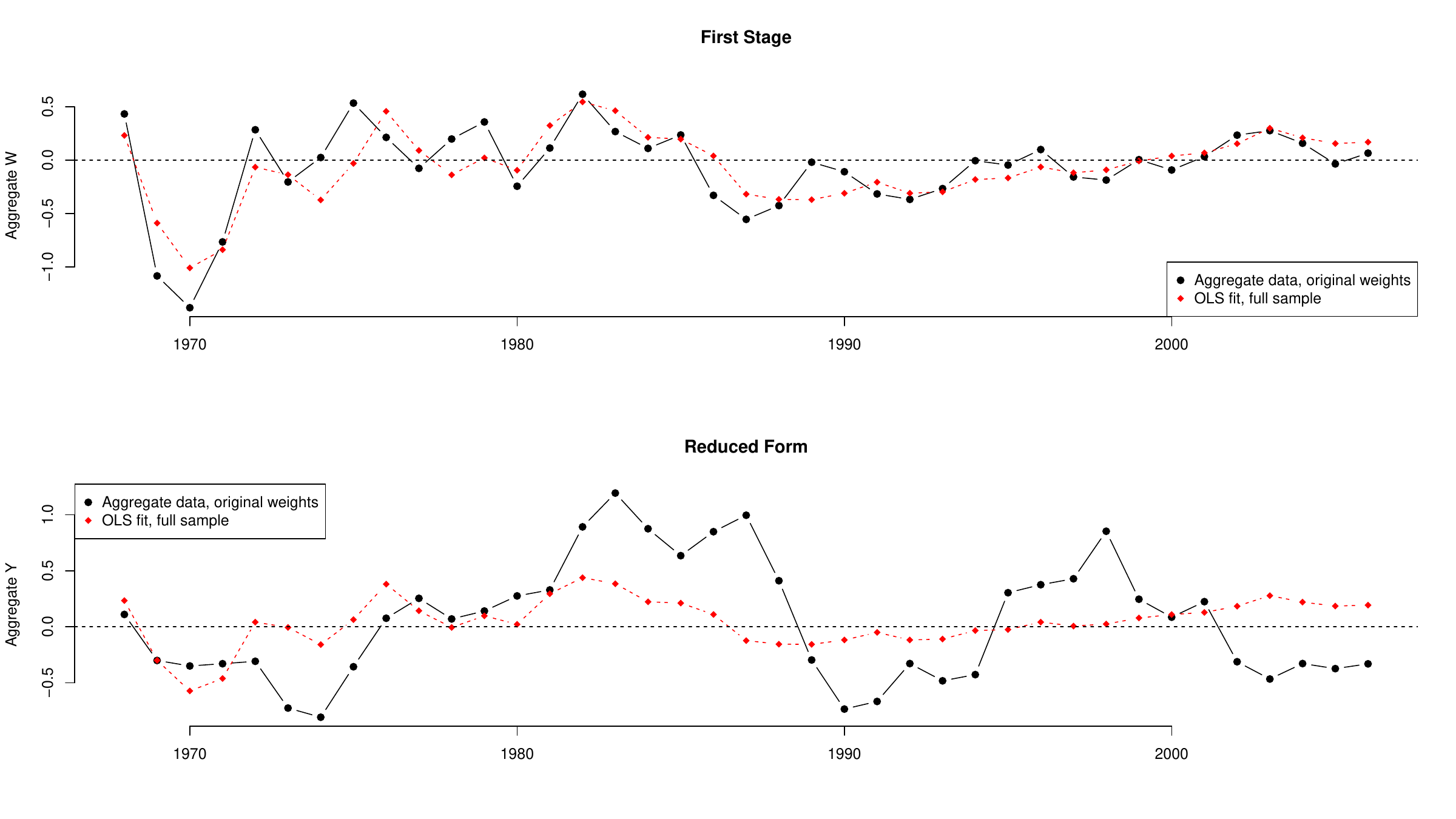}
     \end{subfigure}
     \begin{subfigure}{0.85\textwidth}
         \centering
        \caption*{\textbf{Panel B:} Aggregation Over $n = 48$ States With Robust Weights}
         \includegraphics[width=\textwidth]{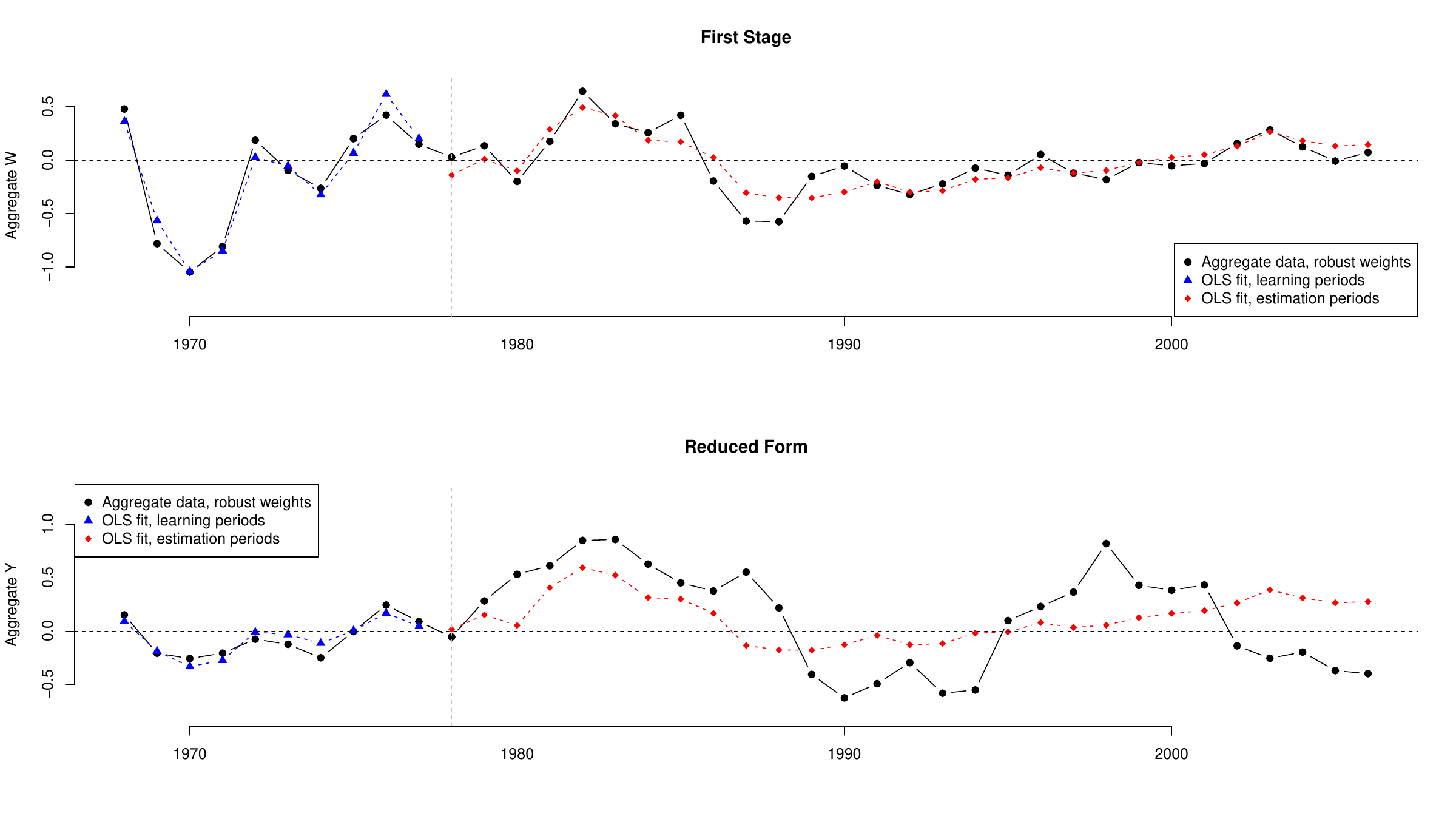}
     \end{subfigure}
     \end{center}
    \renewcommand{\baselinestretch}{0.7}
    \footnotesize{\textit{Notes}: Solid lines represent aggregate data for different weights; dashed lines represent OLS predictions of the aggregate data with the instrument. The mean absolute value of weights is scaled to $1$.}
\end{figure}

Alternatively, $\hat \tau_{TSLS}$ is numerically equal to an IV estimator for the aggregate model
\begin{equation}\label{eq:ts_rep_tsls}
Y_t = \alpha + \tau W_t + \epsilon_t,
\end{equation}
where $Y_t := \frac{1}{n}\sum_{i\le n}Y_{it}(D_i -  \frac{1}{n}\sum_{j\le n}D_j)$ and $W_t :=\frac{1}{n}\sum_{i\le n}W_{it}(D_i - \frac{1}{n}\sum_{j\le n}D_j)$, and we use $Z_t$ as an instrument. Panel A of Figure \ref{fig:ts_graph} shows the time-series interpretation of $\hat \tau_{TSLS}$, plotting the aggregate data $Y_t$ and $W_t$ vis-a-vis the OLS fit based on $Z_t$. Using the residuals from these regressions, we produce the robust standard error estimate for $\hat \tau_{TSLS}$, resulting in $\hat {\text{s.e.}}(\hat \tau_{TSLS}) = 0.70$. This estimate is based on using Algorithm \ref{alg:inf}, which we explain in more detail in Section \ref{sec:stat_prop}. The estimates and standard errors are different from the baseline specification in \cite{nakamura2014fiscal} ($1.43$ and $0.36$, respectively) because we drop the three states and compute a standard error based on aggregate uncertainty instead of clustering at the unit level.\footnote{To construct this error, we use the automatic model selection  \textbf{ARIMA} package in \textbf{R} to estimate the model for $Z_t$. See Section \ref{sec:stat_prop} for details.}

\subsection{New Estimator} 

Representation (\ref{eq:cs_int}) shows that $\hat \tau_{TSLS}$ is a weighted combination of the unit-level coefficients $\{(\hat \pi_i,\hat\delta_i)\}_{i \le n}$ with weights proportional to $D_i -  \frac{1}{n}\sum_{j\le n}D_j$. These weights sum up to zero, meaning that the TSLS estimator subtracts the weighted average of the units with relatively large exposures from those with relatively small ones. This action is reflected in Panel A of Figure \ref{fig:cs_graph}, where we use different colors for states with positive and negative weights.

This particular aggregation scheme is a consequence of the two-way model in \eqref{eq:TSLS_intro}. We eliminate the time fixed effects by averaging over the cross-sectional dimension with the weights that sum up to zero. In applications, these effects capture unobserved aggregate confounders potentially correlated with $Z_t$. For example, in \cite{nakamura2014fiscal} time fixed effects are meant to capture other policy variables that are likely correlated with national procurement. 

This strategy is appropriate only if potential confounders affect all cross-sectional units in the same way (or, at least, in a way that is unrelated to $D_i$). Thus, the main threat to the validity of the TSLS estimator is the presence of aggregate variable $H_t$ with heterogeneous coefficients. A natural model that reflects this is given by
\begin{equation}\label{eq:exm_factor}
    Y_{it} = \alpha_i + \mu_t +\tau W_{it} + \theta_i H_t +\epsilon_{it},
\end{equation}
where $\epsilon_{it}$ is assumed to be uncorrelated with $D_i$ and $Z_t$. As long as $\theta_i$ is correlated with $D_i$ and $H_t$ is correlated with $Z_t$, the TSLS estimator suffers from the omitted-variable bias (OVB) and is invalid. Researchers recognize this threat (e.g., see discussions in \citealp{chodorowstock,guren2018housing}) and address it by including additional aggregate and unit-specific control variables. Of course, in practice, we cannot guarantee that these controls are sufficient to account for all confounders. 

Our estimator complements this strategy by using a more flexible weighting scheme.\footnote{Below, we discuss the basic version of our estimator that does not involve additional controls. We show how to use controls to improve our estimator in Section \ref{sec:covariates}.}  To understand why weighting can help with unobserved confounders, suppose $D_i$ is binary and split all units into two groups accordingly.\footnote{We thank an anonymous referee for suggesting this example.} If the average value of $\theta_i$ varies between these groups, then the TSLS strategy is invalid. However, if there is an overlap in distributions of $\theta_i$ in the two groups, we can correct these differences by reweighting. We cannot follow this strategy directly because $\theta_i$ is unknown. Instead, we can use the observed data to indirectly implement this strategy by searching for weights with specific balancing properties. The algorithm we present next does precisely that.

First, we again compute the state-level first-stage and reduced-form coefficients by estimating the analogs of \eqref{eq:un_spec_fs}, but only using data for periods $t =T_0 +1,\dots, T$. Here, $T_0$ is a user-specified parameter, with a default value $T_0 = \ \frac {T}{3}$. We use  $\left\{\left(\hat \pi_i^{post},\hat\delta_i^{post}\right)\right\}_{i\le n}$ to denote the corresponding OLS estimates.  To estimate the effect, we aggregate these coefficients using weights $\wrob_i$,
\begin{equation}\label{eq:estim}
    \hat \tau_{rob} = \frac{\sum_{i\le n}\hat\delta_i^{post}\wrob_i }{\sum_{i\le n}\hat \pi_i^{post}\wrob_i},
\end{equation}
which we define below. Similar to \eqref{eq:ts_rep_tsls}, this estimator is numerically equal to the time-series IV estimator for the equation 
\begin{equation}\label{eq:ts_our}
Y_t^{rob} = \alpha + \tau W_t^{rob} + \epsilon_t,
\end{equation}
where $Y_t^{rob} := \frac{1}{n}\sum_{i\le n}Y_{it}\wrob_i$ and $W_t^{rob} :=\frac{1}{n}\sum_{i\le n}W_{it}\wrob_i$, we use $Z_t$ as an instrument, and estimate (\ref{eq:ts_our}) using data for $t = T_0+1,\dots, T$. Using weights $\wrob_i$ as opposed to $D_i -  \frac{1}{n}\sum_{j\le n}D_j$ is the main conceptual difference between our estimator and the TSLS. Analogously to $\hat \tau_{TSLS}$, one can compute $\hat \tau_{rob}$ by estimating \eqref{eq:TSLS_intro} by the TSLS for periods $t = T_0+1,\dots, T$, using $\wrob_i Z_t$ as the instrument. 

We construct the weights $\wrob_i$ using the first $T_0$ periods. As discussed before, we want to make units with high values of $D_i$ look similar on average to those with low values of $D_i$. To this end, we residualize the data for each unit with respect to $Z_t$ and look for such a combination. We achieve this goal by solving a quadratic optimization problem:
\begin{equation}\label{eq:opt_unit}
\begin{aligned}
&(\wrob,\hat \eta_{0}^{(w)}, \hat \eta_{z}^{(w)},\hat \eta_{0}^{(y)}, \hat \eta_{z}^{(y)}) = \argmin_{\{w,\eta_{0}^{(w)},\eta_{z}^{(w)},\eta_{0}^{(y)}, \eta_{z}^{(y)}\}} \Bigg{\{}  \frac{\zeta^2\|w\|_2^2}{nT_0} + \\
&\frac{\frac{1}{T_0}\sum_{t\le T_0}\left(\frac{1}{n} \sum_{i\le n} w_iY_{it} - \eta_{0}^{(y)} - \eta_{z}^{(y)}Z_{t}\right)^2}{\hat\sigma^2_{y}} +\frac{\frac{1}{T_0}\sum_{t\le T_0}\left(\frac{1}{n} \sum_{i\le n} w_iW_{it} - \eta_{0}^{(w)} - \eta_{z}^{(w)} Z_t\right)^2}{\hat \sigma^2_{w}} \Bigg{\}}\\
  &\text{subject to: } 
\frac{1}{n} \sum_{i\le n} w_i D_i = 1,\quad
   \frac{1}{n} \sum_{i\le n} w_i = 0,
\end{aligned}
\end{equation}
where $\zeta^2$ is a user-specified regularization parameter, and 
\begin{equation}\label{eq:emp_scale}
\begin{aligned}
     &\hat\sigma^2_{y}:= \min_{\{\alpha_i,\gamma_i,\mu_t\}_{i,t}}\left\{\frac{\sum_{i\le n, t \le T_0}(Y_{it} - \alpha_i - \mu_t - \gamma_i Z_t)^2}{nT_0}\right\},\\
    &\hat\sigma^2_{w}:= \min_{\{\alpha_i,\gamma_i,\mu_t\}_{i,t}}\left\{\frac{\sum_{i\le n, t \le T_0}(W_{it} - \alpha_i - \mu_t - \gamma_i Z_t)^2}{nT_0}\right\}.
\end{aligned}
\end{equation}
As a default value, we use
\begin{equation}\label{eq:reg_par}
    \zeta := \sqrt{\log(T_0)} \hat \sigma, \quad \hat \sigma := \frac{\max\{ \| \hat \epsilon^{(y)}\|_{op},\|\hat \epsilon^{(w)}\|_{op}\}}{\sqrt{nT_0}},
\end{equation}
where $\hat \epsilon^{(y)}$ and $\hat \epsilon^{(w)}$ are $n\times T_0$ matrices of residuals from regressions in (\ref{eq:emp_scale}). The scaling factor $\hat \sigma$ captures the size of the errors and satisfies $\hat \sigma \lesssim \max\{\hat\sigma_{y},\hat\sigma_{w}\} $. The estimation procedure is summarized in Algorithm \ref{alg:const_weights}.

To implement our method, users need to specify two parameters: the size of the learning period $T_0$ and the amount of regularization $\zeta^2$, and we provide the default values for both. Our theoretical results allow for a range of these parameters and only restrict their asymptotic behavior.
In practice, we expect the method to be more sensitive to the choice of $T_0$ rather than $\zeta^2$. If the variation in $Z_t$ is particularly large in specific periods, then inclusion (or exclusion) of these periods might meaningfully change the solution of the optimization problem. In such situations, we advise users to report the sensitivity of the final answer to $T_0$. One can use all $T$ periods to learn the weights and estimate the coefficient. There are multiple reasons, though, that make this choice less attractive, and we briefly discuss them in Section \ref{sec:limit}.
\RestyleAlgo{boxruled}
\LinesNumbered
\begin{algorithm}[t]
 \KwData{$\{Y_{it},W_{it}\}_{it},\{D_i\}_{i\le n},\{Z_t\}_{t\le T},T_0, \zeta$}
 \KwResult{ Estimates $(\hat \pi_{rob},\hat \delta_{rob},\hat\tau_{rob})$}
 Construct the unit weights $\{\wrob_i\}_{i\le n}$ by solving optimization problem (\ref{eq:opt_unit})\;
 \For{$t \leftarrow T_0+1$ \KwTo $T$}{
    Construct $Y_{t}^{rob} = \frac{1}{n}\sum_{i\le n}Y_{it}\wrob_i$, and $W_{t}^{rob} = \frac{1}{n}\sum_{i\le n}W_{it}\wrob_i$\\ 
}
Using the data for $t>T_0$, estimate two regressions by OLS:
\begin{equation*}
    \begin{aligned}
    Y_{t}^{rob} = \eta^{(y)}_{0} + \delta Z_t + \varepsilon_{t}^{(y)},\quad
    W_{t}^{rob} =\eta^{(w)}_{0} + \pi Z_t + \varepsilon_{t}^{(w)}\\
    \end{aligned}
\end{equation*}
and report $\hat \delta_{rob}$, $\hat \pi_{rob}$, $\hat\tau_{rob} := \frac{\hat \delta_{rob}}{\hat \pi_{rob}}$.\
 \caption{Estimation Algorithm}\label{alg:const_weights}
\end{algorithm}

\subsection{Applying Robust Estimator}\label{sec:applying}
To reevaluate the results from \cite{nakamura2014fiscal} with our method, we use the original exposures $D_i$, set $T_0 = 10$ (which corresponds to years 1968-1977), and use the default value for $\zeta$ specified in \eqref{eq:reg_par}.  We then construct $\wrob_i$ and estimate $\hat \tau_{rob}$ using Algorithm \ref{alg:const_weights}. 

Panel B of Figure \ref{fig:cs_graph} plots the cross-sectional representation of our estimator. As before, the points represent the state-level first-stage and reduced-form coefficients but are now estimated using data from 1978 to 2006. The circle size reflects the absolute value of $\wrob_i$, and the colors reflect the sign. Our estimator $\hat \tau_{rob} = 1.72$ equals the slope of the line connecting two centers of mass (blue triangles) for negative and positive weights.  Compared to coefficients from Panel A of Figure \ref{fig:cs_graph}, the first-stage coefficients computed for 1981 to 2006 exhibit less variability. By construction, in Panel A, the states with extreme first-stage coefficients have the largest weights (in absolute value), which is no longer the case in Panel B. Aggregating these state-level coefficients, we get a larger multiplier than before, though still within the range  \cite{nakamura2014fiscal} report for alternative specifications. 

\begin{figure}[t!]
    \begin{center}
        \caption{Scatterplot---Nakamura and Steinsson Original Weights and Robust Weights } \label{fig:scatter_weights}
        \includegraphics[width=0.65\textwidth]{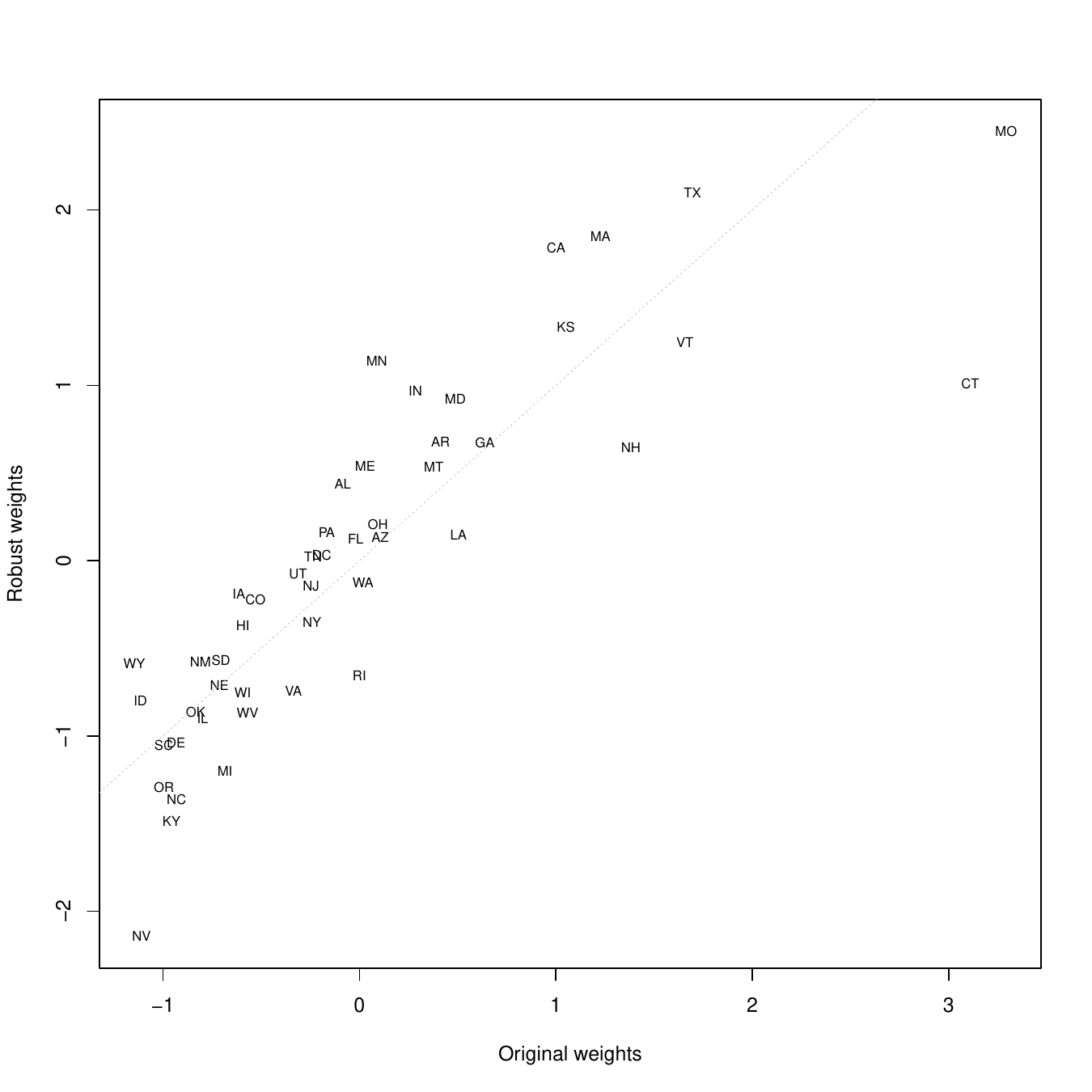}
        \end{center}
        \renewcommand{\baselinestretch}{0.7}
        \footnotesize{\textit{Notes}: $n = 48$ states; state abbreviations are used as labels. The variance of weights is scaled to $1$. }
\end{figure}

There are two differences between Panel A and B of Figure \ref{fig:cs_graph}. The first is the period we use to construct the state-level coefficients; the second is the weighting scheme. If we only change the period but apply the same weights as before, we get a multiplier of $1.71$, with a standard error of $0.91$. The similarity between point estimates is unsurprising, given visually minor differences between the weights, which we plot in Figure \ref{fig:scatter_weights}. The differences are mostly in the tails, with the original weights being extreme for several states. As a result, the standard error of the alternative estimator is $26\%$ higher.

We can see this in Panel B of Figure \ref{fig:ts_graph}, which demonstrates the time-series representation of our estimator. We plot the aggregate data $Y_{t}^{rob}$ and $W_{t}^{rob}$ vis-a-vis two separate OLS predictions for the years 1968 through 1977 (in blue), and 1978 through 2006 (in red). By changing the aggregation scheme, we reduce the variability: there is an $11\%$ reduction in the standard deviation for $W_t$ and a $24\%$ reduction for $Y_t$. Despite this decrease in variability, the aggregate instrument $Z_t$ remains relevant. Focusing only on 1979--2006 period, the $R^2$ for $W_t$ increases from $54\%$ to $74\%$, and for $Y_t$ it increases from $11\%$ to $20\%$. 

The higher -- compared to the original estimate from \cite{nakamura2014fiscal} -- standard error of our estimator, $\hat {\text{s.e.}}(\hat \tau_{rob}) = 0.74$, is explained by the fact that we compute it using Algorithm \ref{alg:inf} rather than by clustering at the unit level as in the original paper. Note that despite a shorter time span, the standard error is comparable to the one we got for the  TSLS estimator.

\subsection{Discussion}

\subsubsection{Intuition}
$\hat \tau_{TSLS}$ and $\hat \tau_{rob}$ rely on aggregating the unit-level first-stage and reduced-form coefficients. As discussed in the introduction, if $Z_t$ is a valid instrument in the sense of \cite{imbens1994identification}, all these coefficients describe interpretable causal effects. The TSLS and our method build on these coefficients by aggregating them in a particular way. This is why we encourage users to produce analogs of Figure \ref{fig:cs_graph} to investigate the importance of alternative aggregation schemes. In applications where the unit-level coefficients exhibit strong association, the aggregation does not play a major role, resulting in similar estimates. If, however, there is significant variation, then the weighting scheme becomes important.  

Our weights are produced by solving the optimization problem (\ref{eq:opt_unit}), and to understand their properties, it is helpful to consider several edge cases. First, if $\zeta$ is equal to infinity, then  $\wrob_i \propto (D_i -  \frac{1}{n}\sum_{j\le n}D_j)$, i.e., we get the same aggregate variables as before, in the TSLS. The resulting estimator is similar but not numerically equal to $\hat \tau_{TSLS}$ because we only use periods $t> T_0$ to estimate the coefficients. 

To understand what happens when $\zeta \ne \infty$, suppose $D_i$ is binary, $D_i \in \{0,1\}$. As discussed, the original aggregation scheme constructs a difference between average exposed ($D_i = 1$) and not-exposed ($D_i = 0$) units. Aggregation with weights $\wrob_i$ has a similar flavor but corresponds to taking a weighted average in both groups. This interpretation follows from examining the constraint in (\ref{eq:opt_unit}). Among all possible weighted averages, we select the one that makes aggregate variables $Y_t^{rob}$ and $W_t^{rob}$ as predictable as possible. 

Motivation for this choice is evident from looking at the first-stage and reduced-form equations that correspond to (\ref{eq:exm_factor}):
\begin{equation*}
\begin{aligned}
    &Y_{it} = \alpha_i^{(y)} + \mu_t^{(y)} + \delta D_iZ_t + \theta_i^{(y)}H_t + u_{it}^{(y)},\\
    &W_{it} = \alpha_i^{(w)} + \mu_t^{(w)} + \pi D_iZ_t + \theta_i^{(w)}H_t + u_{it}^{(w)}.
\end{aligned}
\end{equation*}
Unobserved confounders make the prediction of $Y_{it}$ and $W_{it}$ by $Z_t$ harder, so the weights that eliminate such factors should also make the prediction easier. Terms $u_{it}^{(y)}$ and $u_{it}^{(w)}$ create a statistical challenge: it is possible that instead of eliminating the confounder, the weights produce a combination of errors that compensates $H_t$. To prevent such overfitting, we include the regularization term in (\ref{eq:opt_unit}), which forces the weights to be as uniform as possible. Using this, we can show that $\wrob$ are close to deterministic weights $\wlim$, which minimize
\begin{equation}\label{eq:int_for_or}
\sum_{t\le T_0}\mathbb{E}\left(\frac{1}{n} \sum_{i\le n} w_iY_{it} - \eta_{0}^{(y)} - \eta_{z}^{(y)}Z_{t}\right)^2 + \sum_{t\le T_0}\mathbb{E}\left(\frac{1}{n} \sum_{i\le n} w_iW_{it} - \eta_{0}^{(w)} - \eta_{z}^{(w)} Z_t\right)^2,
\end{equation}
subject to the same constraints. As long as $\frac{1}{n}\sum_{i\le n}\wlim_i\theta_i^{(w)}$ and $\frac{1}{n}\sum_{i\le n}\wlim_i\theta_i^{(y)}$ are small, we can expect $\frac{1}{n}\sum_{i\le n}\wrob_i\theta_i^{(w)}$ and $\frac{1}{n}\sum_{i\le n}\wrob_i\theta_i^{(y)}$ to be negligible as well. 

The expectation in \eqref{eq:int_for_or} is not defined until we describe the randomness behind the observed data. One can think of at least three different sources of such randomness (and their combinations): (a) random sampling of units, (b) random variation in the unit-specific errors $(u_{it}^{(y)},u_{it}^{(w)})$, and (c) random variation in aggregate variables $(Z_t, H_t)$. Among these three, we view the last one as the most appropriate for the analysis in \cite{nakamura2014fiscal} and similar applications. First, the sampling uncertainty is at odds with the fact that we observe all possible states in the US (see the discussion in \citealp{abadie2020sampling,abadie2023should}). Second, probability statements based on $(u_{it}^{(y)},u_{it}^{(w)})$ are disconnected from the original identification idea, which instead relies on the aggregate variation. Guided by this logic, Section \ref{sec:formal_an} focuses on uncertainty coming from $(Z_t, H_t)$.

\subsubsection{Limitations}\label{sec:limit}
In the rest of the paper, we demonstrate the advantages of our algorithm with theory and simulations.  However, as with any method, our approach has limitations and should be used carefully. Below, we discuss the main ones, thus defining practical use cases for our estimator.

Our method is designed for applications where the TSLS regression (\ref{eq:TSLS_intro}) is a priori reasonable, but users worry about potential unobserved confounders. There are multiple reasons why (\ref{eq:TSLS_intro}) might fail, other than omitted variables. For example, the underlying model can be nonlinear, or the dynamic effects of the past treatments can be sizable. In these cases, the TSLS regression, and by extension, our improvement upon it, might be the wrong tool, and researchers should use other methods. We discuss this issue in more detail in Section \ref{sec:formal_an}.

Our algorithm is data-intensive, particularly in the time dimension. Our formal results require the total number of periods to be large and the number of units $n$ to be at least of a similar order. This assumption can be restrictive in some cases, but we believe it is reasonable for applications currently estimating \eqref{eq:TSLS_intro} using TSLS with aggregate instruments.

Finally, for our weights to improve over the TSLS ones, the environment should be sufficiently stable over time. To construct $\wrob$, we use the residuals after projecting out $Z_t$, which behave well when the effect of $Z_t$ does not change over time. We return to this point in Section \ref{sec:het_ef}, discussing heterogeneous effects in more detail.  For our weights to be useful for the second part of the data, any potential confounder like $H_t$ in (\ref{eq:exm_factor}) should have a similar effect in both data parts. This assumption might be too strong in environments where structural breaks are likely. This limitation can potentially be relaxed by using the whole sample to construct the weights. However, our theoretical results, particularly inference, rely on sample splitting. We also believe that sample splitting is a good general practice that protects from potential abuse.\footnote{See \cite{spiess2018optimal} for a formalization of this argument.}

Given these limitations, we recommend that researchers use our technique in situations where the number of observed periods is relatively large, structural breaks are unlikely, and the main potential problem is the presence of unobserved aggregate confounders rather than nonlinearity or dynamics. Our formal results and simulations show that our method either dominates the TSLS or performs similarly under this set of assumptions. 

\section{Theoretical Analysis}\label{sec:formal_an}

This section introduces our formal setup and states three theoretical results.\footnote{All proofs are collected in Appendix \ref{ap:proofs}.} The first theorem shows that our estimator remains consistent even when the TSLS fails. The second guarantees our estimator is asymptotically unbiased and normal under mild technical assumptions. The third theorem justifies conventional inference in situations with sufficient heterogeneity in the outcomes. 

\subsection{Causal model}
We start this section by discussing a causal model that we use to analyze the properties of our estimator. The model is based on two separate assumptions. The first describes the structural relationship between the observed data and underlying potential outcomes \citep{neyman1923applications,rubin1977assignment}. The fundamental property of this model is that it allows for both observed and unobserved aggregate inputs. The second assumption describes the joint behavior of the aggregate variables.  We close this section by discussing the difference between our setup and the more conventional ones. 

\subsubsection{Setup}
We observe $n$ units ($i$ is a generic unit) over $T$ periods ($t$ is a generic period). For each unit, we observe an outcome variable $Y_{it}$, an endogenous policy variable (treatment) $W_{it}$, an aggregate instrument $Z_t$, and a measure of exposure of unit $i$ to this variable $D_i$. We aim to estimate a causal relationship between $Y_{it}$ and $W_{it}$.  To formalize causality, we describe a model of the underlying potential outcomes. In addition to $w_{t}$ (potential value of $W_{it}$) and $z_t$ (potential value of $Z_t$), we also introduce $h_t$, an unobserved aggregate confounder that causally affects both the outcome and the treatment variable. We define $w^t := (\dots, w_{1},\dots, w_t)$, $z^t:=(\dots, z_{1},\dots, z_t)$, and $h^t:= (\dots, h_{1},\dots, h_t)$, and we make our first assumption.

\begin{assumption}\label{as:pot_out}\textsc{(Potential outcomes)}\\
Potential outcomes follow a static linear model:
\begin{equation}\label{eq:main_eq}
\begin{aligned}
&Y_{it}(w^t,h^t) =\alpha^{(y)}_{it} + \tau w_t + \theta^{(y)}_{i}h_t,\\
&W_{it}(h^t, z^t) = \alpha_{it}^{(w)} + \pi_{i}z_t + \theta^{(w)}_{i}h_t.
\end{aligned}
\end{equation}
As a result, the realized outcomes satisfy
\begin{equation}\label{eq:real_outc}
\begin{aligned}
&Y_{it} =\alpha^{(y)}_{it} +\tau W_{it} +  \theta^{(y)}_{i}H_t, \\
&W_{it} =  \alpha_{it}^{(w)} + \pi_{i}Z_t + \theta^{(w)}_{i}H_t.
\end{aligned}
\end{equation}
\end{assumption}

The critical part of this assumption is the presence of unobserved aggregate variable $H_t$. The danger such unobservables present for identification is well recognized in applied work \citep[e.g.,][]{chodorowstock,guren2018housing}. The standard restriction made in the literature is to assume that $\theta_{i}^{(w)}$ and $\theta_{i}^{(y)}$ do not systematically vary over $i$. We do not make this assumption; instead, we allow for such heterogeneity. Following most empirical applications, we focus on contemporaneous treatment effects and assume that only current quantities affect the outcomes. In practice, this assumption is often justified by looking at particular transformations of the available data \citep[e.g.,][]{nakamura2014fiscal}. Finally, to simplify the exposition and focus on a single parameter, we avoid heterogeneity in treatment effects. We relax this in Section \ref{sec:discussion}, where we discuss when the output of our algorithm can be interpreted as a weighted average of individual treatment effects.

Our next assumption restricts the joint distribution of aggregate variables $\{(Z_t,H_t)\}_{t\le T}$.
\begin{assumption}\textsc{(Exogenous variation)}\label{as:des_model}\\
The aggregate variables $(Z_t,H_t)$ follow a time-heterogeneous linear process. In particular, they satisfy
\begin{equation}
\begin{aligned}
&Z_t = \eta_z + \epsilon_t^{(z)},\quad
&H_t = \eta_h + \epsilon_t^{(h)}.\\
\end{aligned}
\end{equation}
For $k\in\{z,h\}$, define $\epsilon^{(k)} := (\epsilon^{(k)}_T,\dots,\epsilon^{(k)}_1)^\top$; there exist $T$-dimensional vectors $\nu^{(z)},\nu^{(h)}$, two upper-triangular matrices $\Lambda^{(z)},\Lambda^{(h)}$, and $\rho_{ag} \in (-1,1)$ such that
\begin{equation}
\begin{aligned}\label{eq:ts_model}
    &\epsilon^{(z)} = \Lambda^{(z)} \nu^{(z)},\quad
    &\epsilon^{(h)} = \Lambda^{(h)} \left(\rho_{ag} \nu^{(z)} + \sqrt{(1-\rho_{ag}^2)} \nu^{(h)}\right).\\
\end{aligned}
\end{equation}
Vectors $\nu^{(z)},\nu^{(h)}$ are independent and have independent components with the uniformly bounded sub-Gaussian norm, and $\mathbb{E}\left[(\nu_t^{(z)})^2\right] = \mathbb{E}\left[(\nu_t^{(h)})^2\right] = 1$. For $k \in \{z,h\}$ and $j\le T$, we have
\begin{equation}
\begin{aligned}
(\Lambda^{(k)})_{jj}\sim 1, \quad 
     \left|(\Lambda^{(k)})_{jl}\right| \lesssim\frac{1}{(l-j)^2}, \text{ for $l>j$}.\\
\end{aligned}
\end{equation}
\end{assumption}
The first part of this assumption restricts the means of $Z_t$ and $H_t$, which are assumed to be constant over time. This is without loss of generality for $H_t$ because its mean can be treated as a part of $\alpha_{it}^{(w)}$ and $\alpha_{it}^{(y)}$, but it is restrictive for $Z_t$. This assumption can be relaxed by considering a parametric model for the mean, for example, allowing for seasonality or secular trends. Fundamentally, our approach relies on researchers knowing how to detrend $Z_t$ and exploit random fluctuations $\epsilon_t^{(z)}$, which is straightforward if the mean is constant. This concept is closely linked to \cite{borusyak2023nonrandom}, who emphasize the importance of demeaning and highlight that the instrument $Z_t$ can be worthless without the ability to demean it.

The second part restricts the distribution of $\epsilon^{(z)}$ and $\epsilon^{(h)}$. These errors are generated by the underlying independent shocks $\nu^{(z)}$ and $\nu^{(h)}$. Since $|\rho_{ag}|$ is less than one, there is variation in $\epsilon^{(h)}_t$ that is not entirely explained by $\epsilon^{(z)}_t$ (and vice versa). Restrictions on the elements of matrices $\Lambda^{(z)}$ and $\Lambda^{(h)}$ exclude persistent cases (e.g., random walks) but allow for time-heterogeneous coefficients, making \eqref{eq:ts_model} particularly appealing. This structure is less flexible than the full linear model:
\begin{equation}
\begin{aligned}\label{eq:var_model}
    &\epsilon^{(z)} =\Lambda^{(z,1)}  \nu^{(1)} +  \Lambda^{(z,2)} \nu^{(2)},\quad
    &\epsilon^{(h)} = \Lambda^{(h,1)}  \nu^{(1)} +  \Lambda^{(h,2)} \nu^{(2)},\\
\end{aligned}
\end{equation}
which would allow $(\epsilon^{(z)},\epsilon^{(h)})$ to be generated by a VAR system with time-heterogeneous coefficients. The additional restrictions we impose with \eqref{eq:ts_model} are needed to have a single parameter $\rho_{ag}$ that governs the correlation, simplifying the exposition and proofs. We also expect our results to hold in the more general model \eqref{eq:var_model}. 

\subsubsection{Discussion}

The model we describe in this section follows the tradition that goes back at least to \cite{neyman1923applications}, in which we treat the observed units and their potential outcomes as fixed and focus explicitly on the randomness introduced by external shocks. As discussed in the previous section, this is natural in applications where the observed units exhaust the population of interest, which is often the case with geographic data. The finite population approach is also commonly motivated by the desire to connect the source of randomness to a well-defined shock while keeping the rest of the structure of the model as flexible as possible. This connection is straightforward in experimental applications, with the shock corresponding to the initial treatment assignment, making this type of analysis particularly appealing.

This logic is more complex once we consider observational studies where we cannot explicitly manipulate the treatment. Still, it is natural, especially in environments with aggregate variables. Recall the setup in \cite{nakamura2014fiscal} discussed in the previous section. In that context, various structural shocks in the US economy (in the sense of \citealp{ramey2016macroeconomic}) affect local-level outcomes through observed and unobserved aggregate variables. Assumption \ref{as:des_model} captures this behavior by modeling $Z_t$ and $H_t$ as functions of the underlying structural shocks $(\nu^{(z)},\nu^{(h)})$. These shocks turn $Z_t$ and $H_t$ into exogenous shifters that, per Assumption \ref{as:pot_out}, serve as inputs for the economic system that determines the local-level variables. The view that aggregate variables are a natural source of quasi-experimental variation is now widely adopted in the methodological literature on shift-share instruments \citep[e.g.,][]{adao2019shift,borusyak2022quasi,borusyak2023nonrandom}. The key assumption distinguishing our model is the possible correlation between the observed and unobserved aggregate variables (e.g., Section 4.2 in \citealp{adao2019shift}). 

As another example suppose  $(Y_{it},W_{it})$ are determined jointly in the local equilibrium:
\begin{equation}
\begin{aligned}
    &Y_{it} = \alpha_{it}^{(y)} + \tau W_{it}  + \theta^{(y)}_{i}H_t\\
    &W_{it} = \alpha_{it}^{(w)} + \gamma Y _{it}  + \pi_iZ_t\\
\end{aligned}
\end{equation}
This structure arises, for example, in \cite{guren2018housing} where $Y_{it}$ is the retail employment in location $i$, period $t$, and $W_{it}$ is the house price. The aggregate variables $Z_t$ and $H_t$ correspond to exogenous demand and supply shifters. Substituting $Y_{it}$ in the expression for $W_{it}$, we get the model (\ref{eq:main_eq}). This example demonstrates the difference between $Z_t$ and $H_t$. The former is a shifter for $W_{it}$ and is excluded from the structural equation for $Y_{it}$. Despite this exclusion restriction, $Z_t$ might be an invalid instrument due to its potential correlation with $H_t$. 

The example above also illustrates the nature of the endogeneity problem we aim to solve with our method. The outcomes $(Y_{it}, W_{it})$ are determined in the local equilibrium, and thus any direct regression of $Y_{it}$ on $W_{it}$ or vise versa does not have any economic meaning.  This feature describes the identification problem in the model, which is unaffected by any sampling assumptions behind $\alpha_{it}^{(w)},\alpha_{it}^{(y)}$, or any separation of these quantities into systematic fixed effects, and idiosyncratic errors, as in \eqref{eq:TSLS_intro}. The key assumption that makes the relationship between $Y_{it}$ and $W_{it}$ meaningful is the exclusion restriction for $Z_t$.

It is also instructive to look at our setup from a more conventional point of view. To this end, consider $Y_{it}(w):= Y_{it}(w,H_t)$ and $W_{it}(z) :=W_{it}(z,H_t)$ -- potential outcomes at realized values of $h_t$.  For each $i$, we can view the potential outcomes  $\{Y_{it}(w), W_{it}(z)\}_{t\le T}$ as a time-series version of the IV model of \cite{imbens1994identification}. In particular, Assumption \ref{as:pot_out} guarantees that the instrument $Z_t$ satisfies the exclusion restriction for each unit. Assumptions \ref{as:pot_out} and \ref{as:des_model}, however, do not guarantee that $Z_t$ is independent of $(Y_{it}(w), W_{it}(z))$. The two assumptions together quantify the extent of this dependence at the unit level: 
\begin{equation}\label{eq:cor}
\begin{aligned}
&\mathbb{E}\left[Y_{it}(w)(Z_t - \mathbb{E}[Z_t])\right] = \theta_i^{(y)} \mathbb{E}[H_t(Z_t - \mathbb{E}[Z_t])]\\
&\mathbb{E}\left[W_{it}(z)(Z_t - \mathbb{E}[Z_t])\right] = \theta_i^{(w)} \mathbb{E}[H_t(Z_t - \mathbb{E}[Z_t])]
\end{aligned}
\end{equation}
Our setup thus relaxes the independence assumption of \cite{imbens1994identification} (Condition 1, (i) in the paper) but imposes a product structure on the correlation. Equation (\ref{eq:cor}) emphasizes the difference between $H_{t}$ and $(\alpha_{it}^{(w)},\alpha_{it}^{(y)})$: heterogeneity in the latter does not lead to a violation of the independence assumption and thus is not crucial for identification. 

By definition, we do not observe  $H_t$ and thus cannot guarantee that it is one-dimensional, potentially making a model with multiple confounders more appropriate. Notationally, this extension is straightforward since we can interpret $(\theta_i^{(y)},\theta_i^{(w)})$ and $H_t$ as $p$-dimensional vectors. When $p$ is large enough, the RHS of (\ref{eq:cor}) can approximate arbitrary covariance between $Y_{it}(w), \, W_{it}(z)$ and $Z_t$. The dimension of $H_t$ does not directly enter Algorithm \ref{alg:const_weights} but makes its analysis more involved. We focus on the one-dimensional case because it transmits our theoretical insights in the simplest form. In Appendix \ref{sec:ap_2}, we establish our main bound, assuming $H_t$ is a vector, and later specialize it to the scalar case. 

\subsection{Heterogeneity}

This section describes two assumptions that restrict heterogeneity in potential outcomes. As discussed in the previous section, in the absence of $H_t$, one can cast our setup into the conventional IV model from \cite{imbens1994identification}, which does not require substantial restrictions on heterogeneity in potential outcomes. To understand why restrictions are necessary when $H_t$ is present, suppose $\theta_{i}^{(w)} \propto \pi_i \propto \theta_i^{(y)}$ and researchers observe $\pi_i$ up to proportionality. In each period, we can project the data on $\{\pi_i\}_{i\le n}$ and its orthogonal complement in $\mathbb{R}^n$. The latter projection depends neither on $Z_t$ nor $H_t$, because $\theta_{i}^{(w)},\theta_i^{(y)}$ are proportional to $\pi_i$. As a result, it is useless for identification without additional assumptions on terms $\alpha_{it}^{(w)}, \alpha_{it}^{(y)}$. The projection on $\{\pi_i\}_{i\le n}$ leaves us with a single time series for the outcome and the endogenous variable. By the proportionality assumption, these time series are necessarily corrupted by the aggregate confounder, making them equally useless for identifying the effect.

The exact proportionality of $\pi_i, \theta_{i}^{(w)},\theta_i^{(y)}$ is an extreme restriction, which is unlikely to hold in any practical application. Once we relax this restriction, the identification results are much more positive. As long as the projection of $\pi_i$ on $\theta_{i}^{(w)}$ and $\theta_i^{(y)}$ is not equal to zero, we can use the residuals from this projection to aggregate units cross-sectionally. This aggregation guarantees that the resulting quantity depends on $Z_t$ but not on $H_t$, thus solving the OVB problem. The conditions we impose below essentially amount to this possibility with two important caveats. First, to implement the strategy described above, we need to know $\{\pi_i\}_{i \le n}$ and $\left\{\theta_{i}^{(w)},\theta_i^{(y)}\right\}_{i \le n}$ up to proportionality, and we do not know these quantities. We assume that $\pi_i$ is directly connected to the observed variation $D_i$ to bypass this issue. Second, we need to guarantee that the terms $\alpha_{it}^{(w)}, \alpha_{it}^{(y)}$ do not ``mask'' important variation, the point we discuss in detail below.  

We start by connecting $\pi_i$ to $D_i$. 
\begin{assumption}\label{as:basic_expos}\textsc{(Linear Exposures)}\\
There exist numbers $(\eta_0,\eta_{\pi})$ with $\eta_{\pi} \ne 0$, such that for every $i$ we have  $\pi_i = \eta_0 + \eta_{\pi}D_i$.
\end{assumption}
This assumption guarantees that the observed variation in the unit-level ``exposure'' $D_i$ is directly connected to the underlying structural parameter $\pi_i$. The linearity is motivated by the empirical practice where researchers often assume that exposures $\pi_i$ are known up to linear transformation (e.g., \citealp{dube2013commodity,nunn2014us}). It can be substantially relaxed:  fundamentally, our theoretical results rely on $D_i$ being strongly correlated with $\pi_i$ after adjusting for other unit-specific coefficients. As a result, one can extend Assumption \ref{as:basic_expos} by explicitly including $\theta_i^{(w)},\theta_i^{(y)}$ or functions of $\{\alpha_{it}^{(w)}, \alpha_{it}^{(y)}\}_{t\le T}$ in the expression for $\pi_i$.

To state our next assumption, we introduce additional notation. For any $T_a >1$, define population analogs of $\hat \sigma_{k,T_0}$ from (\ref{eq:emp_scale}):
\begin{equation}
\begin{aligned}
     &\sigma^2_{y,T_a}:= \min_{\{\alpha_i,\gamma_i,\mu_t\}_{i,t}}\left\{\frac{\sum_{i\le n, t \le T_a}\mathbb{E}[(Y_{it} - \alpha_i - \mu_t - \gamma_i Z_t)^2]}{nT_a}\right\},\\
    &\sigma^2_{w,T_a}:= \min_{\{\alpha_i,\gamma_i,\mu_t\}_{i,t}}\left\{\frac{\sum_{i\le n, t \le T_a}\mathbb{E}[(W_{it} - \alpha_i - \mu_t - \gamma_i Z_t)^2]}{nT_a}\right\}.
\end{aligned}
\end{equation}
For any $T_b > T_a\ge1$, $t\in [T_a,T_b]$, weights $\omega_i$ such that $\sum_{i \le n} \omega_i = 0$, and $k \in \{y,w\}$ define a transformation of the model coefficients:
\begin{equation}
    \alpha_{t, T_a|T_b}^{(k)}(\omega):= \frac{1}{n\sqrt{T_b-T_a+1}}\sum_{i\le n} \omega_i\left(\alpha_{it}^{(k)} -\frac{1}{T_b-T_a+1} \sum_{T_a\le l < T_b} \alpha_{il}^{(k)}\right).
\end{equation}
With this notation, we are ready to introduce our following assumption. 
\begin{assumption}\label{as:over}\textsc{(Residual variation)}\\
For any $T_a>1$ there exist $\{\wlim_{i,T_a}\}_{i \le n}$ such that 
\begin{equation}
\begin{aligned}
    &\sum_{k\in \{y,w\}}\left[\frac{\sum_{t\le T_a}\left( \alpha_{t, 1|T_a}^{(k)}(\wlim_{T_a})\right)^2 + \left(\frac{1}{n}\sum_{i\le n}\wlim_{T_a}\theta_i^{(k)}\right)^2}{\sigma^2_{k,T_a}}\right] \lesssim  \frac{\log(n)}{n}\\
    &\frac{1}{n}\sum_{i\le n}\wlim_{i,T_a}D_i =1, \quad  \frac{1}{n}\sum_{i\le n}\wlim_{i,T_a} = 0,\quad \frac{1}{n}\sum_{i\le n}\left(\wlim_{i,T_a}\right)^2 \lesssim 1
\end{aligned}
\end{equation}
\end{assumption}
To unpack this condition, first suppose that $D_i = \pi_i$ and $\alpha_{it}^{(w)} = \alpha_{it}^{(y)} = 0$. In this case, Assumption \ref{as:over} is satisfied whenever $\pi_i$ is not perfectly predictable by $\theta_i^{(w)},\theta_i^{(y)}$. As we discussed above, this restriction is necessary for identification in our model.  When $\alpha_{it}^{(k)} \ne 0$, Assumption \ref{as:over} imposes additional restrictions.  As we discussed in Section \ref{sec:sec_2}, our statistical results guarantee that the weights $\wrob$ are close to deterministic oracle weights $\wlim$ that optimize the expected version of (\ref{eq:opt_unit}):
\begin{multline}
\frac{1}{T_0 \sigma^2_{y,T_0}}\sum_{t\le T_0}\mathbb{E}\left(\frac{1}{n} \sum_{i\le n} w_iY_{it} - \eta_{0}^{(y)} - \eta_{z}^{(y)}Z_{t}\right)^2 + \\
\frac{1}{T_0 \sigma^2_{w,T_0}}\sum_{t\le T_0}\mathbb{E}\left(\frac{1}{n} \sum_{i\le n} w_iW_{it} - \eta_{0}^{(w)} - \eta_{z}^{(w)} Z_t\right)^2
\end{multline}
subject to appropriate constraints. Using Assumptions \ref{as:pot_out} -- \ref{as:basic_expos} we can compute these expectations, and after concentrating  $\left\{\eta_{0}^{(k)},\eta_{z}^{(k)}\right\}_{k \in\{y,w\}}$, we get 
\begin{multline}\label{eq:oracle_main}
\frac{\sum_{t\le T_0}\left( \alpha_{t, 1|T_0}^{(w)}(\omega)\right)^2 + \kappa^2(T_0)\left(\frac{1}{n}\sum_{i\le n}\omega_i\theta_i^{(w)}\right)^2}{\sigma^2_{w,T_0}} +\\
\frac{\sum_{t\le T_0}\left( \alpha_{t, 1|T_0}^{(y)}(\omega)+\tau \alpha_{t, 1|T_0}^{(w)}(\omega)\right)^2 + \kappa^2(T_0)\left(\frac{1}{n}\sum_{i\le n}\omega_i(\theta_i^{(y)}+\tau \theta_i^{(w)})\right)^2}{\sigma^2_{y,T_0}}
\end{multline}
where $\kappa^2(T_0)$ is strictly positive. Assumption \ref{as:over} guarantees that the oracle problem (\ref{eq:oracle_main}) has a well-behaved solution.

Assumption \ref{as:over} is a sufficient condition directly motivated by our approach. How restrictive are the additional restrictions we impose on top of those necessary for identification? We do not attempt to answer this question fully and instead ask if our sufficient condition holds in statistical models commonly used in empirical and methodological literature. These models put probability structure on unit-specific quantities that we treat as fixed elsewhere. We use this structure to argue that Assumption \ref{as:over} holds for a ``generic'' dataset, i.e., for most of the datasets generated from the models we consider. 
\begin{prop}\label{prop:example}
Suppose that for $k\in \{y,w\}$
\begin{equation}
    \begin{aligned}
     &\alpha_{it}^{(k)} = \alpha_i^{(k)} + \mu_t^{(k)} + L_{it}^{(k)} + \epsilon_{it}^{(k)},\\
    &  (\epsilon_{iT}^{(k)},\dots \epsilon_{i1}^{(k)})^\top = \left(\Sigma^{(k)}\right)^{\frac12}\tilde \epsilon_{i}^{(k)}\\
    &\mathbb{E}[\epsilon_{i}^{(k)}] = \mathbf{0}_{T \times 1},\quad \mathbb{V}[\epsilon_{i}^{(k)}] = \mathcal{I}_{T}
    \end{aligned}
\end{equation}
where $\|\Sigma^{(k)}\|_{op} \lesssim 1$, $T$-dimensional vectors $\epsilon_{i}^{(k)}$ are independent over $i$, with independent uniformly bounded sub-Gaussian components, and for any $t\in\{1,\dots, T\}$  $\frac{\sum_{i\le n}\left(L_{it}^{(k)}\right)^2}{n} \lesssim 1$. In addition, suppose that
\begin{equation}
     D_i = \alpha_i^{(d)} + \epsilon_i^{(d)},\quad
     \mathbb{E}[\epsilon_{i}^{(d)}] = 0,\quad \mathbb{V}[\epsilon_{i}^{(d)}] = \sigma^2_{d}
\end{equation}
where $\frac{\sum_{i\le n} (\alpha_i^{(d)})^2}{n} \lesssim1$ and $\epsilon^{(d)}_i$ are independent over $i$, independent of $ \epsilon_{i}^{(w)}, \epsilon_{i}^{(y)}$, and have uniformly bounded sub-Gaussian norm. 
Finally, suppose for $k\in \{y,w\}$, $\frac{\sum_{i\le n}(\theta_i^{(k)})^2}{n}\lesssim1$ and Assumption \ref{as:des_model} holds. 

Then Assumption \ref{as:over} holds for $\wlim_{i,T_a} \propto (\epsilon_i^{(d)}-\frac{1}{n}\sum_{j\le n}\epsilon_j^{(d)})$ with probability approaching 1, as $n$ approaches infinity.
\end{prop}
This example covers many familiar cases. First, if $\alpha_i^{(d)} = \alpha^{(d)}$, then $D_i$ is as good as randomly assigned -- a situation that rarely holds in applications but serves as a natural benchmark. In this situation, the TSLS weights satisfy Assumption \ref{as:over}. Importantly, this is no longer true if we allow heterogeneity in $\alpha_i^{(d)}$. If $L_{it}^{(k)} \equiv 0$, then we recover a conventional two-way model that is commonly used in applications. In practice, we rarely expect the two-way model to hold exactly, and $L_{it}^{(k)}$ can be viewed as an approximation error. If $L_{it}^{(k)}$ has a low-rank structure, then we recover the interactive fixed-effects model (e.g., \citealp{bai2009panel}). However, this structure is not necessary, and  $L_{it}^{(k)}$ can vary arbitrarily as long as it remains appropriately bounded. 

The result in Proposition \ref{prop:example} relies on the existence of random $\epsilon_i^{(d)}$, suggesting that this variation plays an important identification role. As we explained at the beginning of this section, identification is impossible in the absence of the residual variation in $\pi_i$ that is orthogonal to $\theta_i^{(w)}, \theta_i^{(y)}$, and Proposition \ref{prop:example} generates such variation using random errors. Two important details make this variation different from the one in $Z_t$. First, the result in Proposition \ref{prop:example} holds for a generic parameter realization, allowing us to treat them as fixed elsewhere. Second, the fluctuations $\epsilon_{i}^{(d)}$ are not available to the researcher, because $\alpha_i^{(d)}$ is unknown. Moreover, our analysis does not rely on empirical weights $\wrob$ converging to $\wlim_{T_a}$, instead they are converging to a certain projection of these object, see Theorem \ref{th:limit_beh} for the precise statement. 

The setup of Proposition \ref{prop:example} is motivated by the statistical models commonly used in applications. The class of models it covers is large but by no means exhaustive, and in some applications, one would need to consider a different justification for Assumption \ref{as:over}. For example, if unit exposures $(\pi_i, \theta_i^{(y)}, \theta_i^{(w)})$ are equilibrium objects in some economic model, then to guarantee Assumption \ref{as:over} one needs to restrict the underlying structural parameters of that model. Proposition \ref{prop:example} might not directly apply to such cases, but we expect the underlying intuition to be helpful. 

It is also instructive to compare Assumption \ref{as:over} to conventional restrictions from the interactive fixed effects literature, particularly for models with low-rank regressors. For example, Assumption ID \textit{(ii) -(iii)} in \cite{moon2017dynamic} requires the terms $\alpha_{it}^{(k)}$ to be uncorrelated with other variables in the model and guarantees the presence of residual variation in the analog of $D_i$ after it is projected on the analogs of $\theta_i^{(w)}, \theta_{i}^{(y)}$ (without explicitly modeling the source of this variation). These restrictions imply the analog of Assumption \ref{as:over}, because we can use the residual variation to construct $\wlim_{T_a}$. The model in Proposition \ref{prop:example} is more general because it allows the terms $\alpha_{it}^{(k)}$ to have a systematic component $L_{it}^{(k)}$ with essentially unrestricted spectral properties (apart from boundedness). See also \cite{imbens2023identification} for a related discussion.

\subsection{Statistical Properties}\label{sec:stat_prop}

We now turn to the statistical properties of our estimator. All probability statements in the section except those in Proposition \ref{prop:example_2} refer to the joint distribution of $\{(Z_t, H_t)\}_{t\le T}$. We focus on a particular asymptotic regime characterized by the following assumption.
\begin{assumption}\label{as:asym_regime}\textsc{(Asymptotic regime)}\\
Both $n$ and $T$ increase to infinity and $\frac{T}{n} \rightarrow \gamma_{rat} <\infty$, for $k\in \{y,w\}$ we have
\begin{equation}
\begin{aligned}
    &\frac{1}{n}\sum_{i\le n} \left(\theta_i^{(k)} - \frac{1}{n}\sum_{j\le n} \theta_j^{(k)}\right)^2 \rightarrow \sigma^2_{\theta^{(k)}} >0,\quad
     \frac{1}{n}\sum_{i\le n} \left(D_i - \frac{1}{n}\sum_{j\le n} D_j\right)^2 \rightarrow \sigma^2_{D} >0, \\
     &\frac{\frac{1}{n}\sum_{i\le n} \left(D_i - \frac{1}{n}\sum_{j\le n} D_j\right)\theta_i^{(k)}}{\sigma_{D}\sigma_{\theta^{(k)}}} \rightarrow \rho_{cs}^{(k)}, \quad \frac{1}{n}\sum_{i\le n}\left(\alpha_{it}^{(k)}\right)^2 \rightarrow \left(\alpha_t^{(k)}\right)^2,\\
     &0 < \alpha_{\min}^2\le  \left(\alpha_t^{(k)}\right)^2 \le \alpha_{\max}^2 <\infty
\end{aligned}
\end{equation}
\end{assumption}
With the first part of this assumption, we restrict the analysis to environments where $n$ is comparable to or larger than $T$, which we expect to hold in many applications. The second part implies that the variability in $D_i$ and $\theta_i^{(k)}$ is present in the limit.  For binary $D_i$, this assumption is reasonable if the size of the treated or control group is not too small. The variability in $\theta_i^{(k)}$ implies that $H_t$ is a ``strong'' factor. While common in the theoretical literature on interactive fixed effects (e.g., \citealp{bai2009panel,moon2015linear}), this assumption can be restrictive in some applications, where researchers expect little variability in $\theta_i^{(k)}$. A version of our results holds in environments where $\sigma^2_{\theta^{(k)}} = 0$, thus allowing for weak factors (see the discussion in Appendix \ref{sec:ap_3}).  The restriction on the correlation is innocuous, as it always holds along a subsequence, and we make it to simplify the exposition. Finally, the restriction on $\alpha^{(k)}_{it}$ guarantees that $Y_{it}$ and $W_{it}$ have finite variances.

Assumption \ref{as:asym_regime} describes the limit behavior of unit-specific quantities. For the aggregate variables, we define similar objects for fixed $T_b > T_a \ge 1$ and $k\in \{z,h\}$:
\begin{equation}
\begin{aligned}
    &\sigma_{k,T_a|T_b}:= \sqrt{\frac{1}{T_b-T_a+1}\sum_{T_a \le t < T_b}\mathbb{E}\left[\left(\epsilon_t^{(k)}-\frac{\sum_{T_a \le l < T_b}\epsilon_l^{(k)}}{T_b-T_a+1}\right)^2\right]}\\
   &\rho_{T_a|T_b}:= \frac{\frac{1}{T_b-T_a+1}\sum_{T_a \le t < T_b}\mathbb{E}\left[\left(\epsilon_t^{(z)}-\frac{\sum_{T_a \le l < T_b}\epsilon_l^{(z)}}{T_b-T_a+1}\right)\epsilon_t^{(h)}\right]}{\sigma_{h,T_a|T_b}\sigma_{z,T_a|T_b}}
    \end{aligned}
\end{equation}
As indicated by the indices, these quantities depend on $T_a$ and $T_b$. Assumption \ref{as:des_model} guarantees that $|\rho_{T_a|T_b}| <1$, and $\sigma_{k,T_a|T_b}$ are uniformly bounded from above and below.

Our first result compares probability limits of $\hat \tau_{TSLS}$ and $\hat \tau_{rob}$.
\begin{theorem}\label{th:cons}\textsc{(Consistency)}\\
Suppose Assumptions \ref{as:pot_out} to \ref{as:asym_regime} hold, $\zeta^2 \sim \log(T_0)$, and $\frac{T_0}{T}\rightarrow \gamma_T \in (0,1)$, then 
\begin{equation*}
    \hat \tau_{rob} = \tau + o_p(1)
\end{equation*}
If, in addition, $\left|\rho_{cs}^{(w)} \sigma_{\theta^{(w)}}\rho_{1|T}\sigma_{h,1|T} + \eta_{\pi}\sigma_{D}\sigma_{z,1|T}\right| > c_{\min}>0$, then 
\begin{equation*}
\hat\tau_{TSLS}= \tau +\frac{\rho_{cs}^{(y)} \sigma_{\theta^{(y)}}\rho_{1|T}\sigma_{h,1|T} }{\rho_{cs}^{(w)} \sigma_{\theta^{(w)}}\rho_{1|T}\sigma_{h,1|T} + \eta_{\pi}\sigma_{D}\sigma_{z,1|T}} + o_p(1)
\end{equation*}
\end{theorem}
This result demonstrates that $\hat \tau_{rob}$ remains consistent in the regime where $\hat \tau_{TSLS}$ generally fails. It also formalizes a part of the discussion in Section \ref{sec:limit}. In our model, the only threat to the validity of the TSLS is the presence of omitted variables. As long as either $\rho_{cs}^{(y)}$ or $\rho_{1|T}$ is equal to zero, the TSLS estimator is consistent. In particular, the latter quantity equals zero as long as $\rho_{ag} = 0$. 

Theorem \ref{th:cons} provides a first justification for using our estimator, but it does not describe its distributional properties in large samples. Under current assumptions, we can provide only relatively weak guarantees on the asymptotic behavior of $\hat \tau_{rob}$. In particular, Assumptions \ref{as:over} and \ref{as:asym_regime} imply that there exist weights such that 
\begin{equation*}
    \left|\frac{1}{n}\sum_{i\le n}\wlim_{i,T_0}\theta_i^{(k)}\right| \lesssim \sqrt{\frac{\log(T_0)}{T_0}}
\end{equation*}
however, this property is too weak to achieve asymptotic unbiasedness. To make progress, we impose additional assumptions on $\theta_i^{(y)}$ and $\theta_i^{(w)}$.
\begin{assumption}\label{as:rich_het}\textsc{(Sufficient heterogeneity)}\\
For any $T_a>1$ and $k \in \{y,w\}$, there exists $\{\wlim_{k,i,T_a}\}_{i \le n}$ such that
\begin{equation}
\begin{aligned}
    &\sum_{l\in \{y,w\}}\left[\frac{\frac{1}{T_a}\sum_{t\le T_a}\left( \alpha_{t, 1|T_0}^{(k)}(\wlim_{k,T_a})\right)^2}{\sigma^2_{l,T_a}}\right] + \frac{\left(\frac{1}{n}\sum_{i\le n}\wlim_{k,i,T_a}\theta_i^{(-k)}\right)^2}{\sigma^2_{-k,T_a}} \lesssim \frac{\log(n)}{n},\\
    &\frac{1}{n}\sum_{i\le n}\wlim_{k,i,T_a}\theta_i^{(k)} = \sqrt{\frac{1}{n}\sum_{i\le n} \left(\theta_i^{(k)} - 
    \frac{1}{n}\sum_{j\le n}\theta_j^{(k)}\right)^2}, \quad  \frac{1}{n}\sum_{i\le n}\wlim_{k,i,T_a} = 0,\\
    &\frac{1}{n}\sum_{i\le n}\left(\wlim_{k,i,T_a}\right)^2 \lesssim 1,
\end{aligned}
\end{equation}
where $\{-k\} = \{y, w\} \setminus\{k\}$.
\end{assumption}
This restriction is similar to Assumption \ref{as:over} and requires existence of variation in $\theta_i^{(k)}$ that is not captured by $\alpha_{it}^{(w)},\alpha_{it}^{(y)}$, and $\theta_{i}^{(-k)}$. To justify it, we return to the example from Proposition \ref{prop:example}. 

\begin{prop}\label{prop:example_2}
Suppose conditions of Proposition \ref{prop:example} hold. In addition, suppose for $k\in \{y,w\}$ we have
\begin{equation}
    \theta_i^{(k)} = \alpha_i^{(k)} + \epsilon_i^{(k)}, \quad \mathbb{E}[\epsilon_i^{(k)}] = 0,  \quad \mathbb{V}[\epsilon_i^{(k)}] = \sigma^2_{\theta^{(k)}}
\end{equation}
where $\epsilon_i^{(k)}$ are independent over $i$ and $k$, and have a uniformly bounded sub-Gaussian norm. Then Assumption \ref{as:rich_het} holds with probability one as $n$ approaches infinity.
\end{prop}
To understand why Assumption \ref{as:rich_het} can improve the performance of $\hat \tau_{rob}$, it is useful to consider environments where it fails. In particular, if $\theta_i^{(k)}$ is nearly spanned by $\{\alpha_{it}^{(y)},\alpha_{it}^{(w)}\}_{t \le T_0}$ and $\theta_i^{(-k)}$, but the remaining variation is strongly associated with $\pi_i$, then it is very hard to eliminate it by aggregation, which results in a slow rate of convergence. This problem is mitigated when there is enough variability in $\theta_i^{(k)}$, which is not explained by other variables. Our next result demonstrates these gains by characterizing the asymptotic behavior of $\hat \tau_{rob}$. To state it, we define for arbitrary periods $T_b > T_a>0$ a matrix $\Lambda^{(z)}_{T_a|T_b}$ such that
\begin{equation}
(\epsilon^{(z)}_{T_b},\dots, \epsilon^{(z)}_{T_a})^\top = \Lambda^{(z)}_{T_a|T_b}\nu^{(z)}.
\end{equation}
We also use $\alpha^{(k)}_{T_a, T_b}(\omega)$ to denote the vector $ \left(\alpha_{T_b, T_a|T_b}^{(k)}(\omega),\dots, \alpha_{T_a, T_a|T_b}^{(k)}\right)$.

\begin{theorem}\label{th:limit_beh}\textsc{(Asymptotic behavior)}\\
Suppose Assumption \ref{as:pot_out}-\ref{as:rich_het} hold, $\zeta^2 \sim \log(T_0)$, and $\frac{T_0}{T}\rightarrow \gamma_T \in (0,1)$. Then there exists deterministic weights $\{\wdet_{i,T_0}\}_{i\le n}$ such that $ \frac{1}{\sqrt{n}}\|\wrob_i - \wdet_{i,T_0}\|_2 = o_p(1)$, and 
\begin{equation}\label{eq:as_norm}
      \sqrt{T_1}\left(\hat \tau_{rob} - \tau\right) = \frac{\sigma_{n,T}}{\eta_{\pi} \sigma^2_{z,T_0+1|T}}\xi_n + o_p(1), \quad \mathbb{E}[\xi_n] = 0, \quad \mathbb{V}[\xi_n] = 1,
\end{equation}
where $\sigma_{n,T}:= \left\| \alpha_{T_0+1|T}^{(y)}(\wdet_{T_0}) \Lambda^{(z)}_{T_0+1|T}\right\|_2$. If, in addition, $\frac{\| \alpha_{T_0+1|T}^{(y)}(\wdet_{T_0})\|_{\infty}}{\| \alpha_{T_0+1|T}^{(y)}(\wdet_{T_0})\|_2} = o(1)$, then $\xi_n$ converges in distribution to $\mathcal{N}(0,1)$.
\end{theorem}
This result implies that our estimator is asymptotically unbiased and normal as long as $\sigma_{n,T}$ remains bounded. This condition can be restrictive if $\alpha_{i,t}^{(y)}$ are close to being degenerate. For example, if we view these quantities as being generated from the normal distribution, i.e., $\alpha_{it}^{(y)} \sim \mathcal{N}\left(\alpha^{(y)}_i + \lambda_t^{(y)},\sigma^2_{\alpha^{(y)}}\right)$, then $\sigma_{n,T} =O_p\left(\frac{1}{\sqrt{n}}\right)$ and higher-order terms in (\ref{eq:as_norm}) become important. This lack of uniformity is similar to one analyzed in \cite{menzel2021bootstrap}. In practice, we do not expect the two-way model to hold exactly; rather we view it as an approximation. In this situation, the first term in (\ref{eq:as_norm}) becomes dominant, and we can use Theorem \ref{th:limit_beh} for inference. 

Theorem \ref{th:limit_beh} describes the asymptotic behavior of our estimator in the presence of unobserved confounders. If no such variables exist in the structural equation, i.e., $\theta_i^{(y)} \equiv 0$, then our estimator and the standard TSLS estimator are asymptotically normal under mild technical conditions. In this regime, $\sigma_{n, T}$ can be smaller or larger than its TSLS counterpart  $\left\| \alpha_{1|T}^{(y)}(\omega^{TSLS}) \Lambda^{(z)}_{1|T}\right\|_2$, depending on the underlying complexity of the potential outcomes and differences in the sample sizes.

We conduct inference in several steps, which we summarized in Algorithm \ref{alg:inf}. First, we estimate the variance. We assume that a researcher has access to a consistent estimator for $\Lambda^{(z)}_{T_0+1|T}$, which we denote as $\hat\Lambda^{(z)}_{T_0+1|T}$. In practice, this means that we need to estimate the model for $Z_t$ to conduct inference. In our simulations and empirical analysis, we use off-the-shelf methods to do that, and the resulting estimator performs relatively well.

For $t>T_0$, we construct scaled residuals from the aggregate regression
\begin{equation}
     \hat\alpha_{t, T_0|T}^{(y)}(\wrob):= \frac{Y_{t}^{rob} - \hat \tau_{rob}W_{t}^{rob}}{\sqrt{T_1}}
\end{equation}
and estimate the asymptotic standard error of $\hat\tau_{rob}$:
\begin{equation}
    \hat \sigma_{rob} := \frac{\| \hat \alpha_{T_{0}+1|T}^{(y)}(\wrob) \hat\Lambda^{(z)}_{T_0+1|T}\|_2}{|\hat \pi_{rob}|\frac{1}{T_1}\sum_{T_0 < t < T}\left(Z_t-\frac{\sum_{T_0 < l \le T}Z_l}{T_1}\right)^2}
\end{equation}
With this quantity, we construct a standard asymptotic confidence interval of level $1-\alpha$:
\begin{equation}\label{eq:conf_int}
    \tau \in \hat \tau_{rob} \pm \frac{\hat \sigma_{rob}}{\sqrt{T_1}}z_{1-\alpha/2}
\end{equation}
where $z_{\alpha}$ is $\alpha$-quantile  of the standard normal distribution. Our next result characterizes the asymptotic properties of this interval.
\begin{theorem}\label{th:inference}\textsc{(Inference)}\\
Suppose conditions on Theorem \ref{th:limit_beh} hold, and $\sigma^2_{n,t}\gtrsim 1$. In addition, suppose $\|\hat\Lambda^{(z)}_{T_0+1|T} - \Lambda^{(z)}_{T_0+1|T}\|_{op} = o_p(1)$. Then, the confidence interval (\ref{eq:conf_int}) has asymptotic coverage $1-\alpha$.
\end{theorem}

\RestyleAlgo{boxruled}
\LinesNumbered
\begin{algorithm}[t]
 \KwData{$\{Y_{t}^{rob},W_{t}^{rob},Z_t\}_{t\le T},\hat \tau_{rob},\hat \pi_{rob},\hat \Lambda^{(z)}_{T_0+1|T},\alpha,T_0$}
 \KwResult{ $1-\alpha$ confidence interval}
 \For{$t \leftarrow T_0+1$ \KwTo $T$}{
    Construct $\hat\alpha_{t, T_0|T}^{(y)}(\wrob)= \frac{Y_{t}^{rob} - \hat \tau_{rob}W_{t}^{rob}}{\sqrt{T-T_0}}$\\ 
}
Compute $\hat \sigma_{rob}= \frac{\| \hat \alpha_{T_0|T}^{(y)}(\omega) \hat\Lambda^{(z)}_{T_0+1|T}\|_2}{|\hat \pi_{rob}|\frac{1}{T_1}\sum_{T_0 < t < T}\left(Z_t-\frac{\sum_{T_0 < l \le T}Z_l}{T-T_0}\right)^2}$\;
Report the confidence interval: $\tau \in \hat \tau_{rob} \pm \frac{\hat \sigma_{rob}}{\sqrt{T_1}}z_{1-\alpha/2}$.
 \caption{Inference}\label{alg:inf}
\end{algorithm}

As discussed above, we expect this theorem to be useful in practice whenever the two-way model for $\alpha_{it}^{(k)}$ is only approximately correct. This result focuses on the conventional interval (\ref{eq:conf_int}), which is valid if the first stage is strong, that is, if $\eta_{\pi}$ is large enough. A version of Theorem \ref{th:limit_beh} also holds for the first-stage and reduced-form coefficients and thus can be used to conduct conventional robust inference (see \citealp{andrews2019weak} for a recent survey on robust inference). 

To construct $\hat \sigma_{rob}$, we combine aggregate residuals with the estimator for the parameters of the design model. If $\Lambda^{(z)}$ is diagonal, i.e.,  the variation in $Z_t$ is independent over time, then $\hat \sigma_{rob}$ corresponds to ``clustering at time level.'' However, in practice, this assumption can be too restrictive.  With general $\Lambda^{(z)}$, we need to consider dependence over time, and $\hat \sigma_{rob}$ does that using $\hat \Lambda^{(z)}_{T_0+1|T}$. In the application, we discussed in Section \ref{sec:sec_2} as well as in the Monte-Carlo experiments in Section \ref{sec:sim}, we use the standard automatic model selection \textbf{ARIMA} package in \textbf{R} to estimate the model for $Z_t$ and then simulate from this model to construct $\hat \sigma_{rob}$. We recommend that researchers use the same approach in practice.

\section{Extensions} \label{sec:discussion}

This section discusses three possible extensions of our model and the respective adjustments to the algorithm. We first show how to incorporate covariates in our setting. Secondly, we examine the case of heterogeneous treatment effects as a natural extension. We conclude Section \ref{sec:discussion} by connecting our estimator to the literature on shift-share designs.


\subsection{Additional Information}\label{sec:covariates}

 A typical regression equation estimated in applications will have a more complicated structure than (\ref{eq:TSLS_intro}):
\begin{equation}\label{eq:tsls_adv}
    Y_{it} = \alpha_i + \mu_t(X_i) + \tilde\theta_i^\top \tilde H_t + \tau W_{it} + \epsilon_{it}
\end{equation}
Here, $X_i$ are observed unit-level attributes, for example, region indicators, and $\tilde H_t$ is a vector of observed aggregate variables that we expect to be correlated with $Z_t$. Equation (\ref{eq:tsls_adv}) is estimated by the TSLS using $D_i Z_t$ as an instrument for $W_{it}$ and treating $\alpha_i$ and $\theta_i$ as fixed parameters. Inclusion of $\mu_t(X_i)$ instead of $\mu_t$ and $\tilde \theta_i^\top \tilde H_{t}$ in the equation mitigates the OVB concerns but does not eliminate them. 

Our estimator also allows for unit-level covariates and observed aggregate variables.\footnote{To incorporate time-varying covariates $X_{it}$, we can define $X_i := (X_{i1},\dots, X_{iT})$. Alternatively, and more in line with current empirical practice, we can instead residualize $Y_{it}$ and $W_{it}$ with respect to $X_{it}$.} In particular, we suggest estimating equation (\ref{eq:tsls_adv}) by TSLS using $\wrob_iZ_t$ as instrument for $W_{it}$ and data from periods $T_0+1, \dots T$. The weights $\wrob_i$ then solve an adjusted optimization problem:
\begin{equation}\label{eq:opt_unit_adv}
\begin{aligned}
&\wrob= \argmin_{\{w,\eta_{0}^{(w)},\eta_{z}^{(w)},\eta_{0}^{(y)}, \eta_{z}^{(y)}\}} \Bigg{\{}  \frac{\zeta^2\|w\|_2^2}{nT_0} + 
\frac{\frac{1}{T_0}\sum_{t\le T_0}\left(\frac{1}{n} \sum_{i\le n} w_iY_{it} - \eta_{0}^{(y)} - (\eta_{z}^{(y)})^\top(Z_{t},\tilde H_t)\right)^2}{\hat\sigma^2_{y,T_0}} \\ &+\frac{\frac{1}{T_0}\sum_{t\le T_0}\left(\frac{1}{n} \sum_{i\le n} w_iW_{it} - \eta_{0}^{(w)} - (\eta_{z}^{(w)})^\top(Z_{t},\tilde H_t)\right)^2}{\hat \sigma^2_{w,T_0}} \Bigg{\}}\\
  &\text{subject to: } 
\frac{1}{n} \sum_{i\le n} w_i D_i = 1,\quad
   \frac{1}{n} \sum_{i\le n} w_i = 0, \quad \frac{1}{n} \sum_{i\le n} w_iX_i = 0
\end{aligned}
\end{equation}
The additional constraint guarantees that aggregation eliminates the linear projection of $\theta_i^{(w)}$ and $\theta_i^{(y)}$ on $X_i$. As a typical example, consider a situation where data can be grouped into clusters and where researchers wish to include cluster-specific time fixed effects. This can be achieved using $X_i$ corresponding to cluster indicators, under the natural extension of Assumptions \ref{as:pot_out} to \ref{as:asym_regime}, Theorems \ref{th:cons} and \ref{th:limit_beh} continue to hold for the weights that solve (\ref{eq:opt_unit_adv}).

In some applications, the unit-level variables $Y_{it}$ and $W_{it}$ have different statistical properties; for example, they are measured using a different number of observations. In such situations, researchers commonly use weighted versions of the TSLS. To achieve the same with our algorithm, researchers can estimate (\ref{eq:tsls_adv}) using weighted TSLS with $\wrob_iZ_t$ as an instrument for $W_{it}$. To construct the weights $\wrob_i$, we solve the optimization problem (\ref{eq:opt_unit_adv}) but instead of the standard euclidean norm $\|w\|_2^2$, we use a weighted one:
\begin{equation}
\| \omega\|_{2,A}^2 = \omega^\top A \omega
\end{equation}
where $A$ is a diagonal matrix, and $(A)_i = a_i^2>0$.

\subsection{Heterogeneous Treatment Effects}\label{sec:het_ef}

In applications, it is rarely possible to argue that the treatment effects are constant, and thus Assumption \ref{as:pot_out} can be too restrictive. To address this, we consider a model with heterogeneous effects:
\begin{equation}
Y_{it} = \alpha_{it}^{(y)} + \tau_i W_{it}+ \theta_i^{(y)}H_t, \\
W_{it} = \alpha_{it}^{(w)} + \pi_i Z_{t} + \theta_i^{(w)}H_t
\end{equation}
We also define the reduced form that corresponds to the structural equation above:
\begin{equation}
Y_{it} =  \tilde\alpha_{it}^{(y)} + \tau_i \pi_i Z_t +  \tilde\theta_i^{(y)}H_t
\end{equation}
where $\tilde\alpha_{it}^{(y)} := \alpha_{it}^{(y)} + \tau_i \alpha_{it}^{(w)}$, and $\tilde \theta_i^{(y)}:= \theta_i^{(y)} + \tau_i \theta_i^{(w)}$. For any estimator $\hat \tau(\omega)$ that averages units with arbitrary weights $\omega$ and  constructs the IV ratio from the aggregate regressions, we have
\begin{equation}\label{eq:het_rep}
\hat \tau(\omega) = \frac{\frac{1}{n}\sum_{i\le n} \omega_i\tau_i\pi_i + \text{error}}{\frac{1}{n}\sum_{i\le n} \omega_i\pi_i + \text{error}} = \frac{\frac{1}{n}\sum_{i\le n} \omega_i\tau_i\pi_i}{\frac{1}{n}\sum_{i\le n} \omega_i\pi_i}(1 + \text{error}) + \text{error}
\end{equation}
Our goal in this section is to understand when $\tau(\omega):=\frac{\frac{1}{n}\sum_{i\le n} \omega_i\tau_i\pi_i}{\frac{1}{n}\sum_{i\le n} \omega_i\pi_i}$ has a causal interpretation. We thus ignore the errors in (\ref{eq:het_rep}). Their properties depend on the choice of weights $\omega$ and can be established in the same way as before. 

First, we consider a situation where $\pi_i  = \eta_{\pi} D_i$, for binary $D_i\in\{0,1\}$. For $\omega^{TSLS}$, we get
\begin{equation}
\tau\left(\omega^{TSLS}\right) = \frac{\sum_{i \le n}\tau_i D_i}{\sum_{i \le n}D_i}
\end{equation}
which is an average treatment effect for the exposed group. Using $\wrob$, we get
\begin{equation}
\tau(\wrob) = \frac{1}{n}\sum_{i \le n}\tau_i \wrob_iD_i
\end{equation}
where $\frac{1}{n}\sum_{i \le n}\wrob_iD_i =1$. Without additional restrictions, we cannot interpret $\tau_{rob}$ as a convex combination of treatment effects because the weights $\wrob_i$ can be negative for exposed units. Negative weights lead to extrapolation, which can help with the OVB but at the cost of interpretability.  

This problem is easy to address by adding a non-negativity constraint
\begin{equation}\label{eq:pos_cont}
\omega_i \left(D_i -  \frac{1}{n}\sum_{j\le n}D_j\right) \ge 0
\end{equation}
to the optimization program (\ref{eq:opt_unit}). The resulting $\tau(\wrob)$ is a convex combination of treatment effects by construction. The optimization problem remains convex and can be solved efficiently, even for large datasets. Inequality constraint (\ref{eq:pos_cont}) also acts like a powerful regularizer, improving the statistical properties of the algorithm. To reap these benefits, we need to assume that ``good'' balancing weights that satisfy (\ref{eq:pos_cont}) exist, for example, by adding this restriction to Assumption \ref{as:over}. Overall, in applications with binary $D_i$, we recommend imposing (\ref{eq:pos_cont}) unless the user strongly believes that extrapolation is necessary. 

Many applications do not have a control group with $D_i$ taking arbitrary values, so non-negativity constraints are harder to motivate. However, one can still interpret $\tau(\wrob)$ with additional assumptions. In particular, suppose that
\begin{equation}
D_i = \alpha_i^{(d)} + \epsilon_i^{(d)}
\end{equation}
where $\epsilon_i^{(d)}$ has the same properties as in Proposition \ref{prop:example}. If Assumption \ref{as:basic_expos} holds, we have 
\begin{multline}
\tau(\wrob) = \frac{1}{n}\sum_{i \le n}\tau_i \wrob_iD_i +\frac{\eta_0}{\eta_{\pi}} \left(\frac{1}{n}\sum_{i \le n}\tau_i \wrob_i\right) = \frac{1}{n}\sum_{i \le n}\tau_i \frac{(\epsilon_i^{d})^2}{\sigma^2_{d}} + o_p(1)+ \\
 O_p\left(\frac{\left\|\wrob - \frac{\epsilon^{(d)}}{\sigma^2_{d}}\right\|_2}{\sqrt{n}}\right)
\end{multline}
As long as $\wrob$ converges to $\frac{\epsilon^{(d)}}{\sigma^2_{d}}$, the estimand $\tau(\wrob)$ converges to the average treatment effect. In Appendix \ref{ap:het_effects}, we discuss models where this convergence holds. In this case, our method improves over the TSLS in two ways: it removes the OVB and helps interpretability. 

The heterogeneity we consider in this section is restricted in an important way: we do not allow $\tau_i$ and $\pi_i$ to vary over time. Such variation makes it impossible to project $Z_t$ out when constructing the weights $\wrob$. This problem can be bypassed if the researcher knows that in the initial $T_0$ periods, $\pi_{it} \equiv 0$ for all units. In particular, in applications where $\pi_{it} = (\eta_0+\eta_tD_i)\mathbf{1}_{t>T_0}$, we expect Algorithm \ref{alg:const_weights} to perform well with both cross-sectional and time-series heterogeneity in treatment effects, as long as $\eta_t >0$.

\subsection{Shift-Share Designs}\label{subsec:shift}

This section discusses the relationship between our model and models from the shift-share, or Bartik instruments, literature (\citealp{adao2019shift, borusyak2022quasi, goldsmith2020bartik}). We start by considering an extension of our original framework. Assume that instead of a single aggregate variable, we have $|S|$ of them. In a typical application, these will correspond to industry-level shifters. The following equations are now satisfied for all $i$ and $t$:
\begin{equation}\label{eq:gen_model}
\begin{aligned}
&Y_{it} = \alpha^{(y)}_{it} + \tau W_{it} + \sum_{s\in S}\theta^{(y)}_{is}H_{ts}\\
&W_{it} = \alpha^{(w)}_{it} + \sum_{s\in S} \pi_{its}\gamma_{is}Z_{ts} +  \sum_{s\in S}\theta^{(w)}_{is}H_{ts}
\end{aligned}
\end{equation}
where $s$ is a generic industry, and we observe $\{\gamma_{is}\}_{i,s}$, $\{W_{it},Y_{it}\}_{it}$, $\{Z_{ts}\}_{t,s}$, and $\sum_{i\le n} \gamma_{is} =1$. It is straightforward to see that our model is a special case of this with $|S| = 1$.

The model typically considered in the shift-share literature is a special case of (\ref{eq:gen_model}), with $T=1$, and two additional assumptions: (a) $Z_{ts} = \psi_{ts}^{\top}\mu_t + \epsilon_{ts}$, where $\psi_{ts}$ are known, $\mathbb{E}[\epsilon_{ts}] = 0$, and $\epsilon_{ts}$ are uncorrelated over $s$; and (b) for every $t$, $\{H_{ts}\}_{s\in S}$ is uncorrelated with $\{\epsilon_{ts}\}_{s\in S}$. Identification is achieved by exploiting variation over industries (see \citealp{borusyak2022quasi}). In applications, $T$ is usually not equal to $1$, and the model in differences is often considered. At the same time, the identification argument does not exploit the time dimension and focuses on the variation over industries.

Models of the type (\ref{eq:gen_model}) can be promising because they allow for a combination of two identification arguments: one based on the variation over time and one based on the variation over $s$. In applications, both $|S|$ and $T$ can be modest (especially if we want shifters to be independent over $s$), and thus it is natural to use both sources of variation. The development of strategies that allow for that is an attractive area of future research.

\section{Simulations}\label{sec:sim}

In this section, we illustrate the performance of our estimator in simulations. To make the simulations more realistic, we build them using the dataset from \cite{nakamura2014fiscal}, which we described in Section \ref{sec:sec_2}. In our experiments, we try to capture the spirit of this empirical exercise and investigate how different features of the data-generating process affect the performance of the algorithms. Formally, our simulations are based on the following model:
\begin{equation}\label{eq:sim_dgp}
\begin{aligned}
    &Y_{it} = \beta_i^{(y)} +\mu_t^{(y)}+ L_{it}^{(y)} + \tau W_{it} + \theta_i^{(y)}H_t + \epsilon_{it}^{(y)}\\
   & W_{it} = \beta_i^{(w)} +\mu_t^{(w)}+ L_{it}^{(w)} + \pi_i Z_{t} + \theta_i^{(w)}H_t + \epsilon_{it}^{(w)}
\end{aligned}
\end{equation}
Here, parameters $\{\beta_i^{(y)},\beta_i^{(w)},\mu_t^{(y)},\mu_t^{(w)},L_{it}^{(y)},L_{it}^{(w)},\tau, \pi_i, \theta_i^{(w)}, \theta_i^{(y)}\}_{i\le n,t\le T}$  are fixed, while $\epsilon_{it}^{(y)},\epsilon_{it}^{(w)}$, and $\{Z_t,H_t\}_{t\le T}$ are random.

In Appendix \ref{ap:sim_details}, we describe how exactly we use the data to construct $\{L_{it}^{(y)},L_{it}^{(w)}, \pi_i\}_{i\le n,t\le T}$, and the models for $\{Z_t\}_{t\le T}$ and $\{\epsilon_{it}^{(y)},\epsilon_{it}^{(w)}\}_{i\le n,t\le T}$. Heuristically, we extract the components $L_{it}^{(y)}$ and $L_{it}^{(w)}$ using the SVD decompositions of observed data, while for $\pi_i$, we use the estimated $\hat \pi_{i}$ from Section \ref{sec:sec_2}, which we scale to make the instrument relatively strong.\footnote{The median $F$ statistic for $\hat \tau_{TSLS}$ for the fourth design is equal to $78$.} We make these adjustments to focus on the properties of our estimator in the regime covered by Theorems \ref{th:limit_beh} and \ref{th:inference}. The data are not directly informative about $H_t$ and $\{\theta_{i}^{(w)},\theta_{i}^{(y)}\}_{i\le n}$---we need to make ad hoc choices. We construct $H_t$ as a linear combination of $Z_t$ and an independent random process with the same distribution as $Z_t$. We set $\theta_i^{(w)}$ to be equal to a linear combination of $\hat\pi_i$ and an independent standard normal variable, and we do the same for $\theta_i^{(y)}$. 

Note that this simulation design is not perfectly aligned with our theoretical results in Section \ref{sec:formal_an}. We allow for additional randomness in errors $\{\epsilon_{it}^{(y)},\epsilon_{it}^{(w)}\}_{i\le n,t\le T}$. We make this choice to make the simulation less driven by particular features of the dataset, essentially allowing for matrices that are in the ``neighborhood'' of the original ones. We report the simulation results without these errors in Appendix \ref{ap:ad_sim}, and they largely agree with those in Table \ref{table:sim_res}.

\begin{figure}[t!]
\begin{center}
\caption{Distribution of Errors, $\hat\tau-\tau$}
\label{fig:densities}
\includegraphics[scale = 0.45]{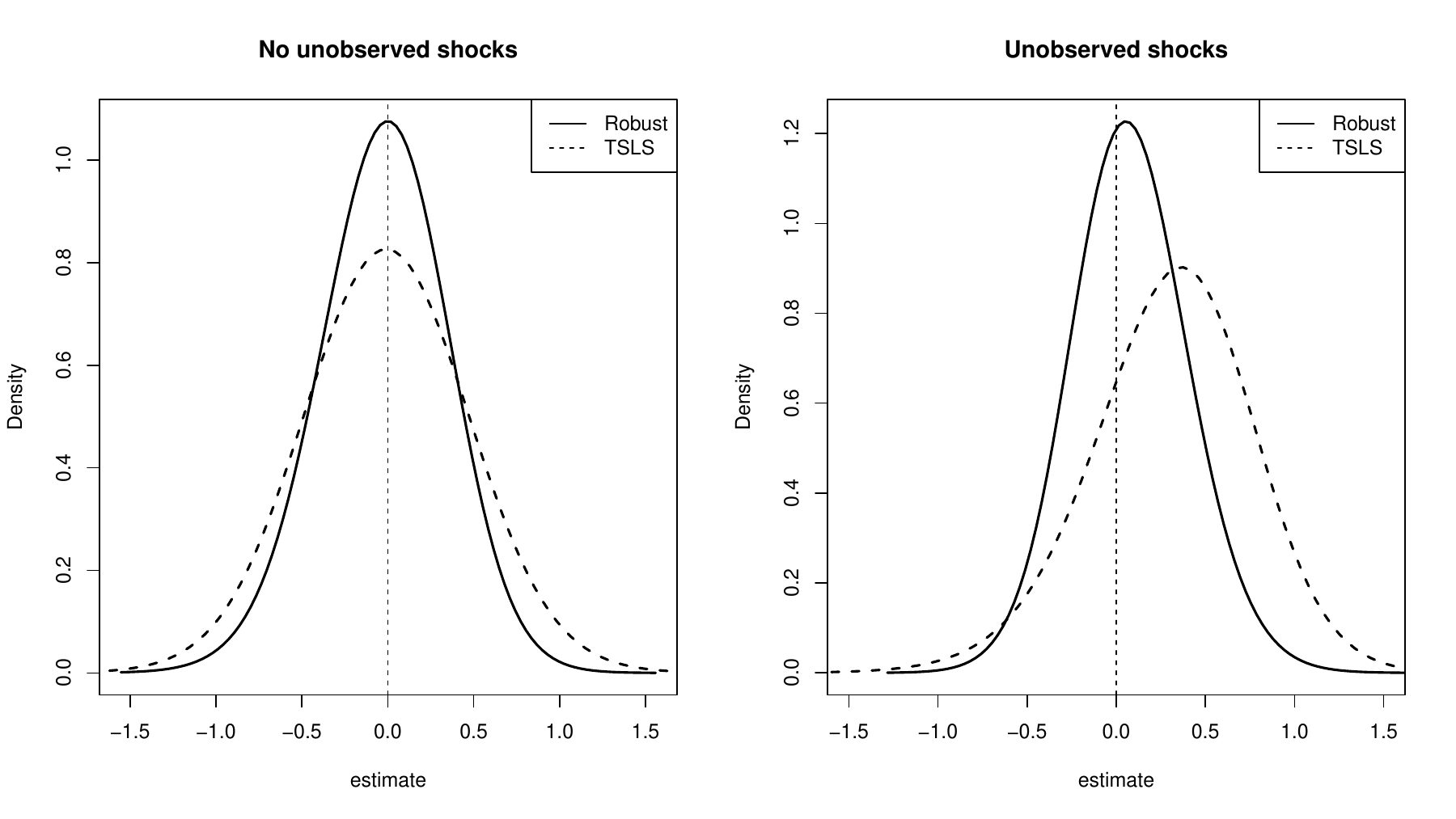} 
\end{center}
        \renewcommand{\baselinestretch}{0.7}
        \footnotesize{\textit{Notes}:  The figure reports the densities for TSLS (dashed line) and our robust estimator (solid line) for $\hat\tau-\tau$. The left figure corresponds to the design in Column (2) of Table \ref{table:sim_res}, with generalized fixed effects and no unobserved aggregate confounders. The right figure corresponds to Column (4) of Table \ref{table:sim_res}, with both generalized fixed effects and unobserved aggregate confounders.}
\end{figure}

We compare the performance of our estimator (as described by Algorithm \ref{alg:const_weights}) with the standard TSLS algorithm from Section \ref{sec:sec_2}. In both cases, we use the data to construct $D_i$ by estimating the next equation by OLS, using data for $t\le \frac{T}{3}$:
\begin{equation}
    W_{it} = \alpha_i + \pi_i Z_t + \varepsilon_{it}
\end{equation}
and set $D_i := \hat \pi_i$. We consider four different designs. In the first design we drop $L_{it}^{(w)},L_{it}^{(y)}$, and $H_t$ from the model (\ref{eq:sim_dgp}). In this case, the TSLS algorithm should perform better than ours because it uses the optimal weights. With the second design, we start to increase the complexity, adding $L_{it}^{(w)}$ and $L_{it}^{(y)}$ back to the model. One can think of this design as a DGP for the data from \cite{nakamura2014fiscal} under which the TSLS approach is justified. Here, we should expect both algorithms to perform well in terms of bias but potentially differ in variance. In the third design, we drop $L_{it}^{(w)},L_{it}^{(y)}$ but add $H_t$. Finally, in the fourth case, we have $L_{it}^{(w)}$ and $L_{it}^{(y)}$, and $H_t$.

\begin{table}[t]
\begin{center}
\caption{Root-Mean-Square Error and Bias in Four Simulation Designs}\label{table:sim_res}
\begin{tabular}{lcc|cc|cc|cc} \hline
& \multicolumn{2}{c}{(1)}& \multicolumn{2}{c}{(2)}& \multicolumn{2}{c}{(3)}& \multicolumn{2}{c}{(4)}\\ 
  &\multicolumn{2}{c}{Basic}& \multicolumn{2}{c}{GFE}& \multicolumn{2}{c}{Agg. Sh.}& \multicolumn{2}{c}{GFE+Agg. Sh.}\\ 
 Estimator & RMSE & Bias & RMSE & Bias & RMSE & Bias & RMSE & Bias \\ 
  \hline
$\hat\pi_{rob}$& 0.01 & 0.00 & 0.04 & 0.00 & 0.05 & 0.04 & 0.17 & 0.13 \\ 
$\hat\pi_{TSLS}$ &0.01 & 0.00 & 0.05 & -0.00 & 0.28 & 0.24 & 0.24 & 0.21 \\ 
  \hline
$\hat\delta_{rob}$ & 0.06 & 0.00 & 0.31 & -0.03 & 0.11 & 0.07 & 0.41 & 0.26 \\ 
$\hat\delta_{TSLS}$ &0.05 & 0.00 & 0.39 & -0.02 & 0.79 & 0.69 & 0.75 & 0.57 \\ 
    \hline
   $\hat \tau_{rob}$ & 0.06 & 0.00 & 0.38 & -0.02 & 0.07 & 0.02 & 0.34 & 0.08 \\ 
$\hat\tau_{TSLS}$ &0.05 & 0.00 & 0.49 & -0.01 & 0.36 & 0.31 & 0.55 & 0.31 \\ 
   \hline \hline
\end{tabular}
\end{center}
\renewcommand{\baselinestretch}{0.7}
\footnotesize{\textit{Notes}: We performed $1000$ replications for each design. The true parameter value $\tau$ is set to $1.43$ to capture the \cite{nakamura2014fiscal} original estimate. Column (1)---first design: no generalized FE, no unobserved confounder. Column (2)---second design: generalized fixed effects, no unobserved confounder. Column (3)---third design: no generalized fixed effects, unobserved confounder. Column (4)---fourth design: generalized fixed effects, unobserved confounder.}
\end{table}

In Table \ref{table:sim_res}, we report results over replication $1000$ for simulations for the case of $\tau =  1.43$, corresponding to the original point estimate obtained by \cite{nakamura2014fiscal}. The results confirm the above intuition: in the simplest case, our estimator is less precise than $\hat \tau_{TSLS}$, although the difference is small. We see sizable gains in RMSE for the second design. In the third case, our estimator eliminates most of the bias, while the TSLS error is dominated by it. Finally, in the most general design, our estimator is nearly unbiased and dominates the TSLS in terms of RMSE. In Figure \ref{fig:densities}, we plot the densities of $\hat \tau - \tau$ over the simulations for the second and fourth designs. These plots demonstrate the gains in variance and bias and show the estimator's overall behavior. Once again, we see that even when TSLS is approximately unbiased, there are gains from using our approach that come from increased precision.

The comparison in Table \ref{table:sim_res} emphasizes the benefits of our approach, with an efficiency loss being present only in the most unrealistic first design. Importantly, these results do not reflect the efficiency cost of not using only part of the data, which we discussed in Section \ref{sec:applying}. This cost is present for both estimators in our simulation because we use the first third of the data to construct $\{D_i\}_{i\le n}$. Formally, this situation is not covered by our theoretical results, which treat $\{D_i\}_{i \le n}$ as fixed quantities. However, the results in Table \ref{table:sim_res} show that the algorithm continues to perform well in this regime, which is relevant for some empirical applications. 

We also investigate the performance of our inference approach as described in Algorithm \ref{alg:inf}. In Table \ref{table:conf}, we report coverage rates for nominal $95\%$ confidence intervals for $\hat\tau_{rob}$ and $\hat \tau_{TSLS}$. We construct $\hat\Lambda^{(z)}_{T_0+1|T}$ by fitting an ARIMA model to the data $\{Z_t\}_{t\le T}$ using the automatic model selection package in \textbf{R}. We see that the coverage is below nominal for all designs and estimators. This is not surprising, given that the sample size is relatively small, and in the third and fourth designs, both estimators are biased. In relative terms, the coverage for $\hat \tau_{rob}$ is closer to the nominal one.

\begin{table}[t]
\begin{center}
\caption{Coverage Rates for $95\%$ Confidence Intervals}\label{table:conf}
\begin{tabular}{lcccc}
\hline
  & (1) & (2) & (3) & (4) \\ 
  &\multicolumn{1}{c}{Basic}& \multicolumn{1}{c}{GFE}& \multicolumn{1}{c}{Agg. Sh.}& \multicolumn{1}{c}{GFE+Agg. Sh.}\\
  \hline
$\hat\tau_{rob}$&  0.91 & 0.86 & 0.80 & 0.84  \\ 
$\hat\tau_{TSLS}$& 0.90 & 0.85 & 0.33 & 0.81 \\ 
   \hline \hline
\end{tabular}
\end{center}
\renewcommand{\baselinestretch}{0.7}
\footnotesize{\textit{Notes}: The table reports coverage rates for $95\%$ confidence intervals based on Algorithm \ref{alg:inf}. Each simulation has $1000$ replications, and the true parameter value $\tau$ is set to $1.43$. Column (1)--first design: no generalized FE, no unobserved confounders. Column (2)--second design: generalized fixed effects, no unobserved shock. Column (3)--third design: no generalized fixed effects, unobserved confounder. Column (4)--fourth design: generalized fixed effects, unobserved confounder.}
\end{table}


\section{Conclusion}\label{sec:concl}

Aggregate instruments provide a natural source of exogenous variation for unit-level outcomes. As a result, they are frequently used to evaluate the effects of local policies. This exercise has two conceptual steps: aggregating unit-level data into a time series and analyzing the aggregated data. We propose a new algorithm for constructing unit weights to produce aggregate outcomes. We use a flexible statistical model to show that our weights eliminate potential unobserved aggregate confounders, leading to a consistent and asymptotically normal estimator. Using data-driven simulations, we demonstrate the superiority of our proposal over the conventional TSLS estimator in various relevant regimes.

\newpage
\bibliographystyle{plainnat}
\bibliography{references}

\newpage
\appendix 
\appendixpagenumbering
\section{Additional Analysis}\label{ap:rob_check}

\begin{figure}[ht] 
    \begin{center}
    \caption{Reduced-Form and First-Stage Coefficients for \cite{nakamura2014fiscal} Data} \label{fig:cs_graph_ap} 
     \begin{subfigure}[b]{0.49\textwidth}
         \caption*{\textbf{Panel A:} Nakamura and Steinsson Weights}
         \centering
         \includegraphics[width=\textwidth]{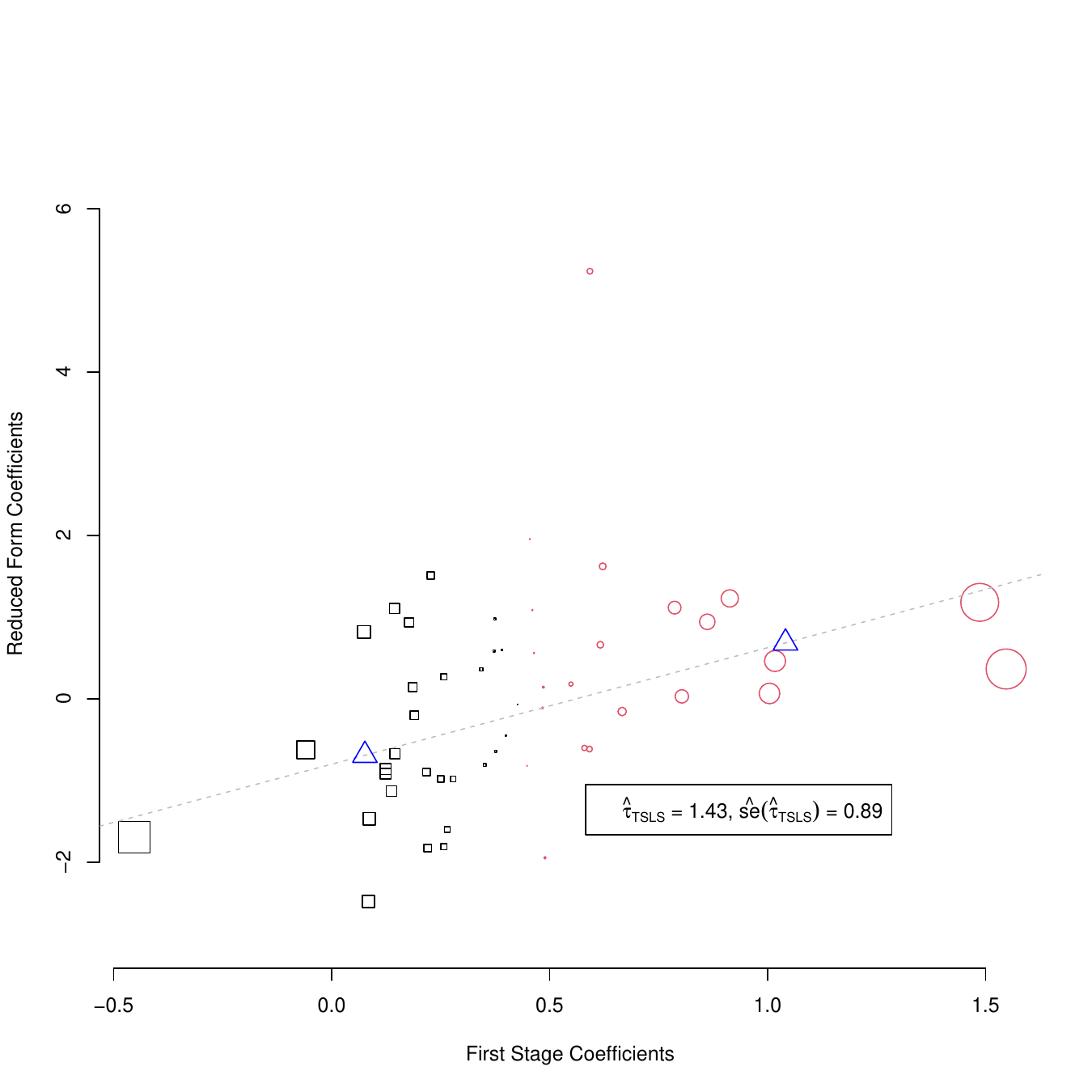}
     \end{subfigure}
     \begin{subfigure}[b]{0.49\textwidth}
        \caption*{\textbf{Panel B:} Robust Weights
        }
         \centering
         \includegraphics[width=\textwidth]{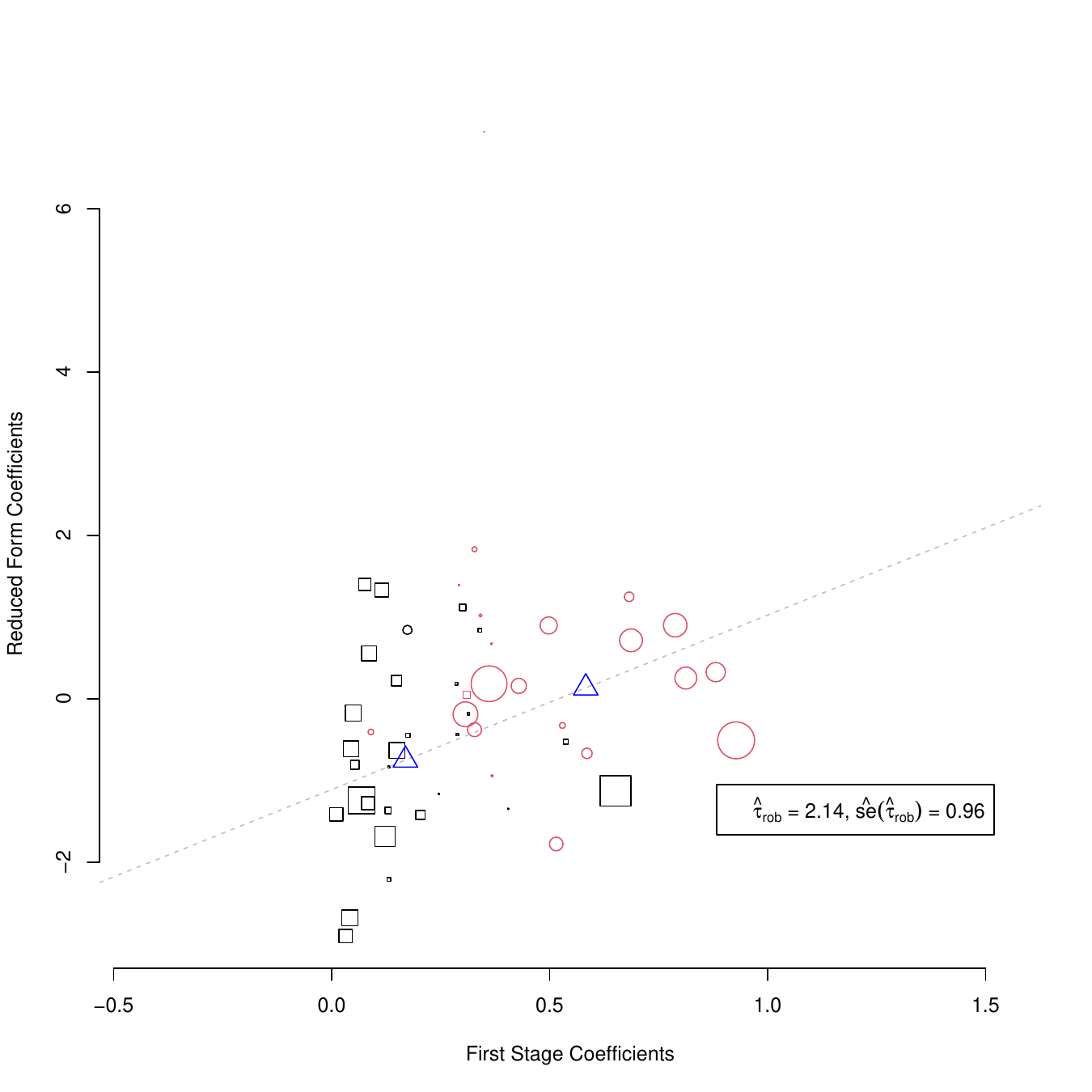}
     \end{subfigure}
     \end{center}
    \renewcommand{\baselinestretch}{0.7}
    \footnotesize{\textit{Notes}: Shape sizes reflect the absolute value of weights; negative weights are printed in black squares, and positive are in red circles.
Panel A presents the results using the whole period of 1968 to 2006 for $n=51$ states. Panel B shows the results from our estimation algorithm. Under our data-splitting procedure, Panel B reports the results for 1978 to 2006, as we use the first third of the data for weight estimation.}
\end{figure}

In this section, we repeat the analysis described in Section \ref{sec:sec_2}, but now for the full original sample. The results are largely similar, and we comment only on the differences. In Figure \ref{fig:cs_graph_ap}, we plot reduced-form and the first-stage coefficients for various periods. Compared with Figure \ref{fig:cs_graph} reported in the main text, we see that states with negative first-stage coefficients receive a large weight in the original exercise, pushing the slope of the line down, which results in a larger coefficient (the same as reported in \cite{nakamura2014fiscal}). We also see that a single state---Alaska---has an extreme reduced-form coefficient, three times larger than the second largest. 

\begin{figure}[t!]
        \begin{center}
        \caption{Aggregate Time-Series for \cite{nakamura2014fiscal} Data} \label{fig:ts_graph_ap}
     \begin{subfigure}{0.85\textwidth}
         \centering
         \caption*{\textbf{Panel A:} Aggregation Over $n = 51$ States With Original Weights}
         \includegraphics[width=\textwidth]{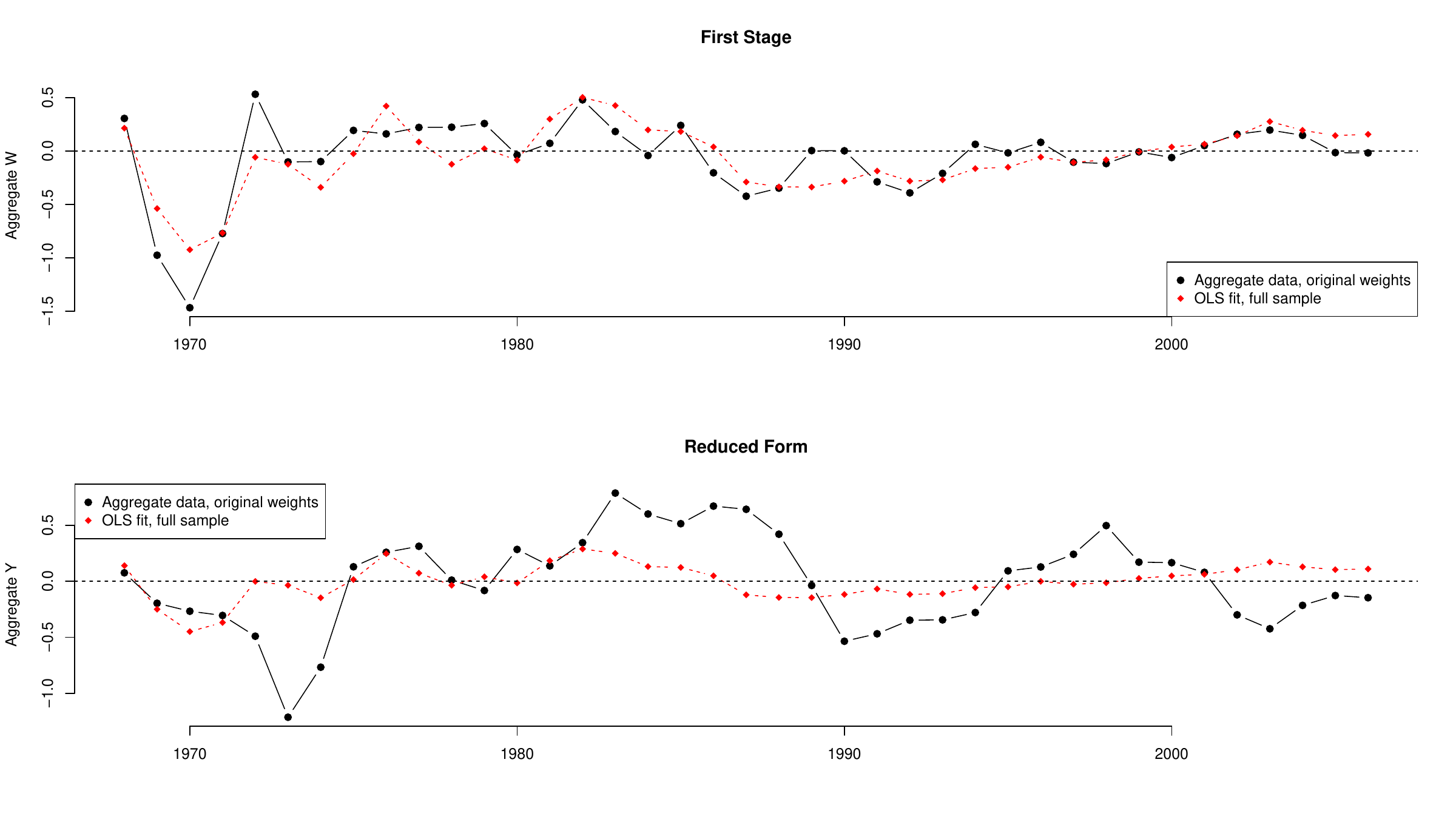}
     \end{subfigure}
     \begin{subfigure}{0.85\textwidth}
         \centering
        \caption*{\textbf{Panel B:} Aggregation Over $n = 51$ States With Robust Weights}
         \includegraphics[width=\textwidth]{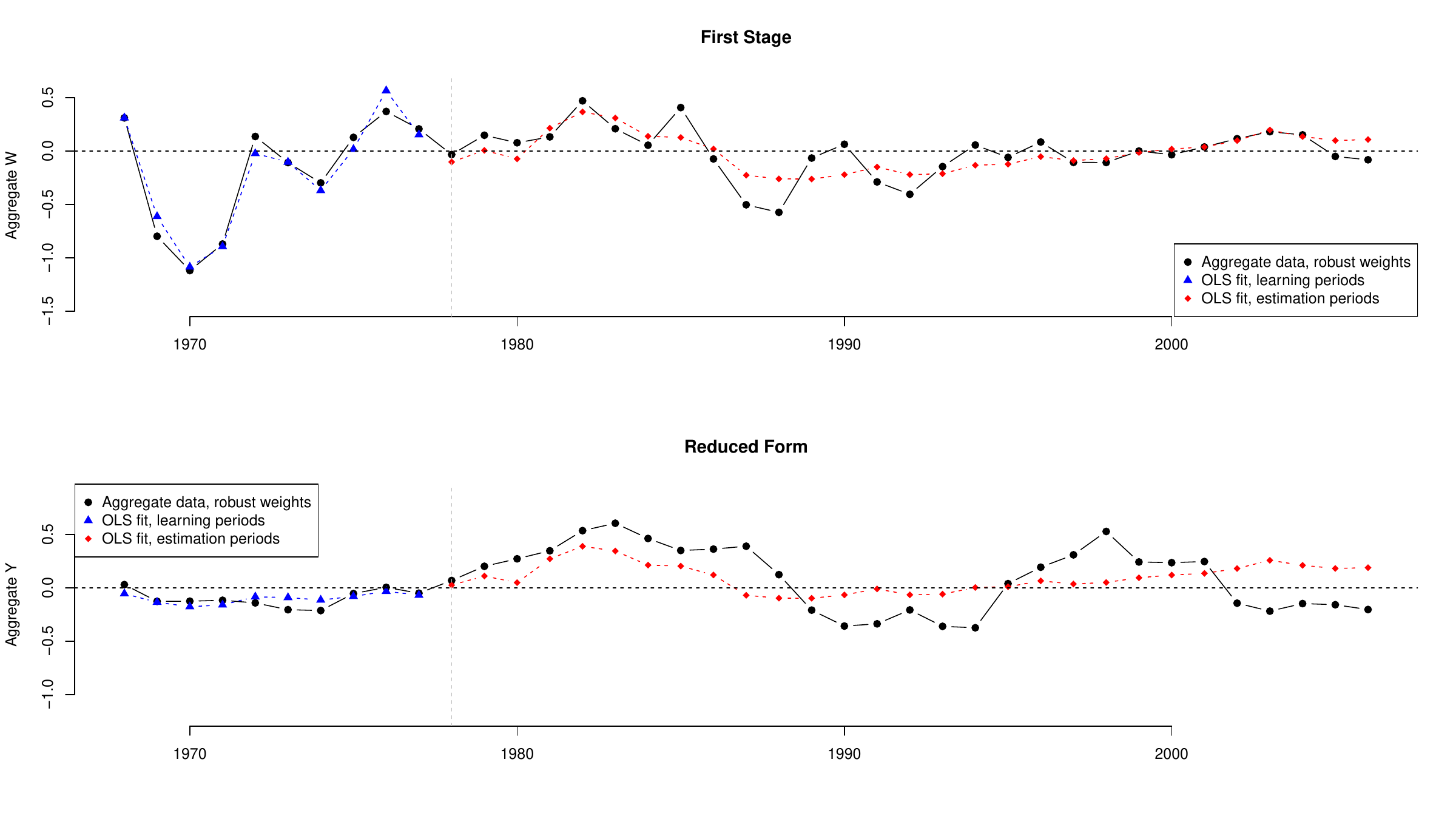}
     \end{subfigure}
     \end{center}
    \renewcommand{\baselinestretch}{0.7}
    \footnotesize{\textit{Notes}: Solid lines represent aggregate data for different weights; dashed lines represent OLS predictions of the aggregate data with the instrument. The mean absolute value of weights is scaled to $1$.}
\end{figure}

\begin{figure}[t!]
    \begin{center}
        \caption{Scatterplot---Nakamura and Steinsson Weights and Robust Weights } \label{fig:scatter_weights_ap}
        \includegraphics[width=0.65\textwidth]{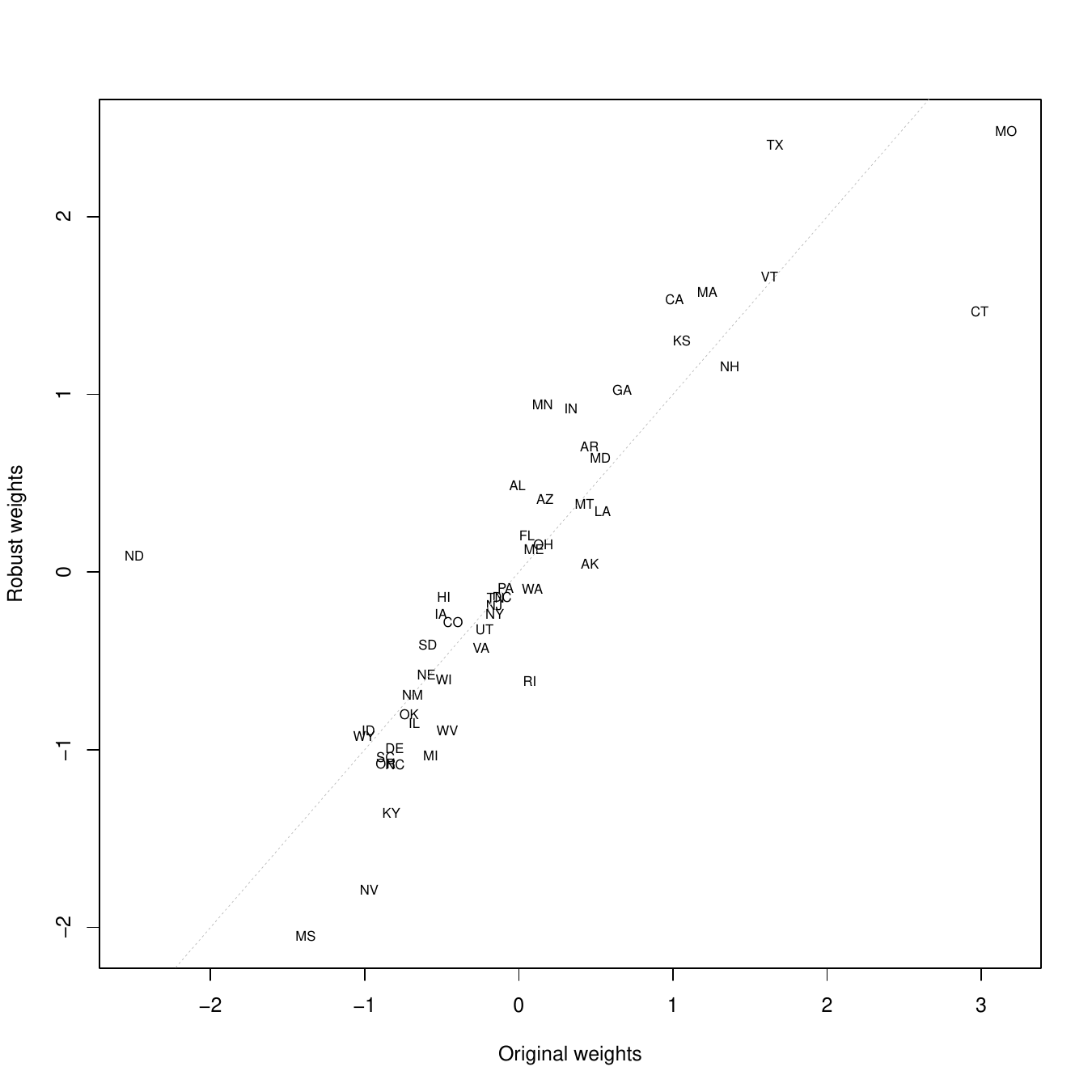}
        \end{center}
        \renewcommand{\baselinestretch}{0.7}
        \footnotesize{\textit{Notes}: $n = 51$; state abbreviations are used as labels. The variance of weights is scaled to $1$. }
\end{figure}

If we compare $\hat \tau_{rob}$ with the estimator that uses the second part of the data but with original weights, they are now different---$2.14$ and $1.83$, respectively. Still, there is a significant difference in estimated standard errors: $31\%$, in favor of the robust estimator. It confirms the intuition from the time-series plots \ref{fig:ts_graph_ap}, where the aggregate data is much better predicted by $Z_t$ if we use robust weights vs. the original ones. The increase in estimated $R^2$ is $57\%$ for the endogenous variable and more than $100\%$ for the outcome variable. Figure \ref{fig:scatter_weights_ap} suggests that these gains come from compressing the distribution of the weights.


\clearpage
\begin{center}
    \LARGE {Online Appendix}
\end{center}
\section{Proofs}\label{ap:proofs}
\footnotesize

In Sections \ref{sec:ap_1} to \ref{sec:ap_3}, we prove the results stated in the main text. In Section \ref{sec:ap_1}, we prove Theorems \ref{th:cons} to \ref{th:inference}, taking the results about the robust weights $\wrob_i$ as given. In Section \ref{sec:ap_2}, we consider an abstract quadratic optimization problem and derive properties of its solution. We then establish a connection between abstract stochastic and deterministic optimization problems.  In Section \ref{sec:ap_3}, we specialize this connection to the probabilistic models from the main text and prove the results about the robust weights. 

We use $\|\cdot\|_2$ to denote the euclidean norm, $\|\cdot\|_{\infty}$ to denote the $\sup$-norm, $\|\cdot \|_{HS}$ to denote the Hilbert-Schmidt norm, and $\|\cdot \|_{op}$ -- the operator norm. For a random vector $X$, we use $\|X\|_{\psi_2}$ to denote its sub-Gaussian norm. We use $\trace{A}$ to denote the trace of a square matrix $A$. For a given set of variables $\{X_i\}_{i \le n}$, we use $\mathbb{P}_n X_i$ to denote their average.  For any $T\ge T_b > T_a\ge 1$, we define two projection matrices:
\begin{equation*}
    \Pi^{f,r}_{T_a|T_b} = \frac{1}{T_b-T_a+1}\mathbf{1}_{T_b-T_a+1}\mathbf{1}_{T_b-T_a+1}^\top,\quad \left(\Pi^{f,r}_{T_a|T_b}\right)^\bot =\mathcal{I}_{T_b-T_a+1}- \Pi^{f,r}_{T_a|T_b}
\end{equation*}
Also, for any $T>T_a>1$ and $k\in \{z,h\}$, we define  $\Lambda^{(k),1|T_a}_{T_a+1|T}$ and $ \Lambda^{(k),T_a+1|T}_{T_a+1|T}$ as submatrices of $\Lambda^{(k)}_{T_a+1|T}$ that correspond to shocks from periods $[1,T_a]$ and $(T_a,T]$, respectively. We also define the matrix that projects out the time fixed effects:
\begin{equation*}
    \Pi^{\bot}_{l,f} = \mathcal{I}_{n} - \frac{1}{n}\mathbf{1}_{n}^\top\mathbf{1}_{n}
\end{equation*}

\subsection{Part I}\label{sec:ap_1}

\subsubsection{Technical Lemmas}

\begin{lemma}\label{lem:proj}
Let $\Pi_p$ be an orthogonal projector on $p$-dimensional subspace or $\mathbb{R}^{T}$ and consider a $T\times n$ matrix $A$ such that $\frac{\|A\|_{op}}{\|A\|_{HS}} = o\left(\frac{1}{\sqrt{p}}\right)$. Then we have
\begin{equation*}
    \frac{\|(\mathcal{I}_T-\Pi) A\|_{HS}}{\|A\|_{HS}} = 1 + o(1)
\end{equation*}
\end{lemma}
\begin{proof}
The result follows from a chain of inequalities:
\begin{equation*}
    \left| \frac{\|(\mathcal{I}_T-\Pi) A\|_{HS}}{\|A\|_{HS}} - 1\right| \le \frac{\|\Pi A\|_{HS}}{\|A\|_{HS}} \le \frac{\|\Pi\|_{HS}\|A\|_{op}}{\|A\|_{HS}} = \sqrt{p}\times o\left(\frac{1}{\sqrt{p}}\right) = o(1)
\end{equation*}
\end{proof}

\begin{lemma}\label{lem:han_wr}
Suppose $\nu^{(z)}$ and $\nu^{(h)}$ are independent, isotropic mean-zero vectors with independent coordinates and subgaussian norms bounded by $1$. Then for any $x <1$ and an absolute constant $c$, we have with probability at least $1- 4\exp\left(- cx^2\frac{\|B\|^2_{HS}}{\|B\|_{op}^2}\right)$
\begin{equation*}
\begin{aligned}
      &\left|(\nu^{(z)})AB\nu^{(z)} - \trace{AB}\right| \le x\frac{\|A\|_{op}\|B\|^2_{HS}}{\|B\|_{op}}\\
    &\left|(\nu^{(z)})AB\nu^{(h)}\right| \le x\frac{\|A\|_{op}\|B\|^2_{HS}}{\|B\|_{op}}
\end{aligned}
\end{equation*}
\end{lemma}
\begin{proof}
Proof follows directly from Hanson-Wright inequality and its proof (e.g., Theorem 6.2.1 in \cite{vershynin2018high}).
\end{proof}

\subsubsection{Theorems in the Main Text}

\noindent\textbf{Proof of Theorem \ref{th:cons}:}
\begin{proof}
We start with the TSLS estimator, which under Assumptions \ref{as:pot_out} and \ref{as:basic_expos} can be represented as
\begin{equation}
    \hat \tau_{TSLS} - \tau = \frac{\frac{1}{\sqrt{T}}\alpha_{1|T}^{(y)}(\omega^{TSLS})\epsilon^{(z)} +\frac{1}{T} \hat\rho(\theta^{(y)},D)\hat \sigma(\theta^{(y)})/\hat \sigma(D)(\epsilon^{(h)})^{\top}(\Pi^{f,r}_{1|T})^{\bot}\epsilon^{(z)}}{\frac{1}{\sqrt{T}}\alpha_{1|T}^{(w)}(\omega^{TSLS})\epsilon^{(z)} + \frac{1}{T}\hat\rho(\theta^{(w)},D)\hat \sigma(\theta^{(w)})/\hat \sigma(D)(\epsilon^{(h)})^{\top}(\Pi^{f,r}_{1|T})^{\bot}\epsilon^{(z)} + \frac{1}{T}\eta_{\pi}(\epsilon^{(z)})^{\top}(\Pi^{f,r}_{1|T})^{\bot}\epsilon^{(z)}}
\end{equation}
We have for $k\in \{y,w\}$
\begin{multline}
    \alpha_{1|T}^{(k)}(\omega^{TSLS})\epsilon^{(z)} =  \alpha_{1|T}^{(k)}(\omega^{TSLS})\Lambda^{(z)}\nu^{(z)} \Rightarrow \|\alpha_{1|T}^{(k)}(\omega^{TSLS})\epsilon^{(z)}\|_{\psi_2} \lesssim \|\alpha_{1|T}^{(k)}(\omega^{TSLS})\Lambda^{(z)}\|_2 \le \\
    \frac{\| \alpha^{(k)}\|_{op}}{\sqrt{Tn}}\frac{\|\omega^{TSLS}\|_2}{\sqrt{n}}\|\Lambda^{(z)}\|_{op} \lesssim1
\end{multline}
where the first implication follows from Assumption \ref{as:des_model}, and the last inequality follows from Assumption \ref{as:asym_regime}. We next apply Lemma \ref{lem:han_wr} for $x = \frac{1}{\sqrt{T}}$, and we use Assumption \ref{as:des_model} and Lemma \ref{lem:proj} to get
\begin{equation}
\begin{aligned}
&\frac{1}{T}(\epsilon^{(h)})^{\top}(\Pi^{f,r}_{1|T})^{\bot}\epsilon^{(z)} = \rho_{1|T}\sigma_{h,1|T}\sigma_{z,1|T}  + O_{p}\left(\frac{1 }{\sqrt{T}}\right),\\ &\frac{1}{T}(\epsilon^{(z)})^{\top}(\Pi^{f,r}_{1|T})^{\bot}\epsilon^{(z)} = \sigma^2_{z,1|T} + O_{p}\left(\frac{1 }{\sqrt{T}}\right)
\end{aligned}
\end{equation}
Using these bounds we get
\begin{equation*}
     \hat \tau_{TSLS} - \tau  = \frac{\hat\rho_{cs}^{(y)} \hat\sigma_{\theta^{(y)}}\rho_{1|T}\sigma_{h,1|T} + O_{p}\left(\frac{1 }{\sqrt{T}}\right)}{\hat\rho_{cs}^{(w)} \hat\sigma_{\theta^{(w)}}\rho_{1|T}\sigma_{h,1|T} + \eta_{\pi}\hat\sigma_{D}\sigma_{z,1|T}+O_{p}\left(\frac{1 }{\sqrt{T}}\right)} =  \frac{\rho_{cs}^{(y)} \sigma_{\theta^{(y)}}\rho_{1|T}\sigma_{h,1|T} + o(1) + O_{p}\left(\frac{1 }{\sqrt{T}}\right)}{\rho_{cs}^{(w)} \sigma_{\theta^{(w)}}\rho_{1|T}\sigma_{h,1|T} + \eta_{\pi}\sigma_{D}\sigma_{z,1|T}+o(1)+O_{p}\left(\frac{1 }{\sqrt{T}}\right)}
\end{equation*}
and the result for the TSLS follows. 

Under Assumptions \ref{as:pot_out} and \ref{as:basic_expos}, we have
\begin{multline}
    \hat \tau_{rob} - \tau=\\ \frac{\frac{1}{\sqrt{T_1}}\alpha_{T_0+1|T}^{(y)}(\wrob)\epsilon_{T_0+1|T}^{(z)} +\frac{1}{T_1} \mathbb{P}_n \wrob_i\theta_i^{(y)}(\epsilon_{T_0+1|T}^{(h)})^{\top}(\Pi^{f,r}_{T_0+1|T})^{\bot}\epsilon_{T_0+1|T}^{(z)}}
    {\frac{\alpha_{T_0+1|T}^{(w)}(\wrob)\epsilon_{T_0+1|T}^{(z)}}{\sqrt{T_1}} + \frac{\mathbb{P}_n \wrob_i\theta_i^{(w)}(\epsilon_{T_0+1|T}^{(h)})^{\top}(\Pi^{f,r}_{T_0+1|T})^{\bot}\epsilon_{T_0+1|T}^{(z)}}{T_1} + \frac{\eta_{\pi}(\epsilon_{T_0+1|T}^{(z)})^{\top}(\Pi^{f,r}_{T_0+1|T})^{\bot}\epsilon_{T_0+1|T}^{(z)}}{T_1}}
\end{multline}
Similar to the result above, we have:
\begin{equation}
\begin{aligned}
&\frac{1}{T_1}(\epsilon^{(h)})^{\top}(\Pi^{f,r}_{T_0+1|T})^{\bot}\epsilon^{(z)} = \rho_{T_0+1|T}\sigma_{h,T_0+1|T}\sigma_{z,T_0+1|T}  + O_{p}\left(\frac{1 }{\sqrt{T_1}}\right)\\ &\frac{1}{T_1}(\epsilon^{(z)})^{\top}(\Pi^{f,r}_{T_0+1|T})^{\bot}\epsilon^{(z)} = \sigma^2_{z,T_0+1|T} + O_{p}\left(\frac{1}{\sqrt{T_1}}\right)
\end{aligned}
\end{equation}
and by (\ref{eq:result_1}), we have
\begin{equation*}
    \mathbb{P}_n \wrob_i\theta_i^{(w)} = o_p(1),\quad \mathbb{P}_n \wrob_i\theta_i^{(y)} = o_p(1), \quad \frac{\|\wrob\|_2}{\sqrt{n}} = O_p(1)
\end{equation*}
By definition $\epsilon_{T_0+1|T}^{(z)} =  \Lambda^{(z),T_0+1|T}_{T_0+1|T} \nu_{T_0+1|T}^{(z)} + \Lambda^{(z),1|T_0}_{T_0+1|T} \nu_{1|T_0}^{(z)}$
and by concentration for anisotropic vectors, we have
\begin{equation}
     \|\Lambda^{(z),1|T_0}_{T_0+1|T} \nu_{1|T_0}^{(z)}\|_2 =  \|\Lambda^{(z),1|T_0}_{T_0+1|T}\|_{HS} + O_p\left(\|\Lambda^{(z),1|T_0}_{T_0+1|T}\|_{op}\right)\le O_p(\|\Lambda^{(z),1|T_0}_{T_0+1|T}\|_{HS})
\end{equation}
Since the weights $\wrob_i$ are independent of $\nu_{T_0+1|T}^{(z)}$ by construction, we have for $k\in \{y,w\}$
\begin{multline}
     \|\alpha_{T_0+1|T}^{(k)}(\wrob)\epsilon_{T_0+1|T}^{(z)}\|_2 \le \|\alpha_{T_0+1|T}^{(k)}(\wrob)\Lambda^{(z),T_0+1|T}_{T_0+1|T} \nu_{T_0+1|T}^{(z)}\|_2 + \|\alpha_{T_0+1|T}^{(k)}(\wrob)\Lambda^{(z),1|T_0}_{T_0+1|T} \nu_{1|T_0}^{(z)}\|_2  = \\
     O_p\left(\frac{\|\alpha^{(k)}\|_{op}}{\sqrt{nT_1}}\frac{\|\wrob\|_2}{\sqrt{n}} \left(\| \Lambda^{(z),T_0+1|T}_{T_0+1|T}\|_{op} + \|\Lambda^{(z),1|T_0}_{T_0+1|T}\|_{HS}\right)\right) = O_p(1)
\end{multline}
The result for $\hat \tau^{rob}$ then follows by combining all the bounds. 
\end{proof}
\noindent\textbf{Proof of Theorem \ref{th:limit_beh}}
\begin{proof}
By (\ref{eq:result_2}) we have:
\begin{equation}
        \mathbb{P}_n \wrob_i\theta_i^{(w)} = o_p\left(\frac{1}{\sqrt{T_0}}\right),\quad \mathbb{P}_n \wrob_i\theta_i^{(y)} = o_p\left(\frac{1}{\sqrt{T_0}}\right), \quad
        \frac{1}{\sqrt{n}}\|\wrob -\wdet_{T_0}\|_2 = o_p(1).
\end{equation}
Using the expansion from the previous theorem, we can conclude that the dominant part of the error is coming from $\alpha_{T_0+1|T}^{(y)}(\wrob)\epsilon_{T_0+1|T}^{(z)}$. We can split this term into two parts:
\begin{equation}
    \alpha_{T_0+1|T}^{(y)}(\wrob)\epsilon_{T_0+1|T}^{(z)} = \alpha_{T_0+1|T}^{(y)}(\wrob - \wdet_{T_0})\epsilon_{T_0+1|T}^{(z)} +  \alpha_{T_0+1|T}^{(y)}(\wdet_{T_0})\epsilon_{T_0+1|T}^{(z)}.
\end{equation}
By a straightforward extension of the argument in the previous proof, we can conclude that the first term is $o_p(1)$.  Using this we get the following expression:
\begin{equation}
    \sqrt{T_1}\left(\hat \tau_{rob} - \tau\right) = \frac{\left\|\alpha_{T_0+1|T}^{(y)}(\wdet_{T_0})\Lambda^{(z)}_{T_0+1|T}\right\|_2}{\eta_{\pi}\sigma^2_{z,T_0+1|T}}\xi_n + o_p(1).
\end{equation}
where $\xi_n$ is a sub-Gaussian centered random variable with unit variance. To justify normality we use the bound:
\begin{equation}
    \frac{\left\|\alpha_{T_0+1|T}^{(y)}(\wdet_{T_0})\Lambda^{(z)}_{T_0+1|T}\right\|_{\infty}}{\left\|\alpha_{T_0+1|T}^{(y)}\wdet_{T_0})\Lambda^{(z)}_{T_0+1|T}\right\|_2} \le \frac{\left\|\left(\Lambda^{(z)}_{T_0+1|T}\right)^{\top}\right\|_{\infty}}{\sigma_{\min}}\frac{\left\|\alpha_{T_0+1|T}^{(y)}(\wdet_{T_0})\right\|_{\infty}}{\left\|\alpha_{T_0+1|T}^{(y)}(\wdet_{T_0})\right\|_2} = o(1)
\end{equation}
where we used the bound
\begin{multline}
    \left\|\alpha_{T_0+1|T}^{(y)}(\wdet_{T_0})\Lambda^{(z)}_{T_0+1|T}\right\|_2^2 = \left\|\alpha_{T_0+1|T}^{(y)}(\wdet_{T_0})\Lambda^{(z),T_0+1|T}_{T_0+1|T}\right\|_2^2 +\\ \left\|\alpha_{T_0+1|T}^{(y)}(\wdet_{T_0})\Lambda^{(z),1|T_0}_{T_0+1|T}\right\|_2^2 \ge \sigma^2_{\min}\|\alpha_{T_0+1|T}^{(y)}(\wdet_{T_0})\|_2^2
\end{multline}
Using Lindeberg's CLT, we can conclude that $\xi_n$ converges in distribution to a standard normal distribution.
\end{proof}
\noindent\textbf{Proof of Theorem \ref{th:inference}}
\begin{proof}
The hypothesis of the theorem guarantees that $\sqrt{T_1}(\hat \tau_{rob} - \tau)$ is asymptotically normal. We thus only need to guarantee that $\hat \sigma_{rob}$ is consistent for the asymptotic standard error. Following the steps of the proof of Theorem \ref{th:cons}, it is straightforward to show that $\hat \pi_{rob}$ is consistent for $\eta_{\pi}$. We also have
\begin{multline}
    \frac{1}{T_1}\sum_{T_0 < t < T}\left(Z_t-\frac{\sum_{T_0 < l \le T}Z_l}{T_1}\right)^2 =\frac{1}{T_1} \left(\Lambda^{(z)}_{T_0+1|T}\nu^{(z)}\right)^{\top}\left(\Pi^{f,r}_{T_0+1|T}\right)^\bot  \Lambda^{(z)}_{T_0+1|T}\nu^{(z)} = \\
    \sigma^2_{z,T_0+1|T} + o_p(1)
\end{multline}
where the last inequality follows from Lemma \ref{lem:han_wr} for $x = o(1)$, Lemma \ref{lem:proj}, Assumption \ref{as:des_model}, and definition of  $\sigma^2_{z,T_0+1|T}$. Finally, we have the following:
\begin{multline}
   \left| \left\|\hat\alpha_{T_0+1|T}^{(y)}(\wdet_{T_0})\hat\Lambda^{(z)}_{T_0+1|T}\right\|_2  - \left\|\alpha_{T_0+1|T}^{(y)}(\wdet_{T_0})\Lambda^{(z)}_{T_0+1|T}\right\|_2\right| \le \left\|\alpha_{T_0+1|T}^{(y)}(\wdet_{T_0}) -\hat\alpha_{T_0+1|T}^{(y)}(\wdet_{T_0}) \right\|_2\left\|\Lambda^{(z)}_{T_0+1|T}\right\|_{op}  \\
    \left(\left\|\alpha_{T_0+1|T}^{(y)}(\wdet_{T_0})\right\|_2 +\left\|\alpha_{T_0+1|T}^{(y)}(\wdet_{T_0}) -\hat\alpha_{T_0+1|T}^{(y)}(\wdet_{T_0}) \right\|_2\right)\|\hat\Lambda^{(z)}_{T_0+1|T} -  \Lambda^{(z)}_{T_0+1|T}\|_{op}  =  \\
    O_p\left(\left\|\alpha_{T_0+1|T}^{(y)}(\wdet_{T_0}) -\hat\alpha_{T_0+1|T}^{(y)}(\wdet_{T_0}) \right\|_2\right) + \\
    o_p\left(\left\|\alpha_{T_0+1|T}^{(y)}(\wdet_{T_0})\right\|_2 +\left\|\alpha_{T_0+1|T}^{(y)}(\wdet_{T_0}) -\hat\alpha_{T_0+1|T}^{(y)}(\wdet_{T_0}) \right\|_2\right)
\end{multline}
The result holds as long as $\left\|\alpha_{T_0+1|T}^{(y)}(\wdet_{T_0})\right\|_2 = O_p(1)$ and $\left\|\alpha_{T_0+1|T}^{(y)}(\wdet_{T_0}) -\hat\alpha_{T_0+1|T}^{(y)}(\wdet_{T_0}) \right\|_2 = o_p(1)$. The second part follows from consistency of $\hat \tau_{rob}$, and (\ref{eq:result_1}) that guarantees $\mathbb{P}_n \wrob_i\theta_i^{(y)} = o_p(1)$. The first part follows from the consistency of $\wrob_i$ and Assumption \ref{as:asym_regime}.
\end{proof}

\textbf{Proof of Proposition \ref{prop:example}:}
\begin{proof}
Consider $\omega_i =\epsilon_i^{(d)} - \mathbb{P}_n\epsilon_i^{(d)}$, it does not satisfy the scale constraint, but as we will see, later it does not matter. By concentration for sub-Gaussian vectors, we have with probability approaching $1$:
\begin{equation}
     \left(\frac{1}{n}\sum_{i\le n}\omega_i\theta_i^{(k)}\right)^2  = \frac{1}{n} |(\epsilon^{(d)})^\top\Pi^{\bot}_{l,f}\Theta^{(k)}|^2\lesssim \frac{\log(n)}{n} \frac{\| \Pi^{\bot}_{l,f}\Theta^{(k)}\|_2^2}{n} \lesssim  \frac{\log(n)}{n}
\end{equation}
Define $\tilde L_{1|T_a}^{(k)} := \frac{\Pi^{\bot}_{l,f}L_{1|T_a}^{(k)}\left(\Pi^{f,r}_{1|T_a}\right)^\bot}{\sqrt{nT_a}}$ and $\tilde E_{1|T_a}^{(k)} := \frac{\Pi^{\bot}_{l,f}E_{1|T_a}^{(k)}\left(\Pi^{f,r}_{1|T_a}\right)^\bot}{\sqrt{nT_a}}$; we have 
\begin{equation}
    \sum_{t\le T_a}\left( \alpha_{t, 1|T_a}^{(k)}(\omega)\right)^2 = \frac{1}{n}\left\| \omega^\top\tilde L_{1|T_a}^{(k)} +\omega^\top\tilde E_{1|T_a}^{(k)}\right\|_2^2 \le \frac{1}{n}\left(\left\| (\epsilon^{(d)})^\top\tilde L_{1|T_a}^{(k)}\right\|_2 + \left\|(\epsilon^{(d)})^\top\tilde E_{1|T_a}^{(k)}\right\|_2\right)^2
\end{equation}
By concentration of anisotropic sub-Gaussian vectors, we have with probability approaching 1,
\begin{equation}
   \left| \left\| (\epsilon^{(d)})^\top\tilde L_{1|T_a}^{(k)}\right\|_2 - \left\|\tilde L_{1|T_a}^{(k)}\right\|_{HS}\right| \lesssim \sqrt{\log(n)} \left\|\tilde L_{1|T_a}^{(k)}\right\|_{op} \lesssim \sqrt{\log(n)}
\end{equation}
By assumption, we also have  $\left\|\tilde L_{1|T_a}^{(k)}\right\|_{HS}\lesssim 1$. Similarly, we have conditionally on $E^{(k)}$ with probability approaching 1,
\begin{equation}
       \left| \left\| (\epsilon^{(d)})^\top\tilde E_{1|T_a}^{(k)}\right\|_2 - \left\|\tilde E_{1|T_a}^{(k)}\right\|_{HS}\right| \lesssim \sqrt{\log(n)} \left\|\tilde E_{1|T_a}^{(k)}\right\|_{op} 
\end{equation}
By concentration of subgaussian random matrices, we have with probability approaching 1,
\begin{equation}
    \left\|\tilde E_{1|T_a}^{(k)}\right\|_{op}\lesssim \frac{\sqrt{n}}{\sqrt{nT_a}} +\frac{\sqrt{T_a}}{\sqrt{nT_a}} \lesssim 1
\end{equation}
and by Hanson-Wright, inequality 
\begin{equation}
     \left\|\tilde E_{1|T_a}^{(k)}\right\|_{HS}^2 = \frac{\trace{\tilde\Sigma_{1|T_a}^{(k)}}}{T_a} + o_p(1)
\end{equation}
It follows that with probability approaching 1, we have
\begin{equation}
    \sum_{t\le T_a}\left( \alpha_{t, 1|T_a}^{(k)}(\omega)\right)^2 \lesssim \frac{\log(n)}{n}.
\end{equation}
Finally, for the denominator, we have 
\begin{multline}
    \sigma^{2}_{k,T_a} \ge  \left\| \tilde L_{1|T_a}^{(k)} +\tilde E_{1|T_a}^{(k)}\right\|_{HS}^2 = \| \tilde L_{1|T_a}^{(k)}\|_{HS} + \|\tilde E_{1|T_a}^{(k)}\|_{HS}^2 + 2 \trace{ \left(\tilde E_{1|T_a}^{(k)}\right)^\top\tilde L_{1|T_a}^{(k)}} \ge \\
    \frac{\trace{\tilde\Sigma_{1|T_a}^{(k)}}}{T_a} + o_p(1)
\end{multline}
Combining all the bounds and using the fact that $\frac{1}{n} \sum_{i\le n} D_i \omega_i = \sigma^2_d + o_p(1)$, we get the result. 
\end{proof}

\textbf{Proof of Proposition \ref{prop:example_2}:}
\begin{proof}
The proof follows the same steps for Proposition \ref{prop:example_2} and is omitted. 
\end{proof}

\subsection{Part II}\label{sec:ap_2}

\subsubsection{Balancing Bounds for Quadratic Problems}
For arbitrary matrices $\{L_k\}_{k=1}^K$ and vector $c$, consider 
\begin{equation}\label{eq:or_prim_prob}
\begin{aligned}
    x_0:= &\arg\min_{x}\left\{\sum_{k=1}^K\|L_kx\|_2^2 + \zeta^2 \| x\|_2^2\right\}\\
    \text{subject to: } & c^\top x = 1\\
\end{aligned}
\end{equation}
and let $V^2(\zeta^2)$ be the optimal value of this program. Our goal is to upper-bound $\|L_kx_0\|_2$, which can be interpreted as a measure of imbalance. A trivial bound is $\|L_kx_0\|_2\le V(\zeta^2)$, and our goal is to establish a better bound under additional conditions on matrices $L_k$.  

Consider a related program
\begin{equation}\label{eq:or_ridge_prob}
    \{\beta_{k,0}\}_{k=1}^K:= \arg\min_{\{\beta_k\}_{k=1}^K}\left\{\left\| c  - \sum_{k=1}^K L_{k}^\top\beta_k \right\|_2^2 + \zeta^2\sum_{k=1}^K\|\beta_k\|_2^2\right\}
\end{equation}
The next lemma describes the balancing properties of $x_0$ in terms of $\{\beta_{0,k}\}_{k=1}^K$ and $V^2(\zeta^2)$.

\begin{lemma}\label{lem:bal_bound}
Suppose $\|c\|_2 \ne 0$, then for any $k$ we have
\begin{equation}\label{eq: bal_bound}
    \|L_k x_0\|_2 = V^2(\zeta^2) \|\beta_{k,0}\|_2
\end{equation}
\end{lemma}
\begin{proof}
Using duality (constraint qualification holds because $\|c\| \ne 0$) we have 
\begin{multline*}
V^2(\zeta^2):=\min_{x: c^\top x = 1}\left\{\sum_{k=1}^K\|L_kx\|_2^2 + \zeta^2 \|x\|_2^2\right\} = \\
   \min_{x,\{t_k\}_{k=1}^K}\max_{\lambda_k \ge  0,\mu}\left\{\sum_{k=1}^K\lambda_k(\|L_{k}x\|_2 -t_k) +
   \sum_{k=1}^Kt_k^2 + \zeta^2 \| x\|_2^2 + \mu(1 - c^\top x)\right\} =\\
       \min_{x,\{t_k\}_{k=1}^K}\max_{\|\beta_k\|_2 \le 1,\lambda_k\ge 0, \mu}\left\{\sum_{k=1}^K\lambda_k(\beta_k^\top L_{k}x -t_k) + 
    \sum_{k=1}^Kt_k^2 + \zeta^2 \| x\|_2^2 + \mu(1 - c^\top x)\right\} = 
\end{multline*}
\begin{multline*}
  =\max_{\|\beta_k\|_2 \le 1,\lambda_k\ge 0,\mu}\min_{x,\{t_k\}_{k=1}^K}\left\{\sum_{k=1}^K\lambda_k(\beta_k^\top L_{k}x -t_k) +
    \sum_{k=1}^Kt_k^2 + \zeta^2 \| x\|_2^2 + \mu(1 - c^\top x)\right\} = \\
\max_{\beta_k,\mu}\left\{-\frac{\mu^2}{4\zeta^2}\left\| c^\top  - \sum_{k=1}^K\beta_k^\top L_{k} \right\|_2^2  - \frac{\mu^2}{4}\left(\sum_{k=1}^K\|\beta_k\|_2^2\right) + \mu
  \right\} =\\
\max_{\beta_k}\left\{\frac{\zeta^2}{\left\| c  - \sum_{k=1}^K L_{k}^\top\beta_k \right\|_2^2 + \zeta^2\left(\sum_{k=1}^K\|\beta_k\|_2^2\right)}
  \right\}
\end{multline*}
We can express $x_0$ in terms of the solution to the dual problem:
\begin{equation*}
    x_0 =\frac{\left(c - \sum_{k=1}^KL_k^\top\beta_{k,0}  \right)}{\left\| c - \sum_{k=1}^KL_k^\top\beta_{k,0} \right\|_2^2 + \zeta^2\left(\sum_{k=1}^K\|\beta_{k,0}\|_2^2\right)}
\end{equation*} 
Using the first-order conditions for the dual problem, we have the following for any $k$:
\begin{equation*}
     \left(c - \sum_{k=1}^KL_k^\top\beta_{k,0}\right)^\top L_k^\top = \zeta^2 \beta_{k,0}^\top\Rightarrow \|L_k x_0\|_2 = V^2(\zeta^2) \|\beta_{k,0}\|_2
\end{equation*}
where for the implication, we used the relationship between $x_0$ and $\beta_0$.
\end{proof}

By construction, the program (\ref{eq:or_ridge_prob}) is invariant to the left rotation of  $L_k$ (the $l_2$ norm of coefficients does not change). By virtue of the SVD decomposition, we can, without loss of generality, assume that each $L_k$ is a product of two matrices,
\begin{equation*}
L_k^\top = U_k D_k
\end{equation*}
where each $D_k$ is a diagonal matrix of size $p_k = \text{rank}(L_k)$, with positive values on the diagonal, and $ U_k^\top U_k = \mathcal{I}_{p_k}$. For a given $k$ let $s_k\in\mathbb{R}^{p_k}$ be a unit vector and define $U(s_k) := U_kD_k s_k$, $\sigma(s_k) := \|U(s_k)\|_2$, $u(s_k) := \frac{1}{\sigma(s_k)}U(s_k)$. Fix $k$ and observe that (\ref{eq:or_ridge_prob}) is equivalent to the following one:
\begin{equation*}
    \left(\{\beta_{0,l}\}_{l\ne k},s_{0,k}, \lambda_{0,k} \right) = \arg\min_{\{\beta_l\}_{l\ne k},s_k, \lambda_k}\left\|c -\sum_{l\ne k}L_l^\top \beta_l- u(s_k)\sigma(s_k)\lambda_k \right\|_2^2 + 
\zeta^2\left(\sum_{l\ne k}\|\beta_l\|_2^2 + \lambda_k^2\right)
\end{equation*}
where $\beta_{0,k} = \lambda s_{0,k}$ and $\|\beta_{0,k}\|_2 = |\lambda_{0,k}|$. For fixed $k$, $s_k$ and $\zeta^2$ define
\begin{equation*}
V^2(s_k, \zeta^2):=\min_{x: u(s_k)^\top x = 1} \left\{\sum_{l\ne k} \|L_kx\|_2^2 + \zeta^2\|x\|_2^2\right\}
\end{equation*}
and let  $x^{\star}(u(s_{0,k}))$ be the solution to this problem. The next lemma connects $\|\beta_{0,k}\|_2$ to $V^2(s_{0,k}, \zeta^2)$, $\zeta^2$, and $\sigma(s_{0,k})$.

\begin{lemma}\label{lem:coef_bound}
Suppose $\|c\|_2 \ne 0$, then the following bound is satisfied for all $k$ such that $p_k>0$:
\begin{equation}\label{eq:coef_bound}
    \|\beta_{0,k}\|_2  \le \|c\|_2\times \|x^{\star}(u(s_{0,k}))\|_2\times \frac{\sigma(s_{0,k})}{V^2(s_{0,k},\zeta^2) + \sigma^2(s_{0,k})}
\end{equation}
\end{lemma}
\begin{proof}
Fix $k$ with $p_k >0$ and stack matrices $\{L_l\}_{l\ne k}$ into a large matrix $L^\top_{-k}$, and define $p_{-k}:= \sum_{l\ne k} p_k$. Using an inversion for block matrices, we get the expression for $\lambda_{0,k}$:
\begin{equation*}
\begin{aligned}
    &\lambda_{0,k} = \frac{c^\top
    \tilde U(s_{0,k})}{U^\top(s_{0,k})\tilde U(s_{0,k}) +  \zeta^2},\\
    &\tilde  U(s_{0,k}) = U(s_{0,k}) - L_{-k}^{\top}(L_{-k}L_{-k}^\top + \zeta^2 \mathcal{I}_{p_{-k}})^{-1}L_{-k} U(s_{0,k})
\end{aligned}
\end{equation*}
Define $\tilde u(s_{0,k}):= \frac{1}{\sigma(s_{0,k})}\tilde U(s_{0,k})$. We have
\begin{equation*}
    \left(\tilde U(s_{0,k})\right)^\top U(s_{0,k}) =\sigma^2(s_{0,k}) \left(\|\tilde u(s_{0,k})\|_2^2 + \zeta^2 \sum_{l\ne k}\|\gamma_{0,l}(s_{0,k})\|_2^2\right) >0
\end{equation*}
where $\{\gamma_{0,l}(s_{0,k})\}_{l\ne k}$ is the solution for the optimization problem:
\begin{equation*}
    \{\gamma_{0,l}(s_{0,k})\}_{l\ne k} = \arg\max_{\{\gamma_l\}_{l\ne k}}\left\{\frac{\zeta^2}{\| u(s_{0,k}) - \sum_{l\ne k}L_l^\top\gamma_l\|_2^2 + \zeta^2 \sum_{l\ne k}\|\gamma_l\|_2^2}\right\}
\end{equation*}
By the same argument as in Lemma \ref{lem:bal_bound}, we have equality between two problems:
\begin{equation*}
   \max_{\{\gamma_l\}_{l\ne k}}\left\{\frac{\zeta^2}{\| u(s_{0,k}) - \sum_{l\ne k}L_l^\top\gamma_l\|_2^2 + \zeta^2 \sum_{l\ne k}\|\gamma_l\|_2^2}\right\} =
    \min_{x: u(s_{0,k})^\top x = 1} \left\{\sum_{l\ne k} \|L_k x\|_2^2 + \zeta^2\|x\|_2^2\right\}
\end{equation*}
Using Lemma \ref{lem:bal_bound}, we get $\|\tilde u(s_{0,k})\|_2 = \frac{\zeta^2\| x^{\star}(u(s_{0,k}))\|_2}{V^2(s_{0,k},\zeta^2)}$. Combining this result with definition of $\lambda_{0,k}$ we get the bound
\begin{equation*}
    |\lambda_{0,k}| \le \frac{\|c\|_2\sigma(s_{0,k}) \|\tilde u(s_{0,k})\|_2}{\zeta^2 + \sigma^2(s_{0,k})\frac{\zeta^2 }{V^2(s_{0,k},\zeta^2)}} \le \|c\|_2\times \| x^{\star}(u(s_{0,k}))\|_2 \times \frac{\sigma(s_{0,k})}{V^2(s_{0,k},\zeta^2) + \sigma^2(s_{0,k})}
\end{equation*}
where we used the CS inequality. Since $|\lambda_{0,k}| = \|\beta_k\|_2$ we get the result.
\end{proof}

The next corollary combines the bounds from Lemmas \ref{lem:bal_bound}, \ref{lem:coef_bound}.
\begin{corollary}\label{cor:final_bound}
Suppose $\|c\|_2 \ne 0$, $V^2(s_{0,k},\zeta^2)\lesssim \zeta^2$, $V^2(\zeta^2)\lesssim V^2(s_{0,k},\zeta^2)$, and $\|c\|_2\lesssim 1$. Then the following holds for all $k$ such that $p_k > 0$:
\begin{equation}\label{eq:final_bound}
    \|L_k x_0\|_2 \lesssim \min\left\{\frac{\zeta^2}{\sigma(s_{0,k})},\sigma(s_{0,k})\right\}
\end{equation}
\end{corollary}
\begin{proof}
Combining the bounds (\ref{eq: bal_bound}) and (\ref{eq:coef_bound}), we have the following for all $k$:
\begin{equation*}
     \|L_k x_0\|_2 \le V^2(\zeta^2) \|\beta_{k,0}\|_2 \lesssim \frac{V^2(s_{0,k},\zeta^2)\sigma(s_{0,k})}{V^2(s_{0,k},\zeta^2)+ \sigma^2(s_{0,k})} \lesssim \min\left\{\frac{\zeta^2}{\sigma(s_{0,k})},\sigma(s_{0,k})\right\}
\end{equation*}
\end{proof}

\subsubsection{Oracle Bound}
This section establishes a connection between a random and a deterministic optimization problem. Consider a $T\times n$ matrix $L$ with a special structure:
\begin{equation*}
\begin{aligned}
  L = H\Theta + \tilde L, \quad
 \mathbb{E}[H] = 0
\end{aligned}
\end{equation*}
where a $k\times n$ matrix $\Theta$ and $T\times n$ matrix $\tilde L$ are deterministic. Let $Z$ be a random $T\times p$ matrix, then define a random variable 
\begin{equation*}
    \mu^2= \frac{\min_{\Phi}\mathbb{E}_{H,Z}\|L - Z\Phi\|^2_{HS}}{ \min_{\Phi}\|L - Z\Phi\|^2_{HS}}
\end{equation*}
where $\Phi$ is a $p\times n$ matrix. Let $A\subseteq \mathbb{R}^n$ be a convex set, define solutions to two programs
\begin{equation}\label{eq:emp_or_problems}
\begin{aligned}
    &x_{1} :=\arg\min_{x \in A,\psi\in \mathbb{R}^p}\Big\{  \mu^2\| Lx-Z\psi\|_2^2 +\zeta^2 \| x\|_2^2\Big\}\\
    &x_{0}:= \arg\min_{x \in A,\psi\in \mathbb{R}^p}\Big\{ \mathbb{E}_{H,Z}\left[ \| Lx-Z\psi\|_2^2\right] +\zeta^2 \| x\|_2^2\Big\}
\end{aligned}
\end{equation}
and define $\delta := x_1 - x_0$. 

Define projection matrices $\Pi := Z\left(Z^\top Z\right)^{-1}Z^\top$ and $\Pi^{\bot} := \mathcal{I}_{T} - \Pi$. By construction, we have
\begin{equation*}
    \min_{\psi \in \mathbb{R}^p}\| Lx-Z\psi\|_2^2 = \|\Pi^{\bot} Lx\|_2^2,
\end{equation*}
and can re-express $x_1$ differently:
\begin{equation*}
    x_{1} :=\arg\min_{x \in A}\Big\{  \mu^2\| \Pi^{\bot}Lx\|_2^2 +\zeta^2 \| x\|_2^2\Big\}
\end{equation*}
Using the definition of $L$, we expand the expectation:
\begin{equation*}
     \mathbb{E}_{H,Z}\left[ \| Lx-Z\psi\|_2^2\right] = \|\tilde Lx\|_2^2 + \mathbb{E}_{H,Z}\left[\|H\Theta x - Z\psi\|_2^2\right]
\end{equation*}
The minimum value of the second part can be expressed differently:
\begin{equation*}
    \min_{\psi\in \mathbb{R}^p}\mathbb{E}_{H,Z}\left[\|H\Theta x - Z\psi\|_2^2\right] =  x^\top \Theta^\top \mathbb{E}_{H,Z}[(H -Z\Psi^{\star})^\top(H -Z\Psi^{\star})]\Theta x
\end{equation*}
where $\Psi^{\star} = \left(\mathbb{E}_{H,Z}[Z^{\top}Z]\right)^{-1}\mathbb{E}_{H,Z}[Z^\top H]$. Similarly for $\hat\Psi = \left(Z^{\top}Z\right)^{-1}\left(Z^\top H\right)$, we can express the empirical value:
\begin{equation*}
    \min_{\psi\in \mathbb{R}^p}\|H\Theta x - Z\psi\|_2^2 =  x^\top \Theta^\top (H -Z\hat\Psi)^\top(H -Z\hat\Psi)\Theta x
\end{equation*}
Define two matrices:
\begin{equation}\label{eq:k_def}
\begin{aligned}
    K:= \mathbb{E}[(H -Z\Psi^{\star})^\top(H -Z\Psi^{\star})], \quad \hat K:=(H -Z\hat \Psi)^\top(H -Z\hat\Psi)
\end{aligned}
\end{equation}
and suppose that a symmetric matrix $K$ is invertible. Define a relative distance between $\hat K$ and $K$:
\begin{equation*}
    E:= K^{-\frac12}(K - \hat K) K^{-\frac12} = \mathcal{I}_k - K^{-\frac12}\hat K K^{-\frac12}
\end{equation*}
and several quantities that govern the behavior of the bound later:
\begin{equation*}
    \begin{aligned}
    &\xi_1^2 := \frac{\|\Pi\tilde L \delta\|_2^2}{\|\delta\|_2^2},\quad
\xi_2 :=  \frac{|x_0^\top\tilde L^\top \Pi \tilde L \delta|}{\|\tilde L \delta\|_2 \|\tilde L x_0\|_2 },\quad
\xi_3 := \frac{ |x_0^\top \Theta^\top H \Pi^{\bot} \tilde L \delta|}{\|K^{\frac12}\Theta x_0\|_2\|\tilde L \delta\|_2},\\
&\xi_4 :=  \frac{|x_0^\top \tilde L^\top \Pi^{\bot} H^{\top}\Theta \delta|} {\|K^{\frac12}\Theta \delta\|_2\|\tilde L x_0\|_2},\quad
\xi_5 :=  \frac{ |x_0^\top \Theta^\top H^\top \Pi^{\bot} H \Theta\delta|}{\|K^{\frac12}\Theta x_0\|_2\|K^{\frac12} \Theta \delta\|_2}
\end{aligned}
\end{equation*}

Define a set $A_1$, on which three inequalities hold:
\begin{equation*}
     \| E\|_{op} \le \frac12, \quad \|\Pi^{\bot} L\delta\|_2^2 \ge \frac{1}{2}\left(\|\Pi^{\bot} H\Theta \delta\|_2^2 + \|\Pi^{\bot}\tilde L \delta\|_2^2\right), \quad |\mu^2-1| \le \frac14
\end{equation*}
In addition, define a set $A_2$, on which another inequality holds:
\begin{equation*}
    \zeta^2 \ge \xi_1^2
\end{equation*}
The next lemma provides a connection between two programs (\ref{eq:emp_or_problems}).
\begin{lemma}\label{lem:oracle_conection}
Suppose matrix $K$ is invertible, then on $A_1\cap A_2 $ we have the following bounds:
\begin{equation}
\begin{aligned}
    &\|K^\frac{1}{2}\Theta \delta\|_2 \lesssim (\|\hat E\|_{op}+\xi_3+|\mu^2-1|\xi_5)\|K^{\frac12}\Theta x_0\|_2 + (\xi_4 + \xi_2+|\mu^2-1|) \|\tilde L x_0\|_2, \\
    &\|\tilde L \delta\|_2 \lesssim (\|\hat E\|_{op}+\xi_3+|\mu^2-1|\xi_5)\|K^{\frac12}\Theta x_0\|_2 + (\xi_4 + \xi_2+|\mu^2-1|) \|\tilde L x_0\|_2, \\
    & \|\delta\|_2 \lesssim\frac{(\|\hat E\|_{op}+\xi_3+|\mu^2-1|\xi_5)\|K^{\frac12}\Theta x_0\|_2 + (\xi_4 + \xi_2+|\mu^2-1|) \|\tilde L x_0\|_2}{\zeta}
\end{aligned}
\end{equation}
\end{lemma}
\begin{proof}
Using first-order conditions for (\ref{eq:emp_or_problems}), we have two inequalities:
\begin{equation*}
    \begin{aligned}
    &\mu^2x_1^\top L^\top\Pi^{\bot} L\delta + \zeta^2 x_1^\top \delta \le 0,\quad 
     &x_0^\top K \delta + x_0^\top \tilde L^\top \tilde L \delta  + \zeta^2 x_0^\top \delta \ge 0.
    \end{aligned}
\end{equation*}
Combining these inequalities, we get:
\begin{multline*}
   \mu^2\|\Pi^{\bot} L\delta\|_2^2 + \zeta^2 \|\delta\|_2^2 + x_0^\top \Theta^\top (\hat K -K) \Theta^\top\delta - x_0^\top\tilde L^\top \Pi \tilde L \delta + 
   \mu^2x_0^\top \Theta^\top H \Pi^{\bot} \tilde L \delta + \mu^2x_0^\top \tilde L^\top \Pi^{\bot} H\Theta \delta + \\
   (\mu^2-1)x_0^\top \tilde L^\top\Pi^{\bot} \tilde L\delta + (\mu^2-1)x_0^\top \Theta^\top H^\top\Pi^{\bot}H \Theta\delta \le 0
\end{multline*}
By CS, we have 
\begin{equation*}
\begin{aligned}
            &\left|(\mu^2-1)x_0^\top \tilde L^\top\Pi^{\bot} \tilde L\delta\right| \le |\mu^2-1|\| \tilde Lx_0\|_2 \| \tilde L\delta\|_2\\
            &\left|(\mu^2-1)x_0^\top \Theta^\top H^\top\Pi^{\bot}H \Theta\delta\right| \le \xi_5|\mu^2-1|\|  K^\frac12\Theta x_0\|_2 \| K^\frac12\Theta\delta\|_2
\end{aligned}
\end{equation*}
By definition, using the fact that $K$ is invertible: 
\begin{equation*}
\begin{aligned}
&\|\Pi^{\bot}H\Theta \delta\|_2^2 = \left(\Theta \delta\right)^\top K \Theta \delta + \left(\Theta \delta\right)^\top K^\frac{1}{2}\hat E K^\frac{1}{2}\Theta \delta \ge \|K^\frac{1}{2}\Theta \delta\|_2^2(1 -\|\hat E\|_{op})\\
&x_0^\top \Theta^\top (\hat K -K) \Theta^\top\delta \ge - \|K^{\frac12}\Theta x_0\|_2 \|\hat E\|_{op}\|K^{\frac12}\Theta^\top\delta\|_2
\end{aligned}
\end{equation*}
By the properties of the projection matrix and the definition of $\xi_1^2$:
\begin{equation*}
    \|\Pi^{\bot}\tilde L \delta\|_2^2 = \|\tilde L \delta\|_2^2 -  \|\Pi\tilde L \delta\|_2^2 = \|\tilde L \delta\|_2^2 - \xi_1^2 \|\delta\|_2^2
\end{equation*}
Combining these results, the definitions of $\xi_2,\xi_3,\xi_4$, we have the following on $A_1\cap A_2$:
\begin{multline*}
    0 \ge \frac{3}{16}\|K^\frac{1}{2}\Theta \delta\|_2^2 + \frac{3}{8} \|\tilde L \delta\|_2^2  + \frac{3}{8}\zeta^2\|\delta\|_2^2 -
\left(\|K^{\frac12}\Theta x_0\|_2\left(\|\hat E\|_{op}+|\mu^2-1|\xi_5\right) + \xi_4 \|\tilde L x_0\|_2\right)\|K^{\frac12}\Theta^\top\delta\|_2 - \\
\left(\|K^{\frac12}\Theta x_0\|_2\xi_3 + (\xi_2 +|\mu^2-1|)\|\tilde L x_0\|_2\right)\|\tilde L \delta\|_2
\end{multline*}
This expression has the following form (for the appropriate $x_1,x_2, x_3, a_1,a_2$):
\begin{multline*}
  0\ge   x_1^2 + x_2^2 + x_3^2 - 2a_1x_1 - 2a_2 x_2 = (x_1 - a_1)^2 + (x_2 - a_2)^2 + x_3^2  -a_1^2 -a_2^2 \Rightarrow \\
  \begin{cases}
  x_1 \le a_1 + \sqrt{a_1^2 + a_2^2}\\
  x_2 \le a_2 +  \sqrt{a_1^2 + a_2^2}\\
  x_3 \le  \sqrt{a_1^2 + a_2^2}.
  \end{cases} \Rightarrow   \begin{cases}
  x_1 \lesssim a_1 + a_2\\
  x_2 \lesssim a_1 + a_2\\
  x_3 \lesssim a_1 + a_2
  \end{cases}
  \end{multline*}
Substituting $x_1,x_2,x_3$ and $a_1,a_2$ we get the result. 
\end{proof}

\subsection{Part III}\label{sec:ap_3}

Lemma \ref{lem:oracle_conection} and Corollary \ref{cor:final_bound} allow us to connect the empirical problem to a deterministic program for which we have a general bound. These results do not restrict the dimension of $H$ as long as their assumptions are satisfied. In particular, we need to guarantee that the hypothesis of Lemma \ref{lem:oracle_conection} holds with high probability and establish high-probability bounds on $\xi_1,\dots, \xi_5$. These guarantees can be established under different assumptions on $H$, and below we prove them for the one-dimensional and sub-Gaussian case. As long as the dimension of $H$ remains bounded, one can establish similar rates at the cost of more elaborate notation.

\subsubsection{One-dimensional sub-Gaussian Noise}
In this section, we analyze the bounds from the previous section under additional assumptions on $H$ and $Z$.
\begin{assumption}\label{as:gen_model_0}
Suppose $L = H\Theta + \tilde L$, where 
\begin{equation*}
\begin{aligned}
        &H = \Lambda_{h}\nu d_h, \quad
        Z = \Lambda_{z}\nu d_z, 
\end{aligned}
\end{equation*}
and $d_h = (1,0)$, $d_z = (\rho, \sqrt{1-\rho^2})$, $\nu$ is a $T\times 2$ matrix, and $\Lambda_h$, $\Lambda_z$ are $T\times T$ matrices.
\end{assumption}
Our goal is to lower-bound the probabilities of sets $A_1,A_2$ and bound $\xi_1^2, \xi_2,\xi_3,\xi_4,\xi_5$ under the tail assumptions on $\nu$, the restrictions on $\Lambda_h, \Lambda_z$, and $\rho$. All asymptotic statements in this section are with respect to $T$ going to infinity.

\begin{assumption}\label{as:gen_model_1}
Random variables $\nu_{tk} = (\nu)_{(t,k)}$ are i.i.d. across $t$ and $k$, and $\|\nu_{t,k}\|_{\psi^2} \lesssim \mathbb{E}[\nu_{t,k}^2]> 0$.
\end{assumption}
\begin{assumption}\label{as:gen_model_2}
$|\rho| <c_{\rho}<1$, $\|\Lambda_z\|_{op}\sim \|\Lambda_h\|_{op}$, $\|\Lambda_h\|_{HS}\sim \|\Lambda_z\|_{HS}$, $\|\Lambda_h\|_{op} = o\left(\|\Lambda_h\|_{HS}\right)$
\end{assumption}
The next lemma establishes properties of $K$ defined in (\ref{eq:k_def}).
\begin{lemma}\label{lem:invert_k}
Suppose Assumptions  \ref{as:gen_model_0}, \ref{as:gen_model_1}, and \ref{as:gen_model_2} hold, then $K = \left(1-\rho^2\frac{\left(\trace{\Lambda_h^\top\Lambda_z}\right)^2}{\|\Lambda_z\|_{HS}^2\|\Lambda_h\|_{HS}^2}\right)\mathbb{E}[\nu_{tk}^2]\|\Lambda_h\|_{HS}^2>0$.
\end{lemma}
\begin{proof}
Result follows from definition of $K$:
\begin{equation*}
    K = \mathbb{E}[\|H\|_2^2] - \frac{\mathbb{E}[H^\top Z]^2}{\left(\mathbb{E}[\|Z\|_2^2]\right)^2 }\mathbb{E}[\|Z\|_2^2] = \mathbb{E}[\|H\|_2^2]\left(1- \frac{\mathbb{E}[H^\top Z]^2}{\mathbb{E}[\|Z\|_2^2]\mathbb{E}[\|H\|_2^2]}\right),
\end{equation*}
and Assumptions \ref{as:gen_model_0}, \ref{as:gen_model_1}, and \ref{as:gen_model_2}.
\end{proof}

\begin{lemma}\label{lem:conc_ineq}
Suppose that Assumptions \ref{as:gen_model_0}, \ref{as:gen_model_1}, and \ref{as:gen_model_2} hold, and $\mathbb{E}[\nu_{tk}^2] = 1$. Suppose $\max\{x_1,x_3\} \lesssim  \frac{\|\Lambda_h\|_{HS}}{\|\Lambda_h\|_{op}}$, and $x_2 \lesssim  \frac{\|\Lambda_z\|_{HS}}{\|\Lambda_z\|_{op}}$. Then with probability at least $1-c\exp(-cx_1^2) - c\exp(-cx_2^2)-c\exp(-cx_3^2)$, we have
\begin{equation}
\begin{aligned}
&\frac{|\|H\|_2^2 - \|\Lambda_h\|_{HS}^2|}{\|\Lambda_h\|_{HS}^2} \le x_1\frac{\|\Lambda_h\|_{op}}{\|\Lambda_h\|_{HS}},\quad
\frac{|\|Z\|_2^2-\|\Lambda_z\|_{HS}^2|}{\|\Lambda_z\|_{HS}^2} \le x_2\frac{\|\Lambda_z\|_{op}}{\|\Lambda_z\|_{HS}},\\
 &\frac{|Z^\top H- \rho\trace{\Lambda_z^\top\Lambda_h}|}{\|\Lambda_h\|_{HS}^2} \le x_3(|\rho| + \sqrt{1-\rho^2}) \frac{\|\Lambda_h\|_{op}}{\|\Lambda_h\|_{HS}}.
\end{aligned}
\end{equation}
\end{lemma}
\begin{proof}
We focus only on the first inequality; the second follows in the same way. By Hanson-Wright inequality the inequality holds with probability at least 
\begin{equation*}
1- 2\exp\left(-c \min\left\{x_1\frac{\|\Lambda_h\|_{op}}{\|\Lambda_h\|_{HS}},x_1^2\left(\frac{\|\Lambda_h\|_{op}}{\|\Lambda_h\|_{HS}}\right)^2\right\}\frac{\|\Lambda_h\|_{HS}^2}{\|\Lambda_h\|_{op}^2}\right) = 1-2\exp(-cx_1^2)
\end{equation*}
where the second equality follows from Assumptions on $x_1$. To analyze the last quantity, we split it into two using Assumption \ref{as:gen_model_0}:
\begin{equation*}
|Z^\top H- \rho\trace{\Lambda_z^\top\Lambda_h}| = |\rho||  d_h\nu^\top\Lambda_z^\top \Lambda_h \nu d_h- \trace{\Lambda_z^\top\Lambda_h}| + \sqrt{1-\rho^2}| \nu_{(1)}^\top\Lambda_z^\top \Lambda_h \nu_{(2)} |
\end{equation*}
For the first quantity we can use Hanson-Wright inequality as before, utilizing Assumption \ref{as:gen_model_2}. For the second one, we can use the argument from \cite{vershynin2018high} Theorem 6.2.1 to make the same conclusion. 
\end{proof}
Define the empirical regression coefficient $\hat \psi = \frac{H^\top Z}{\|Z\|_2^2}$, and its population counterpart $\psi = \frac{\rho\trace{\Lambda_z^\top\Lambda_h}}{\|\Lambda_z\|_{HS}^2}$. The next corollary quantifies the error of $\hat \psi$.

\begin{corollary}\label{cor:cons_reg_coef}
Suppose Assumptions \ref{as:gen_model_0}, \ref{as:gen_model_1}, and \ref{as:gen_model_2} hold, then for any $x\lesssim \frac{\|\Lambda_z\|_{HS}}{\|\Lambda_z\|_{op}}$
\begin{equation}
    |\hat \psi - \psi| \le x \frac{\|\Lambda_z\|_{op}}{\|\Lambda_z\|_{HS}}
\end{equation}
holds with probability at least $1-c\exp(-cx^2)$. In particular, $\hat\psi$ is consistent.
\end{corollary}
\begin{proof}
By construction, $\hat \psi$ is scale-invariant with respect to $\nu$, so we can assume $\mathbb{E}[\nu_{tk}^2] = 1$. The result then follows from applying Lemma \ref{lem:conc_ineq} together with the following expansion:
\begin{equation*}
    \hat \psi - \psi = \frac{Z^\top H}{\|Z\|_2^2} -\psi =  \frac{1}{\|\Lambda_z\|_{HS}^2} 
    \frac{Z^\top H- \rho\trace{\Lambda_z^\top\Lambda_h}  + \frac{\rho\trace{\Lambda_z^\top\Lambda_h}(\|Z\|_2^2-\|\Lambda_z\|_{HS}^2) }{\|\Lambda_z\|_{HS}^2}}{1+\frac{\|Z\|_2^2-\|\Lambda_z\|_{HS}^2}{\|\Lambda_z\|_{HS}^2}}
\end{equation*}
\end{proof}
\begin{lemma}\label{lem:conc_ineq_2}
Suppose Assumptions \ref{as:gen_model_0}, \ref{as:gen_model_1}, and \ref{as:gen_model_2} hold, and $\mathbb{E}[\nu_{tk}^2] =1$. Then for $x_1,x_2 >0$ and a unit vector $u$ inequalities
\begin{equation}
    |H^\top u| \le x_1 \|\Lambda_h\|_{op}, \quad
     |Z^\top u| \le x_2 \|\Lambda_z\|_{op}
\end{equation}
hold with probability at least $1-2\exp(-cx_1^2) -2\exp(-cx_1^2)$.
\end{lemma}
\begin{proof}
We show the first inequality; the second follows in the same way. By concentration for independent sub-Gaussian random variables we have that the inequality holds at least with probability
\begin{equation}
    1- 2\exp\left( -c \frac{x_1^2\|\Lambda_h\|^2_{op}}{\|\Lambda_hu\|_2^2}\right)
\end{equation}
Since $\|\Lambda_hu\|_2^2 \le \|\Lambda_h\|_{op}^2$, the result follows.
\end{proof}
\begin{lemma}\label{lem:bound_4}
Suppose Assumptions \ref{as:gen_model_0}, \ref{as:gen_model_1}, and \ref{as:gen_model_2} hold, then $\mathbb{E}[\{A_1\}] \rightarrow 1$.
\end{lemma}
\begin{proof}
$E$ is scale-invariant, so we can assume that $\mathbb{E}[\nu_{tk}^2] = 1$. Let $\kappa:= \sqrt{(1-\rho^2)}\frac{|\trace{\Lambda_z^\top\Lambda_h)}|}{\|\Lambda_z\|_{HS},\|\Lambda_j\|_{HS}}<1$, by definition 
\begin{equation*}
    E = \frac{\hat K - K}{K}  = 
    \frac{\left(\|H\|_2^2 - \|\Lambda_h\|_{HS}^2\right) - \psi^2\left(\|Z\|_2^2-\|\Lambda_z\|_{HS}^2\right) + (\psi^2-\hat\psi^2)\|Z\|_2^2}{\kappa^2\|\Lambda_h\|_{HS}^2}
\end{equation*}
As a result, if the following inequalities hold, then $|E| \le \frac{3}{8}$:
\begin{equation*}
\begin{aligned}
 &|\|H\|_2^2 - \|\Lambda_h\|_{HS}^2| \le \frac{1}{8}\kappa^2\|\Lambda_h\|_{HS}^2,\\
 &|\|Z\|_2^2-\|\Lambda_z\|_{HS}^2| \le \min\left\{\frac{1}{8\max\{\psi^2,1\}}\kappa^2\|\Lambda_h\|_{HS}^2, \frac{1}{2}\|\Lambda_z\|_{HS}^2\right\}, \quad |\psi^2-\hat\psi^2| \le\frac{\kappa^2 \|\Lambda_{h}\|_{HS}^2}{12\|\Lambda_z\|_{HS}^2}
\end{aligned}
\end{equation*}
By Lemma \ref{lem:conc_ineq} and Corollary \ref{cor:cons_reg_coef}, these inequalities hold with probability approaching 1. 

For the second part of set $A_1$ we have
\begin{multline*}
     \|\Pi^{\bot} L\delta\|_2^2 -\frac{1}{2}\left(\|\Pi^{\bot} H\Theta \delta\|_2^2 + \|\Pi^{\bot}\tilde L \delta\|_2^2\right)  = \frac{1}{2}\|\Pi^{\bot} H\Theta \delta\|_2 +  \frac{1}{2}\|\Pi^{\bot} \tilde L\delta\|_2 + 2\delta^\top \Theta^\top H^\top \Pi^{\bot} \tilde L\delta  = \\
     \frac12\delta^\top \left(\Theta^\top H^\top \Pi^{\bot} H \Theta + \tilde L^\top\Pi^{\bot} \tilde L + 4 \Theta^\top H^\top \Pi^{\bot} \tilde L\right)\delta
\end{multline*}
To guarantee that this expression is nonnegative, we need the underlying matrix to be positive semi-definite: 
\begin{equation*}
    \Theta^\top H^\top \Pi^{\bot} H \Theta + \tilde L^\top\Pi^{\bot} \tilde L + 4 \Theta^\top H^\top \Pi^{\bot} \tilde L\ge 0.
\end{equation*}
By construction, it is enough to check this inequality on $u := \frac{\Theta^\top}{\|\Theta\|_2}$:
\begin{equation*}
     \|\Theta\|_2^2H^\top \Pi^{\bot} H + 4 \|\Theta\|_2H^\top \Pi^{\bot} \tilde L u+u\tilde L^\top\Pi^{\bot} \tilde Lu\ge 0
\end{equation*}
This inequality is satisfied as long as
\begin{equation*}
    \frac{|H^\top \Pi^{\bot} \tilde L u|}{\| \Pi^{\bot}H\|_2\|\Pi^{\bot} \tilde Lu\|_2}\le \frac12
\end{equation*}
which is validated by the following inequalities:
\begin{equation*}
\begin{aligned}
    &\| \Pi^{\bot}H\|_2 \ge \frac12 K^{\frac12}, \quad
    \|\Pi^{\bot} \tilde Lu\|_2 \ge \frac12 \|\tilde Lu\|_2, \quad
    |  H^\top\tilde L u| \le  K^{\frac12}\|\tilde Lu\|_2,\quad 
     |  Z^\top\tilde L u| \le \frac{K^{\frac12}\|\tilde Lu\|_2}{\max\{1,|\psi|\}},\\
      &|\hat \psi-\psi| \le 1
\end{aligned}
\end{equation*}
We have the following expansion:
\begin{equation*}
\begin{aligned}
            \| \Pi^{\bot}H\|^2_2 = \|H\|_2^2 - \hat \psi \|Z\|_2^2,\quad \|\Pi^{\bot} \tilde Lu\|_2^2 = \| \tilde L u\|_2^2 - \frac{(Z^\top \tilde Lu)^2}{\|Z\|_2^2}.
\end{aligned}
\end{equation*}
and thus the result follows from Lemmas \ref{lem:conc_ineq} and \ref{lem:conc_ineq_2}.

For the final part of set $A_1$, we expand $\mu^2$:
\begin{multline*}
    \mu^2 = \frac{\| \tilde L\|_{HS}^2 +K\|\Theta\|_2^2 }{\| \tilde L\|_{HS}^2 + \| \Theta\|_2^2\hat K + 2\Theta^\top\tilde L(H - \hat \psi \tilde Z) } \Rightarrow \\
    \mu^2-1 = \frac{(K-\hat K)\| \Theta\|_2^2 -  2  \Theta_k^\top\tilde L(\tilde H - \hat \psi \tilde Z) }{\| \tilde L\|_{HS}^2 +\|\Theta\|_2^2\hat K+ \| \Theta\|_2^2 (\hat K-K) + 2 \tilde \Theta^\top\tilde L(\tilde H - \hat \psi \tilde Z)} = \frac{e}{1+e}
\end{multline*}
Suppose that $|e| \le \frac{x}{2}$, where $x\le \frac{1}{2}$, then it follows $|\mu^2-1| \le x$. If the next inequalities hold, then $|e|<x$:
\begin{equation*}
    \frac{|\hat K-K|}{K} = |E|\le \frac{x}{2}, \quad \frac{| \Theta^\top\tilde L(\tilde H - \hat \psi \tilde Z)|}{\| \tilde L\|_{HS}^2 +\|\Theta\|_2^2 K} \le \frac{x}{2}
\end{equation*}
From Lemma \ref{lem:conc_ineq}, it follows that $|E| \le x \frac{\| \Lambda_{h}\|_{op}}{\| \Lambda_{h}\|_{HS}}$ with probability at least $1 - c\exp(-x^2)$ for $x \lesssim \frac{| \Lambda_{h}\|_{HS}}{\| \Lambda_{h}\|_{op}}$. By Lemma \ref{lem:conc_ineq_2} and Corollary \ref{cor:cons_reg_coef} next inequality holds with probability at least $1-c\exp(-cx^2)$.
\begin{equation*}
    |\Theta^\top\tilde L(H - \hat \psi Z)| \le x\sqrt{\mathbb{E}[\nu_{tk}^2]}\|\tilde L\|_{op} \| \Theta\|_2\| \Lambda_{h}\|_{op} \Rightarrow
\frac{|\Theta^\top\tilde L( H - \hat \psi Z|}{\| \tilde L\|_{HS}^2 +\| \Theta\|_2^2 K_b} \le x \frac{\|\tilde L\|_{op}\|\Lambda_{h}\|_{op}}{\|\tilde L\|_{HS} \| \Lambda_{h}\|_{HS}}
\end{equation*}
Using appropriate $x$, we get the result.
\end{proof}

\begin{corollary}
Suppose the conditions of Lemma \ref{lem:bound_4} hold, then with probability approaching 1, we have
\begin{equation}
    |\xi_3|\le \sqrt{\frac32}, |\xi_5| \le \frac32
\end{equation}
\end{corollary}
\begin{proof}
Using the results of the previous lemma, we have with probability approaching 1,
\begin{equation*}
    \frac{\|H\Pi^{\bot}\|^2_2}{K} = \frac{K + \hat K - K}{K} = 1 + E \le \frac32 \Rightarrow |\xi_3| = \frac{ |x_0^\top \Theta^\top H \Pi^{\bot} \tilde L \delta|}{\|K^{\frac12}\Theta x_0\|_2\|\tilde L \delta\|_2} \le \frac{\|H\Pi^{\bot}\|_2}{K^{\frac12}}
\end{equation*}
For $\xi_5$, we have
\begin{equation*}
    \frac{ |x_0^\top \Theta^\top H^\top \Pi^{\bot} H \Theta\delta|}{\|K^{\frac12}\Theta x_0\|_2\|K^{\frac12} \Theta \delta\|_2} =  \frac{\|H\Pi^{\bot}\|^2_2}{K}
\end{equation*}
and thus the same conclusion holds.
\end{proof}

\begin{corollary}\label{lem:var_rate}
Let $a_{T}$ be an arbitrary sequence such that $a_{T}\rightarrow \infty$ and $a_T \lesssim \frac{| \Lambda_{h}\|_{HS}}{\| \Lambda_{h}\|_{op}}$. Suppose Assumptions  \ref{as:gen_model_0} to \ref{as:gen_model_2} hold, then with probability approaching 1, we have 
\begin{equation}
    |\mu^2-1| \le a_T \frac{\|\Lambda_h\|_{op}}{\|\Lambda_h\|_{HS}}.
\end{equation}
\end{corollary}
\begin{proof}
Expanding from Lemma \ref{lem:bound_4}, we have $ |\mu^2-1| \le x\frac{\|\Lambda_h\|_{op}}{\|\Lambda_h\|_{HS}}$, with probability at least $1-c\exp(-cx^2)$. Using $x = a_T$, we get the result.
\end{proof}

\begin{lemma}\label{lem:bound_1}
Suppose Assumptions \ref{as:gen_model_0}, \ref{as:gen_model_1}, and \ref{as:gen_model_2} hold, and $\zeta^2 \ge \frac{\|\tilde L^\top \Lambda_z\|^2_{HS}}{\|\Lambda_z\|^2_{HS}}(2+a_T)^2$, where $a_T>0$ is an arbitrary sequence such that $\frac{a_T}{\|\tilde L^\top \Lambda_z\|_{op}}\rightarrow \infty$. Then with probability approaching 1, we have 
\begin{equation}
    \xi_1^2\le \zeta^2
\end{equation}
\end{lemma}
\begin{proof}
By definition of $\Pi$, we have
\begin{equation*}
     \frac{\|\Pi\tilde L \delta\|_2^2}{\|\delta\|_2^2} \le    \| \tilde L^\top\Pi\tilde L\|_{op} = \left(\frac{\|\tilde L^\top \Lambda_z\nu_z\|_2}{\|\Lambda_z \nu_z\|_2}\right)^2
\end{equation*}
where $\nu_z := \nu d_z$. This quantity is scale-invariant, so we can normalize $\mathbb{E}[\nu_{tk}^2] = 1$. We decompose the numerator and denominator
\begin{equation*}
   \frac{\|\tilde L^\top \Lambda_z\nu_z\|_2}{\|\Lambda_z \nu_z\|_2}-  \frac{\|\tilde L^\top \Lambda_z\|_{HS}}{\|\Lambda_z\|_{HS}}= \frac{\|\tilde L^\top \Lambda_z\|_{HS} + e_1}{\|\Lambda_z\|_{HS} +  e_2} - \frac{\|\tilde L^\top \Lambda_z\|_{HS}}{\|\Lambda_z\|_{HS}}= \frac{\|\tilde L^\top \Lambda_z\|_{HS}}{\|\Lambda_z\|_{HS}}(x+1)
\end{equation*}
as long as
\begin{equation*}
    |e_1| \le \frac{x}{2},\quad
    |e_2| \le \frac{\|\Lambda_z\|_{HS}}{2}.
\end{equation*}
By concentration for anisotropic vectors, these inequalities hold with probability at least
\begin{equation*}
    1 - 2\exp\left(-\frac{cx^2}{4\|\tilde L^\top \Lambda_z\|^2_{op}}\right) - 2\exp\left(-\frac{c\|\Lambda_z\|^2_{HS}}{\|\Lambda_z\|^2_{op}}\right)
\end{equation*}
and with the same probability
\begin{equation*}
 \xi_1^2 \le \frac{\|\tilde L^\top \Lambda_z\|^2_{HS}}{\|\Lambda_z\|^2_{HS}}(2+x)^2
\end{equation*}
The result follows by using $x =a_T$ and Assumption \ref{as:gen_model_2}.
\end{proof}
\begin{lemma}\label{lem:bound_2}
Suppose Assumptions \ref{as:gen_model_0}, \ref{as:gen_model_1}, and \ref{as:gen_model_2} hold and let $a_T>0$ be arbitrary sequence that converges to infinity. Then with probability approaching 1,
\begin{equation}
    \xi_2 \le a_T\frac{\|\Lambda_z\|_{op}}{\|\Lambda_{z}\|_{HS}}
\end{equation}
\end{lemma}
\begin{proof}
By construction, $\xi_2$ is scale-invariant, so we can assume that $\mathbb{E}[\nu_{tk}^2] = 1$. By CS inequality,
\begin{equation*}
    \frac{|x_0^\top\tilde L^\top \Pi \tilde L \delta|}{\|\tilde L \delta\|_2 \|\tilde L x_0\|_2 } = \frac{ |x_0^\top\tilde L^\top Z||Z^\top\tilde L \delta|}{\|Z\|_2^2 \|\tilde L \delta\|_2 \|\tilde L x_0\|_2 } \le \frac{|x_0^\top\tilde L^\top Z|}{\|Z\|_2 \|\tilde L x_0\|_2} \le \frac{x\|\Lambda_z\|_{op}}{\|\Lambda_z\|_{HS} }
\end{equation*}
as long as 
\begin{equation*}
    \left|\|Z\|_2 - \|\Lambda_z\|_{HS}\right| \le \frac12 \|\Lambda_z\|_{HS},\quad \frac{|x_0^\top\tilde L^\top Z|}{\|\tilde L x_0\|_2} \le \frac{x\|\Lambda_z\|_{op}}{2}
\end{equation*}
Using concentration properties of anisotropic sub-Gaussian vectors, we conclude that the inequalities hold with probability at least
\begin{equation*}
    1 - 2\exp(-cx^2) - 2\exp(-c\|\Lambda_{z}\|^2_{HS}/\|\Lambda_z\|^2_{op})
\end{equation*}
The result follows by using $x =a_T$ and Assumption \ref{as:gen_model_2}.
\end{proof}

\begin{lemma}\label{lem:bound_3}
Suppose Assumptions \ref{as:gen_model_0}, \ref{as:gen_model_1}, and \ref{as:gen_model_2} hold, and $a_T>0$ is an arbitrary diverging sequence. Then with probability approaching 1,
\begin{equation}
    \xi_4 \le a_T \frac{\|\Lambda_z\|_{op}}{\|\Lambda_h\|_{HS}}
\end{equation}
\end{lemma}
\begin{proof}
$\xi_4$ is scale-invariant, so we can assume $\mathbb{E}[\nu_{tk}^2] =1$. By definition of the projection matrix we have
\begin{equation*}
\begin{aligned}
  &|x_0^\top \tilde L^\top \Pi^{\bot} H| \le |x_0^\top \tilde L^\top H| + |\hat \psi-\psi| |x_0^\top \tilde L^\top Z| + |\psi| |x_0^\top \tilde L^\top Z|
\end{aligned}
\end{equation*}
and thus
$\xi_4 \le cx$, provided the following inequalities hold:
\begin{equation}
    \frac{|x_0^\top \tilde L^\top Z|}{\|\tilde L x_0\|_2} \le \frac{x}{3}\min\left\{1,\frac{1}{|\psi|}\right\}\|\Lambda_h\|_{HS}, \quad \frac{|x_0^\top \tilde L^\top H|}{\|\tilde L x_0\|_2} \le \frac{x}{3}\min\left\{1,\frac{1}{|\psi|}\right\}\|\Lambda_h\|_{HS}, \quad | \hat \psi - \psi| \le 1
\end{equation}
By Corollary \ref{cor:cons_reg_coef}, the last inequality hols with probability approaching 1, and from Lemma \ref{lem:conc_ineq_2} that for $x =z \frac{\max\{\|\Lambda_z\|_{op},\|\Lambda_z\|_{op}\}}{\|\Lambda_h\|_{HS}}$, the first two inequalities hold with probability at least $1- c \exp(- cz^2)$.
\end{proof}

We collect all these statements together in the next theorem.
\begin{theorem}\label{th:build_block}
Let $\{a_{1T},a_{2T}\}$ be arbitrary sequence such that $\frac{a_{1T}}{\|\tilde L^\top \Lambda_z\|_{op}}\rightarrow \infty$, and $a_{2T}\rightarrow \infty$, and $a_{2T}\lesssim\sqrt{T}$; suppose Assumptions \ref{as:gen_model_0}, \ref{as:gen_model_1}, \ref{as:gen_model_2} hold,
\begin{equation*}
\begin{aligned}
        \frac{\|\Lambda_z\|_{op}}{\|\Lambda_z\|_{HS}} \lesssim \frac{1}{\sqrt{T}},\quad
         \frac{\|\Lambda_h\|_{op}}{\|\Lambda_h\|_{HS}} \lesssim \frac{1}{\sqrt{T}},\quad
         \|\tilde L\|_{HS}\lesssim 1,\quad \zeta^2 = \frac{a_{1T}^2}{T}.
\end{aligned}
\end{equation*}
Then with probability approaching 1, we have:
\begin{equation}
\begin{aligned}
    &\max\{\|K^\frac{1}{2}\Theta \delta\|_2,\|\tilde L \delta\|_2\} \lesssim \|K^{\frac12}\Theta x_0\|_2 + \frac{a_{2T}}{\sqrt{T}} \|\tilde L x_0\|_2\\
    & \|\delta\|_2 \lesssim \frac{\sqrt{T}}{a_{1T}}\left(\|K^{\frac12}\Theta x_0\|_2 + \frac{a_{2T}}{\sqrt{T}}\|\tilde L x_0\|_2\right)
\end{aligned}
\end{equation}
\end{theorem}
\begin{proof}
The result follows by combining Lemma \ref{lem:oracle_conection} and the results in this section, and using inequality \begin{equation*}
    |AB|_{HS} \le \min\{\|A\|_{op}\|B\|_{HS},\|A\|_{HS}\|B\|_{op}\}
\end{equation*}
\end{proof}

\subsubsection{Analysis of the Estimator}
Define $Y_{1:T_0}, W_{1:T_0}, Z_{1:T_0}$---the part of the data that corresponds to periods $1,\dots, T_0$. 
Our estimator has the following form:
\begin{equation}
    \begin{aligned}
    \wrob = \arg\min_{\omega, \psi_0, \psi_1}&\left\{\frac{\left\| Y_{1:T_0}^\top \omega\frac{1}{n} - \psi_{0,y} - \psi_{1,y} Z_{1:T_0}\right\|_2^2}{T_0\hat \sigma_{y}^2} + \frac{\left\| W_{1:T_0}^\top \omega\frac{1}{n} -  \psi_{0,w} - \psi_{1,w}  Z_{1:T_0}\right\|_2^2}{T_0\hat \sigma_{w}^2} + \zeta^2_{n,T}\frac{\|\omega\|_2^2}{n}\right\}\\
    \text{ subject to: }& \frac{1}{n} D^\top \omega = 1,\quad \frac{1}{n} \mathbf{1}^{\top}\omega = 0
    \end{aligned}
\end{equation}
Define 
\begin{equation*}
    \begin{aligned}
      &\tilde D := \frac{1}{\sqrt{n}}\Pi^{\bot}_{l,f}D,\\
      &\tilde Y_{1:T_0} := \frac{1}{\sqrt{nT_0}\sigma_{y,T_0}}\Pi^{\bot}_{l,f} Y_{1:T_0}\left(\Pi^{f,r}_{1|T_0}\right)^\bot, \quad \tilde W_{1:T_0} := \frac{1}{\sqrt{nT_0}\sigma_{w,T_0}}\Pi^{\bot}_{l,f} W_{1:T_0}\left(\Pi^{f,r}_{1|T_0}\right)^\bot,\\
      &\tilde H_t = \frac{H_t - \mu_h}{\sqrt{T_0}}, \quad \tilde Z_t = \frac{Z_t - \mu_h}{\sqrt{T_0}}
    \end{aligned}
\end{equation*}
Define 
\begin{equation*}
    \begin{aligned}
    \wrobn = \arg\min_{\omega,\psi_1}&\left\{\mu^2_y\left\| \tilde Y_{1:T_0}^\top \omega - \psi_{1,y} \tilde Z_{1:T_0}\right\|_2^2 + \mu^2_w\left\| \tilde W_{1:T_0}^\top \omega - \psi_{1,w}  \tilde Z_{1:T_0}\right\|_2^2 + \zeta^2_{n,T}\|\omega\|_2^2\right\}\\
    \text{ subject to: }& \tilde D^\top \omega = 1
    \end{aligned}
\end{equation*}
where $\mu_k^2:= \frac{\sigma_{k}^2}{\hat \sigma_{k}^2}$ for $k \in \{y,w\}$. By construction, $\wrobn = \frac{1}{\sqrt{n}}\wrob$. We define the deterministic weights
\begin{equation}
    \begin{aligned}
    \wdetn_{T_0} = \arg\min_{\omega,\psi_1}&\, \left\{\mathbb{E}\left[\left\| \tilde Y_{1:T_0}^\top \omega - \psi_{1,y} \tilde Z_{1:T_0}\right\|_2^2 + \left\| \tilde W_{1:T_0}^\top \omega - \psi_{1,w}  \tilde Z_{1:T_0}\right\|_2^2\right] + \zeta^2_{n,T}\|\omega\|_2^2\right\}\\
    \text{ subject to: }& \tilde D^\top \omega = 1
    \end{aligned}
\end{equation}
and set $\tilde \delta := \wrobn - \wdetn_{T_0}$.

Under Assumption \ref{as:pot_out}, we have
\begin{equation*}
\begin{aligned}
Y_{1:T_0} = L_{y,b} + \Theta_yH_{1:T_0} + \Delta Z_{1:T_0}, \quad W_{1:T_0} = L_{w,b} + \Theta_wH_{1:T_0} + \Pi Z_{1:T_0}\\
\end{aligned}
\end{equation*}
where $(L_{y,1:T_0})_{it} = \alpha_{it}^{(y)} + \tau \alpha_{it}^{(w)}$, $( L_{w,1:T_0})_{it} =  \alpha_{it}^{(w)}$, $(\Theta_y)_i = \theta_i^{(y)} + \tau \theta_i^{(w)}$, $(\Theta_w)_i =\theta_i^{(w)}$, $(\Delta)_i = \tau \pi_i$, and $(\Pi)_{i} = \pi_i$. Without loss of generality, we can drop $ Z_{1:T_0}$ from these expressions (since we later project them out later), and get the expression for $\tilde Y_{1:T_0}$, $\tilde W_{1:T_0}$:
\begin{equation*}
\begin{aligned}
\tilde Y_{1:T_0} = \tilde L_{y,1:T_0} + \tilde\Theta_y \tilde H_{b}^\top\left(\Pi^{f,r}_{1|T_0}\right)^\bot, \quad \tilde W_{1:T_0} = \tilde L_{w,1:T_0} + \tilde\Theta_y \tilde H_{1:T_0}^\top\left(\Pi^{f,r}_{1|T_0}\right)^\bot
\end{aligned}
\end{equation*}
where for $k \in \{y,w\}$ $\tilde\Theta_k = \Pi^{\bot}_{l,f}\Theta_k$, and $ \tilde L_{k,1:T_0} = \frac{1}{\sqrt{nT_0}}\Pi^{\bot}_{l,f}L_{k,1:T_0}\left(\Pi^{f,r}_{1|T_0}\right)^\bot$.  
By Assumption \ref{as:des_model}, we have 
\begin{equation*}
\begin{aligned}
  &\left(\Pi^{f,r}_{1|T_0}\right)^\bot\tilde H_{1:T_0} =\left(\Pi^{f,r}_{1|T_0}\right)^\bot \Lambda_{h,b}\frac{\nu d_h}{\sqrt{T_0}} = \tilde \Lambda_{h,b}\tilde \nu d_h,\\
   &\left(\Pi^{f,r}_{1|T_0}\right)^\bot\tilde Z_{1:T_0} =\left(\Pi^{f,r}_{1|T_0}\right)^\bot \Lambda_{z,b}\frac{\nu d_z}{\sqrt{T_0}} = \tilde \Lambda_{z,b}\tilde \nu d_z
\end{aligned}
\end{equation*}

By Assumption \ref{as:des_model}, we have that Assumptions \ref{as:gen_model_0} to \ref{as:gen_model_2} hold for $\left(\Pi^{f,r}_{1|T_0}\right)^\bot\tilde H_{1:T_0}$ and $\left(\Pi^{f,r}_{1|T_0}\right)^\bot\tilde Z_{1:T_0}$. Define the size of the residual variation in $\tilde H_{1:T_0}$:
\begin{equation*}
\tilde K_{1:T_0} := \min_{\psi}\mathbb{E}\left[\|  \tilde H_{1:T_0} - \psi \tilde Z_{1:T_0}\|_2^2\right]
\end{equation*}
By Lemma \ref{lem:invert_k}, we have $0<\tilde K_{1:T_0}\lesssim 1$. Invoking Theorem \ref{th:build_block} we can conclude 
\begin{equation}\label{eq:main_emp_bound}
\begin{aligned}
  &\max_{k\in y,w}\{\|\tilde K_{1:T_0}^\frac{1}{2}\tilde \Theta_k \tilde \delta\|_2,\|\tilde L_{k,1:T_0} \tilde \delta\|_2\} \lesssim \max_{k\in y,w}\left\{\|\tilde K_{1:T_0}^{\frac12}\tilde \Theta_k \wdetn_{T_0}\|_2 + \frac{\log(T_0)}{\sqrt{T_0}} \|\tilde L_{k,1:T_0} \wdetn_{T_0}\|_2\right\}\\
    & \|\tilde \delta\|_2 \lesssim \frac{\sqrt{T_0}}{\log(T_0)}\max_{k\in y,w}\left\{\|\tilde K_{1:T_0}^{\frac12}\tilde\Theta_k \wdetn_{T_0}\|_2 + \frac{\log(T_0)}{\sqrt{T_0}}\|\tilde L_{k,b}\wdetn_{T_0}\|_2\right\}
 \end{aligned}
\end{equation}
as long as $\zeta^2 = \log(T_0)$.

We can express the problem for $\wdetn_{T_0}$ differently:
\begin{equation}\label{eq:oracle_problem_ap}
    \begin{aligned}
    \wdetn_{T_0} = \arg\min_{\omega}&\, \left\{\sum_{k\in y,w}\left[\left\| \tilde L_{k,1:T_0}^\top \omega\right\|_2^2 + K_{1:T_0}\left(\tilde \Theta_k \omega \right)^2\right] + \zeta^2_{n,T}\|\omega\|_2^2\right\}\\
    \text{ subject to: }& \tilde D^\top \omega = 1,
    \end{aligned}
\end{equation}
Let $V_{T_0}(\zeta^2_{n,T})$ be the value of this program. Assumption \ref{as:over} guarantees that $V_{T_0}(\zeta^2_{n,T}) \lesssim \frac{\log(n)}{n} + \zeta^2_{n,T}$, which under Assumption \ref{as:asym_regime} and $\zeta^2 = \log(T_0)$ implies 
\begin{equation*}
V_{T_0}(\zeta^2_{n,T}) \lesssim \zeta^2_{n,T}
\end{equation*}
It immediately follows that for $k \in \{y,w\}$ $K^{\frac12}_{1:T_0}\left|\tilde \Theta_k \wdetn_{T_0} \right| \lesssim \sqrt{\frac{\log(T_0)}{T_0}}$ and $\left\| \tilde L_{k,1:T_0}^\top\wdetn_{T_0}\right\|_2\lesssim \sqrt{\frac{\log(T_0)}{T_0}}$. Using (\ref{eq:main_emp_bound}), we can conclude that
\begin{equation}
    \max_{k\in \{y,w\}}\left|\mathbb{P}_n \wrob_i\theta_i^{k}\right| = O_p\left( \sqrt{\frac{\log(T_0)}{T_0}}\right)
\end{equation}

We define for $k\in \{y,w\}$:
\begin{equation*}
V^2_{k,T_0}(\zeta_{n,T}^2):=\min_{x: \tilde\Theta_k^\top x = \|\tilde\Theta_k\|_2} \left\{\|\tilde L_{y,1:T_0}x\|_2^2 + \|\tilde L_{w,1:T_0}x\|_2^2 + K_{1:T_0}(\tilde \Theta_{-k}x)^2 + \zeta^2_{n,T}\|x\|_2^2\right\}
\end{equation*}
Assumptions \ref{as:rich_het} implies
\begin{equation}
    V^2_{k,T_0}(\zeta_{n,T}^2) \lesssim \frac{\log(n)}{n} + \zeta_{n,T}^2
\end{equation}
Using Assumption \ref{as:asym_regime}, we conclude $V^2_{k,T_0}(\zeta_{n,T}^2) \lesssim \frac{\log(T_0)}{T_0}$. We can thus invoke Corollary \ref{cor:final_bound} and conclude that
\begin{equation}
    K^{\frac12}_{1:T_0}\left|\tilde \Theta_k \wdetn_{T_0} \right| \lesssim \min\left\{\frac{\log(T_0)}{T_0\|\tilde \Theta_{k}\|_2},\|\tilde \Theta_{k}\|_2\right\}
\end{equation}
Under Assumption \ref{as:asym_regime} $\|\tilde \Theta_{k}\|_2 \sim 1$ and we can conclude using (\ref{eq:main_emp_bound}) that
\begin{equation}
\begin{aligned}
        &\max_{k\in \{y,w\}}\left|\mathbb{P}_n \wrob_i\theta_i^{k}\right| = O_p\left( \frac{\log(T_0)}{T_0}\right),\\
        &\|\tilde \delta\|_2 = O_p\left(\sqrt{\frac{\log(T_0)}{T_0}}\right)
\end{aligned}
\end{equation}

We collect these statements in the following theorem.
\begin{theorem}\label{th:main_ap}
Suppose the conditions of Theorem \ref{th:cons} hold; then we have
\begin{equation}\label{eq:result_1}
     \max_{k\in \{y,w\}}\left|\mathbb{P}_n \wrob_i\theta_i^{k}\right| = O_p\left( \sqrt{\frac{\log(T_0)}{T_0}}\right), \quad \frac{\|\wrob\|_2}{\sqrt{n}} \lesssim 1
\end{equation}
If, in addition, conditions of Theorem \ref{th:limit_beh} hold, then we have
\begin{equation}\label{eq:result_2}
\begin{aligned}
        &\max_{k\in \{y,w\}}\left|\mathbb{P}_n \wrob_i\theta_i^{k}\right| = O_p\left( \frac{\log(T_0)}{T_0}\right),\\
        &\left\|\wrobn-\wdetn_{T_0}\right\|_2 = O_p\left(\sqrt{\frac{\log(T_0)}{T_0}}\right)
\end{aligned}
\end{equation}
\end{theorem}
It is easy to see that a similar result holds in the regime where $\|\tilde \Theta_{k}\|_2 \rightarrow 0$. The worst rate is achieved if $\|\tilde \Theta_{k}\|_2 \sim \sqrt{\frac{\log(T_0)}{T_0}}$, and in this case, there is no improvement in the rate from using our estimator compared to the standard TSLS. If $\|\tilde \Theta_{k}\|_2 \lesssim \sqrt{\frac{\log(T_0)}{T_0}}$, then our estimator performs similarly to the TSLS in terms of rate; otherwise it dominates it.

\section{Heterogeneous Treatment Effects}\label{ap:het_effects}
In this section, we sketch the argument for the convergence of $\omega_{i}^{rob}$ to limit the weights described in Section \ref{sec:het_ef}. It relies on the bounds established in \cite{OLS_skip}, and a formal proof can be completed by verifying the conditions of Theorem 1 in that paper. 

To describe the analysis under heterogeneous treatment effects, we impose additional structure on $D_i$ and $\alpha_{it}^{(k)}$ for $k\in \{y,w\}$:
\begin{equation}
\begin{aligned}
        &\alpha_{it}^{(k)} = (\gamma_i^{(k)})^\top \psi_t + \epsilon_{it}^{(k)},\\
        &D_i = \beta_0 + \beta^\top(\gamma_i^{(w)},\gamma_i^{(y)},\theta_i^{(w)},\theta_i^{(y)}) + \epsilon_i^{(d)}
\end{aligned}
\end{equation}
where $(\epsilon_{it}^{(y)},\epsilon_{it}^{(w)})$ and $\epsilon_i^{(d)}$ satisfy the conditions of Proposition \ref{prop:example}.

Using the dual for the oracle problem (\ref{eq:oracle_problem_ap}), we get that $\wdetn_{T_0}$ is proportional to the residual
\begin{equation}
    \wdetn_{T_0} \propto \left(\tilde D - \sum_{k\in \{y,w\}}\tilde L_{k,1:T_0} \tilde a_{1}^{(k)} - K^{\frac12}_{1:T_0}\tilde  {\Theta}_{k}\tilde a_{2}^{(k)}\right)
\end{equation}
where $(\tilde a_{1}^{(y)},\tilde a_{2}^{(y)},\tilde a_{1}^{(w)},\tilde a_{2}^{(w)})$ solves the optimization problem:
\begin{equation}
\min_{ \{a_1^{(k)}, a_2^{(k)}\}_{k\in \{y,w\}}}\left\| \tilde D - \sum_{k\in \{y,w\}}\tilde L_{k,1:T_0} a_{1}^{(k)} - K^{\frac12}_{1:T_0}\tilde  {\Theta}_{k}a_{2}^{(k)}\right\|_2^2 + \zeta^2_{n,T}\left(\sum_{k\in \{y,w\}}\|a_1^{(k)}\|_{2}^2 + (a_2^{(k)})^2\right)
\end{equation}
Next, consider the expected version of this problem:
\begin{equation}\label{eq:ya_oracle}
    \min_{ \{a_1^{(k)}, a_2^{(k)}\}_{k\in \{y,w\}}}\mathbb{E}\left[\left\| \tilde D - \sum_{k\in \{y,w\}}\tilde L_{k,1:T_0} a_{1}^{(k)} - K^{\frac12}_{1:T_0}\tilde  {\Theta}_{k}a_{2}^{(k)}\right\|_2^2\right] + \zeta^2_{n,T}\left(\sum_{k\in \{y,w\}}\|a_1^{(k)}\|_{2}^2 + (a_2^{(k)})^2\right)
\end{equation}
where the expectation is now with respect to the errors in $(\epsilon_{it}^{(y)},\epsilon_{it}^{(w)})$ and $\epsilon_i^{(d)}$, and define 
\begin{equation}
   \wlimn_{T_0} \propto \left(\tilde D - \sum_{k\in \{y,w\}}\tilde L_{k,1:T_0} \check a_{1}^{(k)} - K^{\frac12}_{1:T_0}\tilde  {\Theta}_{k}\check a_{2}^{(k)}\right)
\end{equation}
For $\zeta^2 = \log(T_0)$, the results in \cite{OLS_skip} guarantee  $\|\wlimn_{T_0} - \wdetn_{T_0}\|_{2} = o_p(1)$. Finally, as long as $\zeta^2_{n,T}$ converges to zero, the solution to (\ref{eq:ya_oracle}) itself converges to the solution of the unpenalized regression problem:
\begin{equation*}\label{eq:ols_oracle}
        \min_{ \{a_1^{(k)}, a_2^{(k)}\}_{k\in \{y,w\}}}\mathbb{E}\left[\left\| \tilde D - \sum_{k\in \{y,w\}}\tilde L_{k,1:T_0} a_{1}^{(k)} - K^{\frac12}_{1:T_0}\tilde  {\Theta}_{k}a_{2}^{(k)}\right\|_2^2\right]
\end{equation*}
and thus residuals are equal (up to proportionality) to $\epsilon_{i}{(d)}$. As a result, $\frac{\left\|\wrob-\frac{\epsilon_{i}^{(d)}}{\sigma^2_d}\right\|_2}{n} = o_p(1)$.

\section{Simulation Details}\label{ap:sim_details}

Our simulations are based on the following model:
\begin{equation}\label{eq:sim_dgp_ap}
\begin{aligned}
    &Y_{it} = \beta_i^{(y)} +\mu_t^{(y)}+ L_{it}^{(y)} + \tau W_{it} + \theta_i^{(y)}H_t + \epsilon_{it}^{(y)},\\
   & W_{it} = \beta_i^{(w)} +\mu_t^{(w)}+ L_{it}^{(w)} + \pi_i Z_{t} + \theta_i^{(w)}H_t + \epsilon_{it}^{(w)}.
\end{aligned}
\end{equation}
Here parameters $\{\beta_i^{(y)},\beta_i^{(w)},\mu_t^{(y)},\mu_t^{(w)},L_{it}^{(y)},L_{it}^{(w)},\tau, \pi_i, \theta_i^{(w)}, \theta_i^{(y)}\}_{i,t}$  are fixed, while $\epsilon_{it}^{(y)},\epsilon_{it}^{(w)}$, and $\{Z_t,H_t\}_{t\le T}$ are random.

We set the treatment effect equal to the original estimate $\tau = 1.43$; we estimate unit-level regressions by OLS:
\begin{equation}\label{eq:cleaning}
\begin{aligned}
    &\tilde Y_{it} = \alpha_i^{(y)} + \delta_i Z_t + \varepsilon^{(y)}_{it},\\
    &\tilde W_{it} = \alpha_i^{(w)} + \pi_i Z_t + \varepsilon^{(w)}_{it}
\end{aligned}
\end{equation}
and we use estimated $\hat \pi_i$ scaled by $\frac{\|L_{it}^{(w)}\|_{F}}{\|\hat\pi_i Z_t\|_{F}}=2.7$ in (\ref{eq:sim_dgp}).

For $k\in\{y,w\}$, let $E^{(k)}$ be the $n\times T$ matrix of residuals from (\ref{eq:cleaning}): $(E^{(k)})_{it} := \hat \varepsilon_{it}^{(k)}$. We construct $L^{(k)}_{it}$ by solving
\begin{equation}\label{eq:l_construction}
L^{(k)} := \argmin_{M,\text{rank}(M) = 13}\sum_{it}\left(E^{(k)}_{it}-M_{it}\right)^2
\end{equation}
which implies that $L^{(k)}$ simply sets all but the $13$ largest singular values of $E^{(k)}$ to zero. We use the residuals $E^{(k)} - L^{(k)}$ to construct the covariance matrix:
\begin{equation}
    \Sigma := \frac{1}{nT}\sum_{it}\begin{pmatrix}
    \left(E^{(y)}_{it} - L^{(y)}_{it}\right)^2 &\left(E^{(y)}_{it} - L^{(y)}_{it}\right)\left(E^{(w)}_{it} - L^{(w)}_{it}\right)\\
\left(E^{(y)}_{it} - L^{(y)}_{it}\right)\left(E^{(w)}_{it} - L^{(w)}_{it}\right) & \left(E^{(w)}_{it} - L^{(w)}_{it}\right)^2
    \end{pmatrix}
\end{equation}
and we generate $(\epsilon_{it}^{(y)},\epsilon_{it}^{(w)})$ from $\mathcal{N}(0,\Sigma)$. 

We estimate the model for $Z_t$ by fitting an ARIMA model to the data $\{Z_t\}_{t\le T}$  using the automatic model selection package in R, which delivers an MA(2) model with coefficients $(1.15,  0.53)$. We set $H_t$ to
\begin{equation}
    H_t = 0.5Z_t + \sqrt{1-0.25}\tilde Z_t
\end{equation}
where $\tilde Z_t$ has the same distribution as $Z_t$ and is independent of it. Exposures $\theta_i^{(w)}$ and $\theta_i^{(y)}$ are defined as
\begin{equation}
\begin{aligned}
    &\theta_i^{(w)} = 0.2\pi_i + \sqrt{1-0.2^2}\xi_i^{(w)},\\
    &\theta_i^{(y)} = 3\left(0.3\pi_i + \sqrt{1-0.3^2}\xi_i^{(y)}\right)
\end{aligned}
\end{equation}
where $\xi_i^{(w)}$ and $\xi_i^{(y)}$ are independent realizations of standard normal random variables.
\subsection{Alternative simulations}\label{ap:ad_sim}
We also report the results for the model where we simply use the original data as fixed, without simulating any additional errors $\epsilon_{it}^{(y)}, \epsilon_{it}^{(w)}$, and instead setting rank to $T$ in the problem \eqref{eq:l_construction}. We report these results in Table \ref{table:sim_res_alt}, and they largely agree with those in Table \ref{table:sim_res}. One obvious difference is that now the results for the first design do not have any errors in it because the two-way model holds exactly.

\begin{table}[t]
\begin{center}
\caption{Root-Mean-Square Error and Bias in Four Simulation Designs}\label{table:sim_res_alt}
\begin{tabular}{lcc|cc|cc|cc} \hline
& \multicolumn{2}{c}{(1)}& \multicolumn{2}{c}{(2)}& \multicolumn{2}{c}{(3)}& \multicolumn{2}{c}{(4)}\\ 
  &\multicolumn{2}{c}{Basic}& \multicolumn{2}{c}{GFE}& \multicolumn{2}{c}{Agg. Sh.}& \multicolumn{2}{c}{GFE+Agg. Sh.}\\ 
$\hat\pi_{rob}$ & 0.00 & 0.00 & 0.04 & -0.00 & 0.04 & 0.03 & 0.17 & 0.13 \\ 
  $\hat\pi_{TSLS}$ & 0.00 & 0.00 & 0.05 & -0.00 & 0.29 & 0.25 & 0.26 & 0.21 \\ 
    \hline
$\hat\delta_{rob}$  & 0.00 & -0.00 & 0.29 & 0.01 & 0.05 & 0.04 & 0.39 & 0.25 \\ 
$\hat\delta_{TSLS}$& 0.00 & -0.00 & 0.36 & 0.01 & 0.81 & 0.71 & 0.74 & 0.57 \\ 
  \hline
   $\hat \tau_{rob}$ & 0.00 & -0.00 & 0.35 & 0.02 & 0.02 & 0.00 & 0.32 & 0.07 \\ 
$\hat\tau_{TSLS}$ & 0.00 & -0.00 & 0.44 & 0.03 & 0.34 & 0.31 & 0.51 & 0.28 \\ 
   \hline \hline
\end{tabular}
\end{center}
\renewcommand{\baselinestretch}{0.7}
\footnotesize{\textit{Notes}: We performed $1000$ replications for each design. The true parameter value $\tau$ is set to $1.43$ to capture the \cite{nakamura2014fiscal} original estimate. Column (1)---first design: no generalized FE, no unobserved confounder. Column (2)---second design: generalized fixed effects, no unobserved confounder. Column (3)---third design: no generalized fixed effects, unobserved confounder. Column (4)---fourth design: generalized fixed effects, unobserved confounder.}
\end{table}

\end{document}